\newcommand*{\ATLASLATEXPATH}{}
\pdfoutput=1
\documentclass[cernpreprint, texlive=2016,UKenglish,titlesize=small]{\ATLASLATEXPATH atlasdoc}
\DeclareOldFontCommand{\rm}{\normalfont\rmfamily}{\mathrm}
\DeclareOldFontCommand{\sf}{\normalfont\sffamily}{\mathsf}
\DeclareOldFontCommand{\tt}{\normalfont\ttfamily}{\mathtt}
\DeclareOldFontCommand{\bf}{\normalfont\bfseries}{\mathbf}
\DeclareOldFontCommand{\it}{\normalfont\itshape}{\mathit}
\DeclareOldFontCommand{\sl}{\normalfont\slshape}{\@nomath\sl}
\DeclareOldFontCommand{\sc}{\normalfont\scshape}{\@nomath\sc}
\DeclareRobustCommand*\cal{\@fontswitch\relax\mathcal}
\DeclareRobustCommand*\mit{\@fontswitch\relax\mathnormal}

\usepackage{\ATLASLATEXPATH atlaspackage}
\usepackage{multirow}
\usepackage{bm}

\usepackage{dcolumn} 
\newcolumntype{d}[1]{D{.}{.}{#1}}
\newcolumntype{P}[1]{D{,}{\,\pm\,}{#1}}
\usepackage{\ATLASLATEXPATH atlasbiblatex}

\usepackage{\ATLASLATEXPATH atlascontribute}

\usepackage{\ATLASLATEXPATH atlasphysics}
\graphicspath{{}}

\usepackage{strong2l_paper-defs}

\newcommand{\zstar}{\ensuremath{Z^{(*)}}\xspace}

\AtlasTitle{Search for new phenomena using the invariant mass distribution of same-flavour opposite-sign dilepton pairs in events with missing transverse momentum in $\boldmath\sqrt{s}=13$~\TeV\ $\boldmath pp$ collisions with the ATLAS detector}

\author{The ATLAS Collaboration}

\AtlasRefCode{SUSY-2016-33}
\usepackage{longtable}
\PreprintIdNumber{CERN-EP-2018-053}

\AtlasJournalRef{Eur. Phys. J. C 78 (2018) 625}
\AtlasDOI{DOI:10.1140/epjc/s10052-018-6081-9}

\AtlasAbstract{%
A search for new phenomena in final states containing an $e^+e^-$ or $\mu^+\mu^-$ pair,
jets, and large missing transverse momentum is presented. 
This analysis makes use of proton--proton collision data with an integrated luminosity of \lumi, 
collected during 2015 and 2016 at a centre-of-mass energy $\sqrt{s}$ = 13~TeV with the ATLAS detector at the Large Hadron Collider.
The search targets the pair production of supersymmetric coloured particles (squarks or gluinos) and their decays into final states containing an $e^+e^-$ or $\mu^+\mu^-$ pair
and the lightest neutralino ($\tilde{\chi}_1^0$) via one of two 
next-to-lightest neutralino ($\tilde{\chi}_2^0$) decay mechanisms: $\tilde{\chi}_2^0 \rightarrow Z \tilde{\chi}_1^0$, where the $Z$ boson decays leptonically 
leading to a peak in the dilepton invariant mass distribution around the $Z$ boson mass; 
and $\tilde{\chi}_2^0 \rightarrow \ell^+\ell^- \tilde{\chi}_1^0$ with no intermediate $\ell^+\ell^-$ resonance,
yielding a kinematic endpoint in the dilepton invariant mass spectrum. 
The data are found to be consistent with the Standard Model expectation.
Results are interpreted using simplified models, and 
exclude gluinos and squarks with masses as large as~1.85~\TeV\ and 1.3~\TeV\ at 95\% confidence level, respectively.
}

\hypersetup{pdftitle={ATLAS document},pdfauthor={The ATLAS Collaboration}}

\begin{document}

\maketitle

\tableofcontents

\section{Introduction}
\label{sec:intro}
Supersymmetry (SUSY)~\cite{Golfand:1971iw,Volkov:1973ix,Wess:1974tw,Wess:1974jb,Ferrara:1974pu,Salam:1974ig}
is an extension to the Standard Model (SM) that introduces partner particles (called {\it sparticles}), 
which differ by half a unit of spin from their SM counterparts. 
For models with R-parity conservation~\cite{Farrar:1978xj}, strongly produced sparticles would be pair-produced and are expected to decay into quarks or gluons, 
sometimes leptons, and the lightest SUSY particle (LSP), which is stable. 
The LSP is assumed to be weakly interacting and thus is not detected, 
resulting in events with potentially large missing transverse momentum ($\boldsymbol{\mathit{p}}_{\text{T}}^{\text{miss}}$, with magnitude \met).
In such a scenario the LSP could be a dark-matter candidate~\cite{Goldberg:1983nd,Ellis:1983ew}.

For SUSY models to present a solution to the SM hierarchy problem~\cite{Sakai:1981gr,Dimopoulos:1981yj,Ibanez:1981yh,Dimopoulos:1981zb}, 
the partners of the gluons (gluinos, \gluino), 
top quarks (top squarks, $\tilde{t}_{\mathrm{L}}$ and $\tilde{t}_{\mathrm{R}}$) and Higgs bosons (higgsinos, $\tilde{h}$) should be close to the \TeV\ scale.
In this case, strongly interacting sparticles could be produced at a high enough rate to be detected by the experiments at the Large Hadron Collider (LHC). 

Final states containing same-flavour opposite-sign (SFOS) lepton pairs may arise from the cascade decays of squarks and gluinos via several mechanisms. 
Decays via intermediate neutralinos (${\tilde{\chi}}_{i}^{0}$), 
which are the mass eigenstates formed from the linear superpositions of higgsinos and the superpartners of the electroweak gauge bosons, 
can result in SFOS lepton pairs being produced in the decay $\tilde{\chi}_{2}^{0} \to \ell^{+}\ell^{-} \tilde{\chi}_{1}^{0}$. 
The index $i=1,\ldots,4$ orders the neutralinos according to their mass from the lightest to the heaviest.
In such a scenario the lightest neutralino, \chionezero, is the LSP\@.
The nature of the \chitwozero decay depends on the mass difference
$\Delta m_\chi \equiv m_{\tilde{\chi}_{2}^{0}} - m_{\tilde{\chi}_{1}^{0}}$, 
the composition of the charginos and neutralinos, and on whether there are additional sparticles
with masses less than $m_{\tilde{\chi}_{2}^{0}}$ that could be produced in the decay.
In the case where $\Delta m_\chi>m_Z$, SFOS lepton pairs may be produced in the decay 
$\tilde{\chi}_{2}^{0} \to Z \tilde{\chi}_{1}^{0} \to \ell^{+}\ell^{-} \tilde{\chi}_{1}^{0}$, 
resulting in a peak in the invariant mass distribution at $m_{\ell\ell} \approx m_Z$. 
For $\Delta m_\chi < m_Z$, the decay $\tilde{\chi}_{2}^{0} \to Z^* \tilde{\chi}_{1}^{0} \to \ell^{+}\ell^{-} \tilde{\chi}_{1}^{0}$
leads to a rising \mll\ distribution with a kinematic endpoint (a so-called ``edge''), 
the position of which is given by $m_{\ell\ell}^{\text{max}}=\Delta m_\chi < m_Z$, 
below the $Z$ boson mass peak.
In addition, if there are sleptons ($\tilde{\ell}$, the partner particles of the SM leptons) with masses less than $m_{\tilde{\chi}_{2}^{0}}$, 
the \chitwozero could follow the decay $\tilde{\chi}_{2}^{0} \to \tilde{\ell}^{\pm}\ell^{\mp} \to \ell^{+}\ell^{-} \tilde{\chi}_{1}^{0}$,
also leading to a kinematic endpoint, but with a different position given by 
$m_{\ell\ell}^{\mathrm{max}} = \sqrt{ (m^2_{\chitwozero}-m^2_{\tilde{\ell}})(m^2_{\tilde{\ell}}-m^2_{\chionezero}) / m^2_{\tilde{\ell}}}$. 
This may occur below, on, or above the $Z$ boson mass peak, depending on the value of the relevant sparticle masses.
In the two scenarios with a kinematic endpoint, if $\Delta m_\chi$ is small, production of leptons with low transverse momentum (\pt) is expected, motivating a search to specifically target low-\pt\ leptons.
Section~\ref{sec:susy} and Figure~\ref{fig:models} provide details of the signal models considered.

This paper reports on a search for SUSY, where either an on-$Z$ mass peak or an edge 
occurs in the invariant mass distribution of SFOS $ee$ and $\mu\mu$ lepton pairs.
The search is performed using \lumi\ of $pp$ collision data at $\sqrt{s}=13$~\TeV\ recorded during 2015 and 2016 by the ATLAS detector at the LHC. 
In order to cover compressed scenarios, i.e.\@ where $\Delta m_\chi$ is small, 
a dedicated ``low-\pt\ lepton search'' is performed in addition to the relatively ``high-\pt\ lepton searches'' in this channel, which have been  
performed previously by the CMS~\cite{Sirunyan:2017qaj} and ATLAS~\cite{SUSY-2016-05} collaborations.
Compared to the $14.7~\ifb$ ATLAS search~\cite{SUSY-2016-05}, 
this analysis extends the reach in $m_{\tilde g/\tilde q}$ by several hundred~\GeV\ and improves the sensitivity of the search into the compressed region.
Improvements are due to the optimisations for $\sqrt{s}=13$~\TeV\ collisions and to the addition of the low-\pt\ search, 
which lowers the lepton \pt\ threshold from $>25~\GeV$ to $>7~\GeV$.

\section{ATLAS detector}
\label{sec:detector}
The ATLAS detector~\cite{PERF-2007-01} is a general-purpose detector with almost $4\pi$ coverage in solid angle.\footnote{ATLAS uses a right-handed coordinate system with its origin at the nominal interaction point (IP) in the centre of the detector and the $z$-axis along the beam pipe. The $x$-axis points from the IP to the centre of the LHC ring, and the $y$-axis points upward. Cylindrical coordinates $(r,\phi)$ are used in the transverse plane, $\phi$ being the azimuthal angle around the $z$-axis. The pseudorapidity is defined in terms of the polar angle $\theta$ as $\eta=-\ln\tan(\theta/2)$ and the rapidity is defined as $y=1/2 \cdot\ln[(E+p_{\text{z}})/(E-p_{\text{z}})])$, where $E$ is the energy and $p_{\text{z}}$ the longitudinal momentum of the object of interest. The opening angle between two analysis objects in the detector is defined as $\Delta R=\sqrt{(\Delta y)^2+(\Delta \phi)^2}$.}
The detector comprises an inner tracking detector, a system of calorimeters, and a muon spectrometer.

The inner tracking detector (ID) is immersed in a 2~T magnetic field provided by a superconducting solenoid and allows charged-particle tracking out to $|\eta|=2.5$.
It includes silicon pixel and silicon microstrip tracking detectors inside a straw-tube tracking detector.
In 2015 a new innermost layer of silicon pixels was added to the detector and this improves tracking and $b$-tagging performance~\cite{ATLAS-TDR-19}.

High-granularity electromagnetic and hadronic calorimeters cover the region $|\eta|<4.9$.
All the electromagnetic calorimeters, as well as the endcap and forward hadronic calorimeters, 
are sampling calorimeters with liquid argon as the active medium and lead, copper, or tungsten as the absorber.
The central hadronic calorimeter is a sampling calorimeter with scintillator tiles as the active medium and steel as the absorber.

The muon spectrometer uses several detector technologies to provide precision tracking out to $|\eta|=2.7$ and triggering in $|\eta|<2.4$, 
making use of a system of three toroidal magnets.

The ATLAS detector has a two-level trigger system, with the first level implemented in custom hardware and the second level implemented in software.
This trigger system reduces the output rate to about 1~kHz from up to 40~MHz \cite{Aaboud:2016leb}.

\section{SUSY signal models}
\label{sec:susy}

SUSY-inspired simplified models are considered as signal scenarios for this analysis. 
In all of these models, squarks or gluinos are directly pair-pro\-duced, 
decaying via an intermediate neutralino, $\tilde{\chi}_2^0$, into the LSP ($\tilde{\chi}_1^0$).
All sparticles not directly involved in the decay chains considered are assigned very high masses, such that they are decoupled.
Three example decay topologies are shown in Figure~\ref{fig:models}.  
For all models with gluino pair production, 
a three-body decay for  $\tilde{g}\to q \bar{q} \tilde{\chi}_2^0$ is assumed.
Signal models are generated on a grid over a two-dimensional space, 
varying the gluino or squark mass and the mass of either the $\tilde{\chi}_2^0$ or the $\tilde{\chi}_1^0$.

The first model considered with gluino production,
illustrated on the left of Figure~\ref{fig:models}, is the so-called slepton model, 
which assumes that the sleptons are lighter than the $\tilde{\chi}_{2}^{0}$. 
The $\tilde{\chi}_{2}^{0}$ then decays either as $\tilde{\chi}_{2}^{0} \to \tilde{\ell}^{\mp}\ell^{\pm}; \tilde{\ell} \to \ell\tilde{\chi}_{1}^{0}$ 
or as $\tilde{\chi}_{2}^{0} \to \tilde{\nu}\nu; \tilde{\nu} \to \nu\tilde{\chi}_{1}^{0}$, the two decay channels having equal probability. 
In these decays, $\tilde{\ell}$ can be $\tilde{e}$, $\tilde{\mu}$ or $\tilde{\tau}$ and $\tilde{\nu}$ can be $\tilde{\nu}_e$, $\tilde{\nu}_\mu$ or $\tilde{\nu}_\tau$ with equal probability. 
The masses of the superpartners of the left-handed leptons are set to the average of the \chitwozero\ and \chionezero\ masses, 
while the superpartners of the right-handed leptons are decoupled. 
The three slepton flavours are taken to be mass-degenerate. 
The kinematic endpoint in the invariant mass distribution of the two final-state leptons in this decay chain can occur at any mass, 
highlighting the need to search over the full dilepton mass distribution. 
The endpoint feature of this decay topology provides a generic signature for many models of beyond-the-SM (BSM) physics.

In the \zstar\ model in the centre of Figure~\ref{fig:models} the $\tilde{\chi}_{2}^{0}$ from the gluino decay then decays as $\tilde{\chi}_{2}^{0} \to Z^{(*)}\tilde{\chi}_{1}^{0}$. 
In both the slepton and \zstar\ models, the $\tilde{g}$ and \chionezero\ masses are free parameters that are varied to produce the two-dimensional grid of signal models. 
For the gluino decays, $\tilde{g}\to q \bar{q} \tilde{\chi}_2^0$, both models have equal branching fractions for $q=u,d,c,s,b$. 
The \chitwozero\ mass is set to the average of the gluino and \chionezero\ masses.
The mass splittings are chosen to enhance the topological differences between these simplified models and other models with only one intermediate particle between the gluino and the LSP~\cite{wacker}.

Three additional models with decay topologies as illustrated in the middle and right diagrams of Figure~\ref{fig:models}, 
but with exclusively on-shell $Z$ bosons in the decay, are also considered.
For two of these models, the LSP mass is set to 1~\GeV, 
inspired by SUSY scenarios with a low-mass LSP (e.g. generalised gauge mediation~\cite{Dine:1981gu,AlvarezGaume:1981wy,Nappi:1982hm}).
Sparticle mass points are generated across the $\tilde{g}$--$\tilde{\chi}_2^0$ (or $\tilde{q}$--$\tilde{\chi}_2^0$) plane. 
These two models are referred to here as the $\tilde{g}$--$\tilde{\chi}_2^0$ on-shell and $\tilde{q}$--$\tilde{\chi}_2^0$ on-shell models, respectively.
The third model is based on topologies that could be realised in the 19-parameter phenomenological supersymmetric Standard Model (pMSSM)~\cite{Berger:2008cq,Djouadi:1998di}
with potential LSP masses of 100~\GeV\ or more.
In this case the $\tilde{\chi}_2^0$ mass is chosen to be 100~\GeV\ above the $\tilde{\chi}_1^0$ mass,
which can maximise the branching fraction to $Z$ bosons.
Sparticle mass points are generated across the $\tilde{g}$--$\tilde{\chi}_1^0$ plane, and this model is thus referred to as the $\tilde{g}$--$\tilde{\chi}_1^0$ on-shell model. 
For the two models with gluino pair production, the branching fractions for $q=u,d,c,s$ are each 25\%.   
For the model involving squark pair production, the super-partners of the $u$-, $d$-, $c$- and $s$-quarks have the same mass, 
with the super-partners of the $b$- and $t$-quarks being decoupled. 
A summary of all signal models considered in this analysis can be found in Table~\ref{tab:models}.

\begin{figure*}[htb]
\centering
\includegraphics[width=0.27\textwidth]{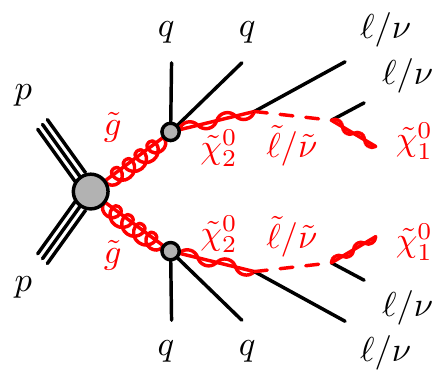}
\hspace{1cm}
\includegraphics[width=0.27\textwidth]{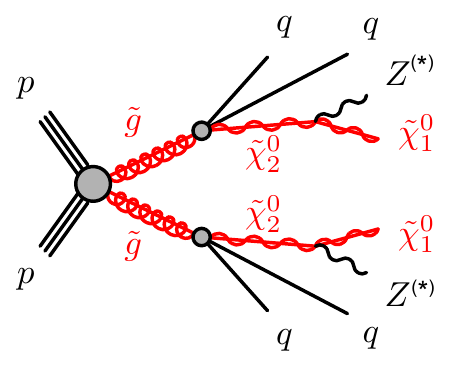}
\hspace{1cm}
\includegraphics[width=0.27\textwidth]{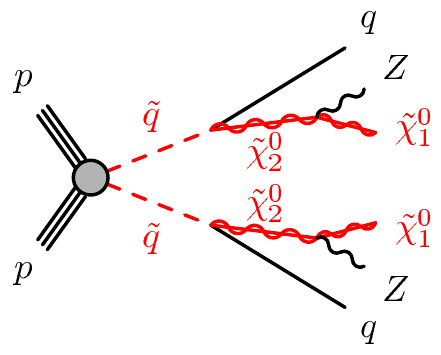}
\caption{Example decay topologies for three of the simplified models considered. 
The left two decay topologies involve gluino pair production, 
with the gluinos following an effective three-body decay for $\tilde{g}\to q \bar{q} \tilde{\chi}_2^0$, 
with $\tilde{\chi}_{2}^{0} \to \tilde{\ell}^{\mp}\ell^{\pm} / \tilde{\nu}\nu$ for the ``slepton model'' (left) and $\tilde{\chi}_2^0\rightarrow \zstar \tilde{\chi}_1^0$ in the \zstar, $\tilde{g}$--$\tilde{\chi}_2^0$ or $\tilde{g}$--$\tilde{\chi}_1^0$ model (middle).  
The diagram on the right illustrates the $\tilde{q}$--$\tilde{\chi}_2^0$ on-shell model, where squarks are pair-produced, 
followed by the decay $\tilde{q}\to q \tilde{\chi}_2^0$, with $\tilde{\chi}_2^0\rightarrow Z \tilde{\chi}_1^0$.}
\label{fig:models}
\end{figure*}

\begin{table}[htbp!]
\caption{
Summary of the simplified signal model topologies used in this paper.
Here $x$ and $y$ denote the $x$--$y$ plane across which the signal model masses are varied to construct the signal grid.
For the slepton model, the masses of the superpartners of the left-handed leptons are given by $[m(\chitwozero)+m(\chionezero)]/2$,
while the superpartners of the right-handed leptons are decoupled.
}
\begin{center}
    \begin{tabular}{lcccccc}
      \hline
      Model                               & Production mode        & Quark flavours            & $m(\tilde{g})/m(\tilde{q})$ & $m(\tilde{\chi}^{0}_{2})$  & $m(\tilde{\chi}^{0}_{1})$ \\
      \hline\hline
      slepton                             & $\tilde{g}\tilde{g}$   &  $u$, $d$, $c$, $s$, $b$  & $x$                         & $[m(\tilde{g})+m(\chionezero)]/2$        & $y$           \\
      \zstar\                             & $\tilde{g}\tilde{g}$   &  $u$, $d$, $c$, $s$, $b$  & $x$                         & $[m(\tilde{g})+m(\chionezero)]/2$        & $y$           \\
      $\tilde{g}$--$\chitwozero$ on-shell & $\tilde{g}\tilde{g}$   &  $u$, $d$, $c$, $s$       & $x$                         & $y$                              & 1~\GeV        \\            
      $\tilde{q}$--$\chitwozero$ on-shell & $\tilde{q}\tilde{q}$   &  $u$, $d$, $c$, $s$       & $x$                         & $y$                              & 1~\GeV        \\
      $\tilde{g}$--$\chionezero$ on-shell & $\tilde{g}\tilde{g}$   &  $u$, $d$, $c$, $s$       & $x$                         & $m(\tilde{\chi}^0_1)+100$~\GeV   & $y$           \\
      \hline
\end{tabular}
\end{center}
\label{tab:models}
\end{table}

\section{Data and simulated event samples}
\label{sec:data}
The data used in this analysis were collected by ATLAS during 2015 and 2016, 
with a mean number of additional $pp$ interactions per bunch crossing ({\it pile-up}) of approximately 14 in 2015 and 25 in 2016,
and a centre-of-mass collision energy of 13~\TeV.  
After imposing requirements based on beam and detector conditions and data quality, the data set corresponds to an integrated luminosity of \lumi.
The uncertainty in the combined 2015 and 2016 integrated luminosity is $\pm2.1$\%. Following a methodology similar to that detailed in Ref.~\cite{DAPR-2013-01}, 
it is derived from a calibration of the luminosity scale using $x$--$y$ beam-separation scans performed in August 2015 and May 2016. 

For the high-\pt\ analysis, data events were collected using single-lepton and dilepton triggers~\cite{Aaboud:2016leb}. 
The dielectron, dimuon, and electron--muon triggers have $\pt$ thresholds in the range 12--24~\GeV\ for the higher-$\pt$ lepton.
Additional single-electron (single-muon) triggers are used, with $\pt$ thresholds of 60 (50)~\GeV, to increase the trigger efficiency for events with high-\pt leptons.
Events for the high-\pT\ selection are required to contain at least two selected leptons with $\pt>25$~\GeV.
This selection is fully efficient relative to the lepton triggers with the \pt\ thresholds described above.

For the low-\pT\ analysis, 
triggers based on \MET\ are used in order to increase efficiency for events where the \pt\ of the leptons is too low for the event to be selected by the single-lepton or dilepton triggers.
The \MET\ trigger thresholds varied throughout data-taking during 2015 and 2016, with the most stringent being 110~\GeV.
Events are required to have $\MET>200~\GeV$, making the selection fully efficient relative to the \MET\ triggers with those thresholds.

An additional control sample of events containing photons was collected using a set of single-photon triggers with
\pt thresholds in the range 45--140~\GeV. 
All photon triggers, except for the one with threshold $\pt>120$~\GeV\ in 2015, or the one with $\pt>140$~\GeV\ in 2016, were prescaled.
This means that only a subset of events satisfying the trigger requirements were retained.
Selected events are further required to contain a selected photon with $\pt>50$~\GeV.

\label{sec:mc}
Simulated event samples are used to aid in the estimation of SM backgrounds, validate the analysis techniques, optimise the event selection, and provide predictions for SUSY signal processes.  
All SM background samples 
used are listed in Table~\ref{tab:MC}, 
along with the parton distribution function (PDF) set, the configuration of underlying-event and hadronisation parameters (underlying-event tune) and the cross-section calculation order in \alphas\ used to normalise the event yields for these samples. 
      
The $\ttbar+W$, $\ttbar+Z$, and $\ttbar+WW$ processes were generated at leading order (LO) in \alphas\ with the {\sc NNPDF2.3LO} PDF set~\cite{Ball:2012cx} 
using {\sc MG5\_aMC@NLO} v2.2.2~\cite{Alwall:2014hca}, 
interfaced with {\sc Pythia} 8.186~\cite{Sjostrand:2007gs} with the {\sc A14} underlying-event tune~\cite{ATL-PHYS-PUB-2014-021} to simulate the parton shower and hadronisation.
Single-top and \ttbar\ samples were generated using {\sc Powheg Box v2}~\cite{PowhegBOX1,PowhegBOX2,PowhegBOX3} with
{\sc Pythia} 6.428~\cite{Sjostrand:2006za} used to simulate the parton shower, hadronisation, and the underlying event. 
The {\sc CT10} PDF set \cite{Lai:2010vv} was used for the matrix element, and the {\sc CTEQ6L1} PDF set with corresponding {\sc Perugia2012}~\cite{perugiatunes} tune for the parton shower. 
In the case of both the {\sc MG5\_aMC@NLO} and {\sc Powheg} samples, 
the {\sc EvtGen} v1.2.0 program~\cite{EvtGen} was used for properties of the bottom and charm hadron decays. 
Diboson and \dyjets\ processes were simulated using the \sherpa\ $2.2.1$ event generator. 
Matrix elements were calculated using Comix~\cite{Gleisberg:2008fv} and OpenLoops~\cite{Cascioli:2011va} 
and merged with {\sc Sherpa}'s own internal parton shower~\cite{Schumann:2007mg} using the ME+PS@NLO prescription~\cite{Hoeche:2012yf}.
The {\sc NNPDF3.0nnlo}~\cite{Ball:2014uwa} PDF set is used in conjunction with dedicated parton shower tuning developed by the Sherpa authors.
For Monte Carlo (MC) closure studies of the data-driven \dyjets estimate (described in Section~\ref{sec:zjets}), 
\gjets\ events were generated at LO with up to four additional partons using \sherpa\ $2.1$, 
and are compared with a sample of \dyjets\ events with up to two additional partons at NLO (next-to-leading order) and up to four at LO generated using \sherpa\ $2.1$.
Additional MC simulation samples of events with a leptonically decaying vector boson and photon ($V\gamma$, where $V=W,Z$) were generated at LO using \sherpa\ $2.2.1$. 
Matrix elements including all diagrams with three electroweak couplings were calculated with up to three partons.
These samples are used to estimate backgrounds with real \met\ in $\gjets$ data samples.

The SUSY signal samples were produced at LO using {\sc MG5\_aMC@NLO} with the {\sc NNPDF2.3LO} PDF set, interfaced with {\sc Pythia} 8.186.
The scale parameter for CKKW-L matching~\cite{CKKW,CKKWL} was set at a quarter of the mass of the gluino.
Up to one additional parton is included in the matrix element calculation. 
The underlying event was modelled using the {\sc A14} tune for all signal samples, and {\sc EvtGen} was adopted to describe the properties of bottom and charm hadron decays.
Signal cross-sections were calculated at NLO in \alphas, including resummation of soft gluon emission at next-to-leading-logarithmic accuracy (NLO+NLL)~\cite{Beenakker:1996ch,Kulesza:2008jb,Kulesza:2009kq,Beenakker:2009ha,Beenakker:2011fu}. 

All of the SM background MC samples were passed through a full ATLAS detector simulation~\cite{:2010wqa} using $\textsc{Geant}$4~\cite{Agostinelli:2002hh}. 
A fast simulation~\cite{:2010wqa}, in which a parameterisation of the response of the ATLAS electromagnetic and hadronic calorimeters is combined with $\textsc{Geant}$4 elsewhere,
was used in the case of signal MC samples. 
This fast simulation was validated by comparing a few signal samples to some fully simulated points.

Minimum-bias interactions were generated and overlaid on top of the hard-scattering process to simulate the effect of multiple $pp$ interactions occurring during the same (in-time) or a nearby (out-of-time) bunch-crossing.
These were produced using {\sc Pythia} 8.186 with the {\sc A2} tune~\cite{ATLAS:2012uec} and {\sc MSTW 2008} PDF set~\cite{Martin:2009iq}.
The MC simulation samples were reweighted such that the distribution of the average number of interactions per bunch crossing matches the one observed
in data.

\begin{table*}[ht]
\begin{center}
\caption{Simulated background event samples used in this analysis with the corresponding matrix element and parton shower generators,
cross-section order in \alphas\ used to normalise the event yield, underlying-event tune and PDF set.  
}
\scriptsize
\begin{tabular}{l c c c c c }
\hline
Physics process &  Generator  & Parton & Cross-section & Tune & PDF set\\
                &             & Shower &              &      & \\
\noalign{\smallskip}\hline\noalign{\smallskip}
$t\bar{t}+W$ and $t\bar{t}+Z$~\cite{ATL-PHYS-PUB-2016-005,Garzelli:2012bn}& {\sc MG5\_aMC@NLO}        & {\sc Pythia} 8.186 & NLO \cite{Campbell:2012,Lazopoulos:2008} & {\sc A14} & NNPDF2.3LO\\
$t\bar{t}+WW$~\cite{ATL-PHYS-PUB-2016-005}      & {\sc MG5\_aMC@NLO}          & {\sc Pythia} 8.186 & LO \cite{Alwall:2014hca} & {\sc A14}  &  NNPDF2.3LO\\
$t\bar{t}$~\cite{ATL-PHYS-PUB-2016-004}         & {\sc Powheg Box v2} r3026   & {\sc Pythia} 6.428 & NNLO+NNLL \cite{ttbarxsec1,ttbarxsec2}          &\sc{Perugia2012}     &NLO CT10\\
Single-top ($Wt$)~\cite{ATL-PHYS-PUB-2016-004}  & {\sc Powheg Box v2} r2856   & {\sc Pythia} 6.428 & Approx. NNLO \cite{Kidonakis:2010b}& \sc{Perugia2012}    &NLO CT10\\ 
$WW$, $WZ$ and $ZZ$~\cite{ATL-PHYS-PUB-2016-002} & \sherpa\ 2.2.1 & \sherpa\ 2.2.1 & NLO \cite{diboson1,diboson2} & \sherpa\ default & {\sc NNPDF3.0nnlo} \\
$Z/\gamma^{*}(\rightarrow \ell \ell)$ + jets~\cite{ATL-PHYS-PUB-2016-003}& \sherpa\ 2.2.1           & \sherpa\ 2.2.1  &NNLO \cite{DYNNLO1,DYNNLO2}       & \sherpa\ default     &{\sc NNPDF3.0nnlo}\\
\gjets & \sherpa\ 2.1.1 & \sherpa\ 2.1.1 & LO~\cite{sherpa} & \sherpa\ default & NLO CT10 \\
$V(=W,Z)\gamma$ & \sherpa\ 2.1.1 & \sherpa\ 2.1.1 & LO~\cite{sherpa} & \sherpa\ default & NLO CT10 \\
\noalign{\smallskip}\hline\noalign{\smallskip}
\end{tabular}
\label{tab:MC}
\end{center}
\end{table*}

\section{Object identification and selection}
\label{sec:objects}
Jets and leptons selected for analysis are categorised as either ``baseline'' or ``signal'' objects according to various quality and kinematic requirements.
Baseline objects are used in the calculation of missing transverse momentum, and to resolve ambiguity between the analysis objects in the event,
while the jets and leptons used to categorise the event in the final analysis selection must pass more stringent signal requirements.

Electron candidates are reconstructed using energy clusters in the electromagnetic calorimeter matched to ID tracks.
Baseline electrons are required to have $\pt>10$~\GeV\ ($\pt>7$~\GeV) 
in the case of the high-\pt\ (low-\pt) lepton selection. 
These must also satisfy the ``loose likelihood'' criteria described in Ref.~\cite{ATLAS-CONF-2016-024} and reside within the region $|\eta|=2.47$.
Signal electrons are required to satisfy the ``medium likelihood'' criteria of Ref.~\cite{ATLAS-CONF-2016-024}, and those entering the high-\pt\ selection are further required to have $\pt>25$~\GeV.
Signal-electron tracks must pass within $|z_0\sin\theta| = 0.5$~mm of the primary vertex\footnote{The primary vertex in each event is defined as the reconstructed vertex~\cite{ATL-PHYS-PUB-2015-026} with the highest $\sum p_{\text{T}}^2$,
where the summation includes all particle tracks with $\pt>400$~\MeV\ associated to the vertex.}, where $z_0$ is the longitudinal impact parameter with respect to the primary vertex.  
The transverse-plane distance of closest approach of the electron to the beamline, divided by the corresponding uncertainty, must be $|d_0/\sigma_{d_0}|<5$.
These electrons must also be isolated from other objects in the event,
according to a \pt-dependent isolation requirement, 
which uses calorimeter- and track-based information to obtain 95\% efficiency at $\pt=25$~\GeV\ for $Z\rightarrow ee$ events, rising to 99\% efficiency at $\pt=60$~\GeV. 

Baseline muons are reconstructed from either ID tracks matched to muon segments (collections of hits in a single layer of the muon spectrometer) 
or combined tracks formed in the ID and muon spectrometer~\cite{PERF-2015-10}. 
They are required to satisfy the ``medium'' selection criteria described in Ref.~\cite{PERF-2015-10}, 
and for the high-\pt (low-\pt) analysis must satisfy $p_{\text{T}}>10$~\GeV\ ($p_{\text{T}}>7$~\GeV) and $|\eta|<2.5$.
Signal muon candidates are required to be isolated and have $|z_0\sin\theta| < 0.5$~mm and $|d_0/\sigma_{d_0}|<3$; those entering the high-\pT\ selection are further required to have $\pT>25$~\GeV.
Calorimeter- and track-based isolation criteria are used to obtain 95\% efficiency at $\pt=25$~\GeV\ for $Z\rightarrow\mu\mu$ events, rising to 99\% efficiency at $\pt=60$~\GeV~\cite{PERF-2015-10}.

Jets are reconstructed from topological clusters of energy~\cite{PERF-2014-07} in the calorimeter using the anti-$k_{t}$ algorithm~\cite{Cacciari:2008gp,N3Myth} with a radius parameter of 0.4 by making use of utilities within the \texttt{FastJet} package~\cite{Fastjet}.
The reconstructed jets are then calibrated to the particle level by the application of a jet energy scale (JES) derived from 13~\TeV\ data and simulation~\cite{PERF-2016-04}.
A residual correction applied to jets in data is based on studies of the $\pT$ balance between jets and well-calibrated objects in the MC simulation and data~\cite{ATL-PHYS-PUB-2015-015}.
Baseline jet candidates are required to have $\pt>20$~\GeV\ and reside within the region $|\eta|=4.5$.
Signal jets are further required to satisfy $\pt>30$~\GeV\ and reside within the region $|\eta|=2.5$.
Additional track-based criteria designed to select jets from the hard scatter and reject those
originating from pile-up are applied to signal jets with $\pt<60$~\GeV\ and $|\eta|<2.4$.
These are imposed by using the jet vertex tagger described in Ref.~\cite{ATLAS-CONF-2014-018}.
Finally, events containing a baseline jet that does not pass jet quality requirements are vetoed in order to remove events impacted by detector noise and non-collision backgrounds~\cite{Aad:2013zwa,ATLAS-CONF-2015-029}.
The MV2C10 boosted decision tree algorithm~\cite{PERF-2012-04,ATL-PHYS-PUB-2016-012} identifies jets containing $b$-hadrons ($b$-jets) by using quantities such as
the impact parameters of associated tracks and positions of any good reconstructed secondary vertices.
A selection that provides 77\% efficiency for tagging $b$-jets in simulated \ttbar\ events is used. 
The corresponding rejection factors against jets originating from $c$-quarks, tau leptons, and light quarks and gluons in the same sample for this selection are 6, 22, and 134, 
respectively.
These tagged jets are called $b$-tagged jets.

Photon candidates are required to satisfy the ``tight'' selection criteria described in Ref.~\cite{Aaboud:2016yuq}, have $\pt>25$~\GeV\ and reside within the region $|\eta|=2.37$,
excluding the calorimeter transition region $1.37<|\eta|<1.6$.
Signal photons are further required to have $\pt>50$~\GeV\ and to be isolated from other objects in the event, 
according to \pt-dependent requirements on both track-based and calorimeter-based isolation. 

To avoid the duplication of analysis objects, an overlap removal procedure is applied using baseline objects.
Electron candidates originating from photons radiated off of muons are rejected if they are found to share an inner detector track with a muon.
Any baseline jet within $\Delta R=0.2$ of a baseline electron is removed, unless the jet is $b$-tagged.
For this overlap removal, a looser 85\% efficiency working point is used for tagging $b$-jets.
Any electron that lies within $\Delta R<\mathrm{ min } (0.04+(10~\GeV)/\pt ,0.4)$ from a remaining jet is discarded.
If a baseline muon either resides within $\Delta R=0.2$ of, or has a track associated with, a remaining baseline jet, that jet is removed unless it is $b$-tagged.
Muons are removed in favour of jets with the same \pt-dependent $\Delta R$ requirement as electrons.
Finally, photons are removed if they reside within $\Delta R=0.4$ of a baseline electron or muon, and any jet within $\Delta R=0.4$ of any remaining photon is discarded.

The missing transverse momentum $\boldsymbol{\mathit{p}}_{\text{T}}^{\text{miss}}$ is defined as the negative vector sum of the transverse momenta of all baseline electrons,
muons, jets, and photons~\cite{Aaboud:2018tkc}.
Low momentum contributions from particle tracks from the primary vertex that are not associated with reconstructed analysis objects are included in the calculation of $\boldsymbol{\mathit{p}}_{\text{T}}^{\text{miss}}$.

Signal models with large hadronic activity are targeted by placing additional requirements on the quantity \HT, defined as the scalar sum of the $\pt$ values
of all signal jets.  For the purposes of rejecting \ttbar\ background events, the $\mttwo$ \cite{Lester:1999tx,Barr:2003rg} variable is used, defined as an extension of the transverse mass $m_{\text{T}}$ for the case of two missing particles:

\begin{equation*}
m_{\text{T}}^2\left(\boldsymbol{\,\mathit{p}}_{\text{T},\ell a},\boldsymbol{\mathit{p}}_{\text{T}}^{\text{miss}}\right) = 2 \times \left( p_{\text{T},\ell a} \times \met - \boldsymbol{\mathit{p}}_{\text{T},\ell a} \cdot \boldsymbol{\mathit{p}}_{\text{T}}^{\text{miss}} \right),
\end{equation*}

\begin{equation*}
\mttwo^2 = \min_{\mathbf{x}_{\text{T,1}}+\mathbf{x}_{\text{T,2}}=\boldsymbol{\mathit{p}}_{\text{T}}^{\text{miss}}} \left[ \max \{ m_{\text{T}}^2\left(\,\boldsymbol{\mathit{p}}_{\text{T},\ell 1},\mathbf{x}_{\text{T,1}}\right) , m_{\text{T}}^2\left(\,\boldsymbol{\mathit{p}}_{\text{T},\ell 2},\mathbf{x}_{\text{T,2}}\right) \} \right],
\end{equation*}

\noindent where $\boldsymbol{\mathit{p}}_{\text{T},\ell a}$ is the transverse-momentum vector of the highest \pt ($a=1$) or second highest \pt ($a=2$) lepton, and $\mathbf{x}_{\text{T,b}}$ ($b=1,2$) are two vectors representing the possible momenta of the invisible particles that minimize the \mttwo\ in the event. For typical \ttbar\ events, the value of \mttwo\ is small, while for signal events in some scenarios it can be relatively large.

All MC samples have MC-to-data corrections applied to take into account small differences between data and MC simulation in identification, reconstruction and trigger efficiencies.
The $\pt$ values of leptons in MC samples are additionally smeared to match the momentum resolution in data.

\section{Event selection}
\label{sec:selection}
This search is carried out using signal regions (SRs) designed to select events where heavy new particles decay into an ``invisible'' LSP, 
with final-state signatures including either a $Z$ boson mass peak or a kinematic endpoint in the dilepton invariant mass distribution.
In order to estimate the expected contribution from SM backgrounds in these regions, 
control regions (CRs) are defined in such a way that they are enriched in the particular SM process of interest and have low expected contamination from events potentially arising from SUSY signals. 
For signal points not excluded by the previous iteration of this analysis~\cite{SUSY-2016-05}, the signal contamination in the CRs is $<5$\%, with the exception of models with $m_{\tilde g}<600$~\GeV\ in the higher-\met\ CRs of the \lowpt\ search where it can reach 20\%.
To validate the background estimation procedures, various validation regions (VRs) are defined so as to be analogous but orthogonal to the CRs and SRs,
by using less stringent requirements than the SRs on variables used to isolate the SUSY signal, 
such as \mttwo, \met\ or \HT.
VRs with additional requirements on the number of leptons are used to validate the modelling of 
backgrounds in which more than two leptons are expected.
The various methods used to perform the background prediction in the SRs are discussed in Section~\ref{sec:bg}.

Events entering the SRs must have at least two signal leptons (electrons or muons), 
where the two highest-\pT\ leptons in the event are used when defining further event-level requirements. 
These two leptons must have the same-flavour (SF) and oppositely signed charges (OS). 
For the high-\pt\ lepton analysis, in both the edge and on-$Z$ searches, the events must pass at least one of the leptonic triggers, 
whereas \met\ triggers are used for the \lowpt\ analysis so as to select events containing softer leptons.
In the cases where a dilepton trigger is used to select an event,
the two leading (highest \pt) leptons must be matched to the objects that triggered the event. 
For events selected by a single-lepton trigger, at least one of the two leading leptons must be matched to the trigger object in the same way.
The two leading leptons in the event must have $\pt>\{50,25\}$~\GeV\ to pass the high-\pt\ event selection,
and must have $\pt>\{7,7\}$~\GeV, while not satisfying $\pt>\{50,25\}$~\GeV, to be selected by the \lowpt\ analysis.

Since at least two jets are expected in all signal models studied,
selected events are further required to contain at least two signal jets.
Furthermore, for events with a \met\ requirement applied, the minimum azimuthal opening angle between either of the two leading jets and the ${\boldsymbol p}_{\mathrm{T}}^{\mathrm{miss}}$,
$\Delta\phi(\text{jet}_{12},{\boldsymbol p}_{\mathrm{T}}^{\mathrm{miss}})$, is required to be greater than 0.4 so as to remove events with \met\ arising from jet mismeasurements.

The selection criteria for the CRs, VRs, and SRs are summarised in Tables~\ref{tab:regions-edge}~and~\ref{tab:regions-lowpt}, for the high- and low-\pT\ analyses respectively.
The most important of these regions are shown graphically in Figure~\ref{fig:region_diagrams}.

\begin{table}[htbp]
\begin{center}
 \caption{Overview of all signal, control and validation regions used in the high-$\pt$ edge and on-$Z$ searches.
 The flavour combination of the dilepton pair is denoted by either ``SF'' for same-flavour or ``DF'' for different-flavour.
All regions require at least two opposite-charge leptons with $\pT>\{50,25\}~\GeV$, with the exception of the three $\gamma$ CRs, which require zero leptons and one photon, and the diboson CRs (VR-WZ and VR-ZZ).
Unlike the rest of the regions, the diboson CRs do not include a lepton-charge requirement.
More details are given in the text.
The main requirements that distinguish the control and validation regions from the signal regions are indicated in bold.
Most of the kinematic quantities used to define these regions are discussed in the text.
}
\resizebox{1\textwidth}{!}{
 \begin{tabular}{lrrcrccccc} 
   \noalign{\smallskip}\hline\noalign{\smallskip}
     {\bf High-\pt\ }   &  \multicolumn{1}{c}{\bf \met}   & \multicolumn{1}{c}{\bf $\HT$}  &  {\bf $n_{\text{jets}}$}  & \multicolumn{1}{c}{\bf $m_{\ell\ell} $} & {\bf \mttwo} &  {\bf SF/DF}  & {\bf $n_{b\text{-jets}}$}  &  {\bf $\Delta\phi(\text{jet}_{12},{\boldsymbol p}_{\mathrm{T}}^\mathrm{miss})$ }  & {\bf \mll\ windows} \\
     {\bf regions} &  {\bf [\GeV]} & {\bf [\GeV]} &                           & {\bf [\GeV]}          & {\bf [\GeV]}              &              &                &                           \\
   \noalign{\smallskip}\hline\noalign{\smallskip}
   \multicolumn{2}{l}{Signal regions} &&&&&&& \\
   \noalign{\smallskip}\hline\noalign{\smallskip}
   SR-low     &  $> 250$  &  $> 200$    &  $\geq 2$  & $>12$     & $>70$   &  SF  & $-$  & $>0.4$ & 10 \\
   SR-medium  &  $> 400$  &  $> 400$    &  $\geq 2$  & $>12$     & $>25$   &  SF  & $-$  & $>0.4$ & 9  \\
   SR-high    &  $> 200$  &  $> 1200$   &  $\geq 2$  & $>12$     & $-$     &  SF  & $-$  & $>0.4$ & 10 \\
   \noalign{\smallskip}\hline\noalign{\smallskip}
   \multicolumn{2}{l}{Control regions} &&&&&&&    \\
   \noalign{\smallskip}\hline\noalign{\smallskip}
   CR-FS-low         &  $> 250$          &  $> 200$   &  $\geq 2$ & $>12$ & $>70$    &  {\bf DF}  & $-$  &  $>0.4$ & $-$ \\
   CR-FS-medium      &  $> 400$          &  $> 400$   &  $\geq 2$ & $>12$ & $>25$    &  {\bf DF}  & $-$  &  $>0.4$ & $-$ \\
   CR-FS-high        &  $> 100$          &  $> 1100$  &  $\geq 2$ & $>12$ & $-$      &  {\bf DF}  & $-$  &  $>0.4$ & $-$ \\
   CR$\gamma$-low    &  $-$              &  $> 200$   &  $\geq 2$ & $-$   & $-$      &  {\bf $0\ell$, $1\gamma$ } & $-$  &  $-$ & $-$ \\
   CR$\gamma$-medium &  $-$              &  $> 400$   &  $\geq 2$ & $-$   & $-$      &  {\bf $0\ell$, $1\gamma$ } & $-$  &  $-$ & $-$ \\
   CR$\gamma$-high   &  $-$              &  $> 1200$  &  $\geq 2$ & $-$   & $-$      &  {\bf $0\ell$, $1\gamma$ } & $-$  &  $-$ & $-$ \\
   CRZ-low    &  $< 100$              &  $> 200$   &  $\geq 2$ & $>12$   & $>70$    &  SF & $-$  &  $-$ & $-$ \\
   CRZ-medium &  $< 100$              &  $> 400$   &  $\geq 2$ & $>12$   & $>25$    &  SF & $-$  &  $-$ & $-$ \\
   CRZ-high   &  $< 100$              &  $> 1200$  &  $\geq 2$ & $>12$   & $-$      &  SF & $-$  &  $-$ & $-$ \\
   \noalign{\smallskip}\hline\noalign{\smallskip}
   \multicolumn{2}{l}{Validation regions} &&&&&&& \\
   \noalign{\smallskip}\hline\noalign{\smallskip}
   VR-low            &  $\mathbf{100}$--$\mathbf{200}$  &  $> 200$   &  $\geq 2$ & $>12$  & $>70$    &  SF  & $-$ & $>0.4$ & $-$\\
   VR-medium         &  $\mathbf{100}$--$\mathbf{200}$  &  $> 400$   &  $\geq 2$ & $>12$  & $>25$    &  SF  & $-$ & $>0.4$ & $-$\\
   VR-high           &  $\mathbf{100}$--$\mathbf{200}$  &  $> 1200$  &  $\geq 2$ & $>12$  & $-$      &  SF  & $-$ & $>0.4$ & $-$\\
   VR-$\Delta\phi$-low      &  $> 250$             &  $> 200$   &  $\geq 2$ & $>12$  & $>70$    &  SF  & $-$  &  $\mathbf{<0.4}$ & $-$\\
   VR-$\Delta\phi$-medium   &  $> 400$             &  $> 400$   &  $\geq 2$ & $>12$  & $>25$    &  SF  & $-$  &  $\mathbf{<0.4}$ & $-$\\
   VR-$\Delta\phi$-high     &  $> 200$             &  $> 1200$  &  $\geq 2$ & $>12$  & $-$      &  SF  & $-$  & $\mathbf{<0.4}$ & $-$\\
   VR-WZ            &  $\mathbf{100}$--$\mathbf{200}$   & {\bf $>200$} & $\geq 2$       & $>12$ &  $-$  & $\bm{3\ell}$   &    {\bf $0$}    & $>0.4$  &  $-$  \\
   VR-ZZ            &  {\bf $<50$}          & {\bf $>100$} & {\bf $\geq 1$} & $>12$ &  $-$  & $\bm{4\ell}$   &    {\bf $0$}    & $>0.4$  &  $-$   \\
   \noalign{\smallskip}\hline\noalign{\smallskip}
\end{tabular}
} 
 \label{tab:regions-edge}
\end{center}
\end{table}

\begin{table}[htbp]
\begin{center}
 \caption{Overview of all signal, control and validation regions used in the \lowpt\ edge search.
 The flavour combination of the dilepton pair is denoted by either ``SF'' for same-flavour or ``DF'' for different-flavour.
The charge combination of the leading lepton pairs is given as ``SS'' for same-sign or ``OS'' for opposite-sign.
All regions require at least two leptons with $\pt>\{7,7\}~\GeV$, with the exception of CR-real and CR-fake, which require {\it exactly} two leptons,
and the diboson CRs (VR-WZ-\lowpt\ and VR-ZZ-\lowpt).
More details are given in the text.
The main requirements which distinguish the control and validation regions from the signal regions are indicated in bold. 
The \lowpt\ SR selection is explicitly vetoed in VR-WZ-\lowpt\ and VR-ZZ-\lowpt to ensure orthogonality.
When applied, the $m_{\text{T}}$ requirement is checked for the two leading leptons.}
\resizebox{1\textwidth}{!}{
 \begin{tabular}{lrrccccccccc} 
   \noalign{\smallskip}\hline\noalign{\smallskip}
     {\bf Low-\pt }   &  \multicolumn{1}{c}{\bf \met}   & \multicolumn{1}{c}{\bf $\ptll$}  &  {\bf $n_{\text{jets}}$} & {\bf $n_{b\text{-jets}}$}  & {\bf $m_{\ell\ell} $} &  {\bf SF/DF}  & {\bf OS/SS}  &{\bf $\Delta\phi(\text{jet}_{12},{\boldsymbol p}_{\mathrm{T}}^\mathrm{miss})$ }  & {\bf $m_{\text{T}}$}   & {\bf \mll\ windows} \\
     {\bf regions} &  {\bf [\GeV]} & {\bf [\GeV]} &  &  & {\bf [\GeV]} &  &  &  & {\bf [\GeV]} & \\
   \noalign{\smallskip}\hline\noalign{\smallskip}
   \multicolumn{2}{l}{Signal regions} &&&&&&& \\
   \noalign{\smallskip}\hline\noalign{\smallskip}
   SRC      &  $> 250$  &  $< 20$  & $\geq 2$  & $-$ & $>30$         &  SF  & OS & $>0.4$  &$-$ & 6 \\
   SRC-MET  &  $> 500$  &  $< 75$  & $\geq 2$  & $-$ & $>4,\notin[8.4,11]$         &  SF  & OS & $>0.4$  &$-$ & 6 \\
   \noalign{\smallskip}\hline\noalign{\smallskip}
   \multicolumn{2}{l}{Control regions} &&&&&&   \\
   \noalign{\smallskip}\hline\noalign{\smallskip}
   CRC      &  $> 250$  &  $< 20$  & $\geq 2$  & $-$ & $>30$         & {\bf DF}    & OS & $>0.4$ & $-$ & $-$ \\
   CRC-MET  &  $> 500$  &  $< 75$  & $\geq 2$  & $-$ & $>4,\notin[8.4,11]$         & {\bf DF}    & OS & $>0.4$ & $-$ & $-$ \\
   CR-real  &  $-$      &  $-$     &  $\geq 2$ & $-$ & $\mathbf{81}$--$\mathbf{101}$       & $2\ell$ SF  & OS  &  $-$  & $-$ & $-$\\
   \multirow{2}{*}{CR-fake$\left.\rule{0cm}{5mm}\right\{$}          &  \multirow{2}{*}{$\bm{<125}$}   &   \multirow{2}{*}{$-$}  &  \multirow{2}{*}{$-$} &  \multirow{2}{*}{$-$} & $>4,\notin[8.4,11]$  &  {\bf $2\ell$ $\mu e$} & \multirow{2}{*}{{\bf SS}} &  \multirow{2}{*}{$-$} & \multirow{2}{*}{$-$} &  \multirow{2}{*}{$-$}\\
    &  &  &  & & $>4,\notin[8.4,11],{\bf\boldsymbol \notin[81,101]}$ & {\bf $2\ell$ $\mu\mu$}  & & & &  \\
   \noalign{\smallskip}\hline\noalign{\smallskip}
   \multicolumn{2}{l}{Validation regions} &&&&&&& \\
   \noalign{\smallskip}\hline\noalign{\smallskip}
   VRA          &  {\bf 200}--{\bf 250} &  {\bf $<20$   }  &  $\geq 2$ & $-$  & $>30$  &  SF  & OS  & $>0.4$ & $-$  & $-$\\
   VRA2         &  {\bf 200}--{\bf 250} &  {\bf $>20$   }  &  $\geq 2$ & $-$  & $>4,\notin[8.4,11]$  &  SF  & OS  & $>0.4$  & $-$ & $-$\\      
   VRB          &  {\bf 250}--{\bf 500} &  $\mathbf{20}$--$\mathbf{75}$   &  $\geq 2$ & $-$  & $>4,\notin[8.4,11]$  &  SF  & OS  & $>0.4$  & $-$ & $-$\\
   VRC          &  {\bf 250}--{\bf 500} &  {\bf $>75$   }  &  $\geq 2$ & $-$  & $>4,\notin[8.4,11]$  &  SF  & OS  & $>0.4$  & $-$ & $-$\\
   VR-WZ-\lowpt\    &  {\bf $>200$}  &  $-$   &  $\bm{\geq 1}$ & {\bf 0}  & $>4,\notin[8.4,11]$  &  $\bm{3\ell}$  & $-$  & $>0.4$  & $-$ & $-$\\
   VR-ZZ-\lowpt\    &  {\bf $>200$}   &  $-$   &  $-$   & {\bf 0} & $>4,\notin[8.4,11]$  &  $\bm{4\ell}$  & $-$  & $>0.4$ & $-$ & $-$\\
   VR-$\Delta\phi$      &  {\bf $>250$}  &  $-$             &  $\geq 2$ & $-$  & $>4,\notin[8.4,11]$  &  SF  & OS  & $\bf <0.4$ &$-$  & $-$\\ 
   VR-fakes     &  {\bf $>225$}& $-$                &  $\geq 2$  & $-$ & $>4,\notin[8.4,11]$  &  {\bf DF}  & OS  & $>0.4$ & {\bf $\ell_1, \ell_2<100$}  &$-$\\
   VR-SS     &  {\bf $>225$}& $-$                &  $\geq 2$  & $-$ & $>4,\notin[8.4,11]$  &  SF  & SS  & $>0.4$ & {\bf $\ell_1, \ell_2<100$}  &$-$\\
   \noalign{\smallskip}\hline\noalign{\smallskip}
\end{tabular}
} 
 \label{tab:regions-lowpt}
\end{center}
\end{table}

\begin{figure}[hbtp]
\centering
\includegraphics[width=.8\textwidth]{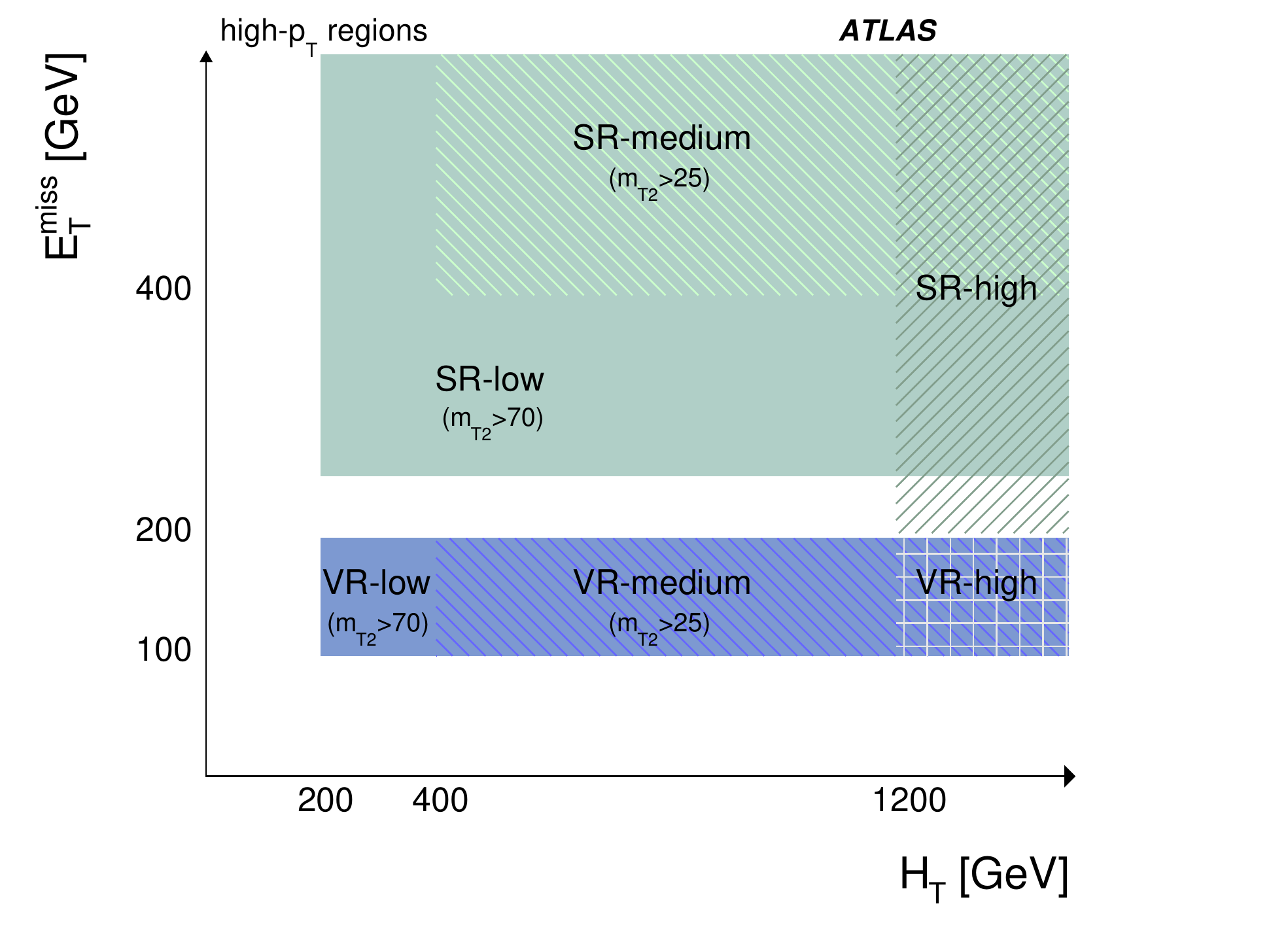}\\
\includegraphics[width=.8\textwidth]{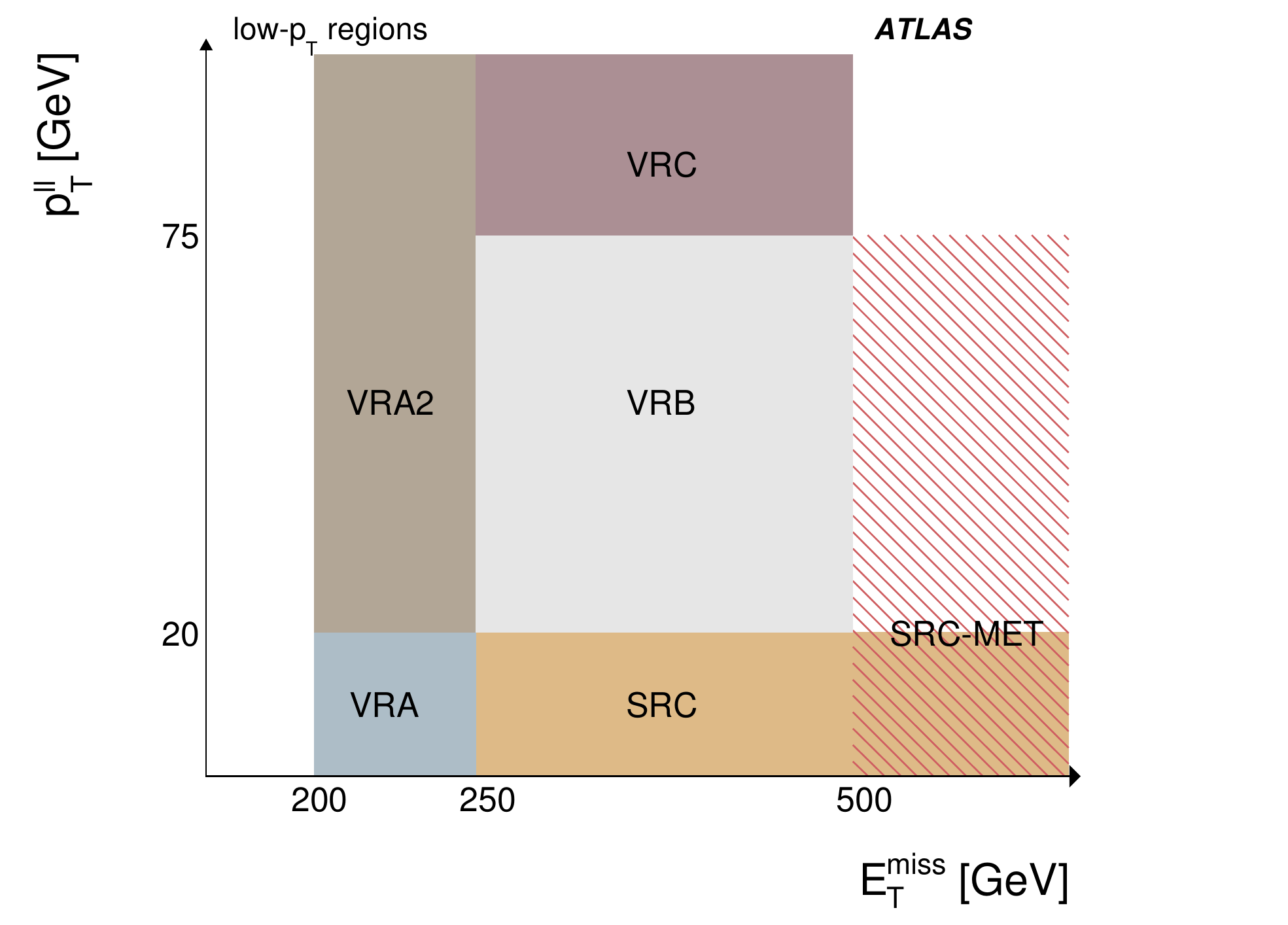}
\caption{
Schematic diagrams of the main validation and signal regions for the high-\pt\ (top) and \lowpt\ (bottom) searches.
Regions where hatched markings overlap indicate the overlap between various regions. 
For each search (high-\pt\ or \lowpt), the SRs are not orthogonal; in the case of high-\pt, the VRs also overlap.
In both cases, as indicated in the diagrams, there is no overlap between SRs and VRs.
\label{fig:region_diagrams}
}
\end{figure}

For the high-\pt\ search, the leading lepton's \pt\ is required to be at least $50$~\GeV\ to reject additional background events while retaining high efficiency for signal events.
Here, a kinematic endpoint in the \mll\ distribution is searched for in three signal regions. 
In each case, it is carried out across the full \mll\ spectrum,
with the exception of the region with $\mll<12$~\GeV, which is vetoed to reject 
low-mass Drell--Yan (DY) events, $\Upsilon$ and other dilepton resonances. 
Models with low, medium and high values of $\Delta m_{\tilde{g}} = m_{\tilde{g}} - m_{\tilde{\chi}^0_1}$ 
are targeted by selecting events with $\HT >200, 400$ and $1200$~\GeV\ to enter SR-low, SR-medium and SR-high, respectively.
Requirements on \met\ are also used to select signal-like events, with higher \met\ thresholds probing models with higher LSP masses. 
For SR-low and SR-medium a cut on \mttwo\ of $>70$~\GeV\ and $>25$~\GeV, respectively, is applied to reduce backgrounds from top-quark production. 
In order to make model-dependent interpretations using the signal models described in Section~\ref{sec:susy}, 
a profile likelihood~\cite{statforumlimits} fit to the \mll\ shape is performed in each SR separately, 
with \mll\ bin boundaries chosen to ensure a sufficient number of events for a robust background estimate
in each bin and maximise sensitivity to target signal models. 
The \mll\ {\it bins} are also used to form 29 non-orthogonal \mll\ {\it windows} 
to probe the existence of BSM physics or to assess model-independent upper limits on the number of possible signal events.       
These windows are chosen so that they are sensitive to a broad range of potential kinematic edge positions. 
In cases where the signal could stretch over a large \mll\ range, the exclusive bins used in the shape fit potentially truncate the lower-\mll\ tail, and so are less sensitive.
Of these windows, ten are in SR-low, nine are in SR-medium and ten are in SR-high. 
A schematic diagram showing the \mll\ bin edges in the SRs and the subsequent \mll\ windows is shown in Figure~\ref{fig:bin_diagrams}. 
More details of the \mll\ definitions in these windows are given along with the results in Section~\ref{sec:result}.
Models without light sleptons are targeted by windows with $\mll<81$~\GeV\ for $\Delta m_\chi < m_Z$,
and by the window with $81<\mll<101$~\GeV\ for $\Delta m_\chi > m_Z$.
The on-$Z$ bins of the SRs, with bin boundaries $81<\mll<101$~\GeV, are each considered as one of the 29 \mll\ windows, 
having good sensitivity to models with on-shell $Z$ bosons in the final state. 

For the low-\pt\ search, events are required to have at least two leptons with \pt$>7$~\GeV. 
Orthogonality with the high-\pt\ channel is imposed by rejecting events that satisfy the lepton \pt\ requirements of the high-\pt\ selection. 
In addition to this, events must have \mll$>4$~\GeV, excluding the region between $8.4$ and $11$~\GeV, in order to exclude the $J/\psi$ and $\Upsilon$ resonances. 
To isolate signal models with small $\Delta m_\chi$, the \lowpt\ lepton SRs place upper bounds on the \ptll\ (\pt\ of the dilepton system) of events entering the two SRs, 
SRC and SRC-MET\@. 
SRC selects events with a maximum \ptll\ requirement of $20$~\GeV, targeting models with small $\Delta m_\chi$. 
SRC-MET requires \ptll$<75$~\GeV\ and has a higher \met\ threshold ($500$~\GeV\ compared with $250$~\GeV\ in SRC), 
maximising sensitivity to very compressed models.
Here the analysis strategy closely follows that of the high-\pt\ analysis, 
with a shape fit applied to the \mll\ distribution performed independently in SRC and SRC-MET\@. 
The \mll\ bins are used to construct \mll\ windows from which model-independent assessments can be made. 
There are a total of 12 \mll\ windows for the \lowpt\ analysis, six in each SR\@.

\begin{figure}[hbtp]
\centering
\includegraphics[width=.8\textwidth]{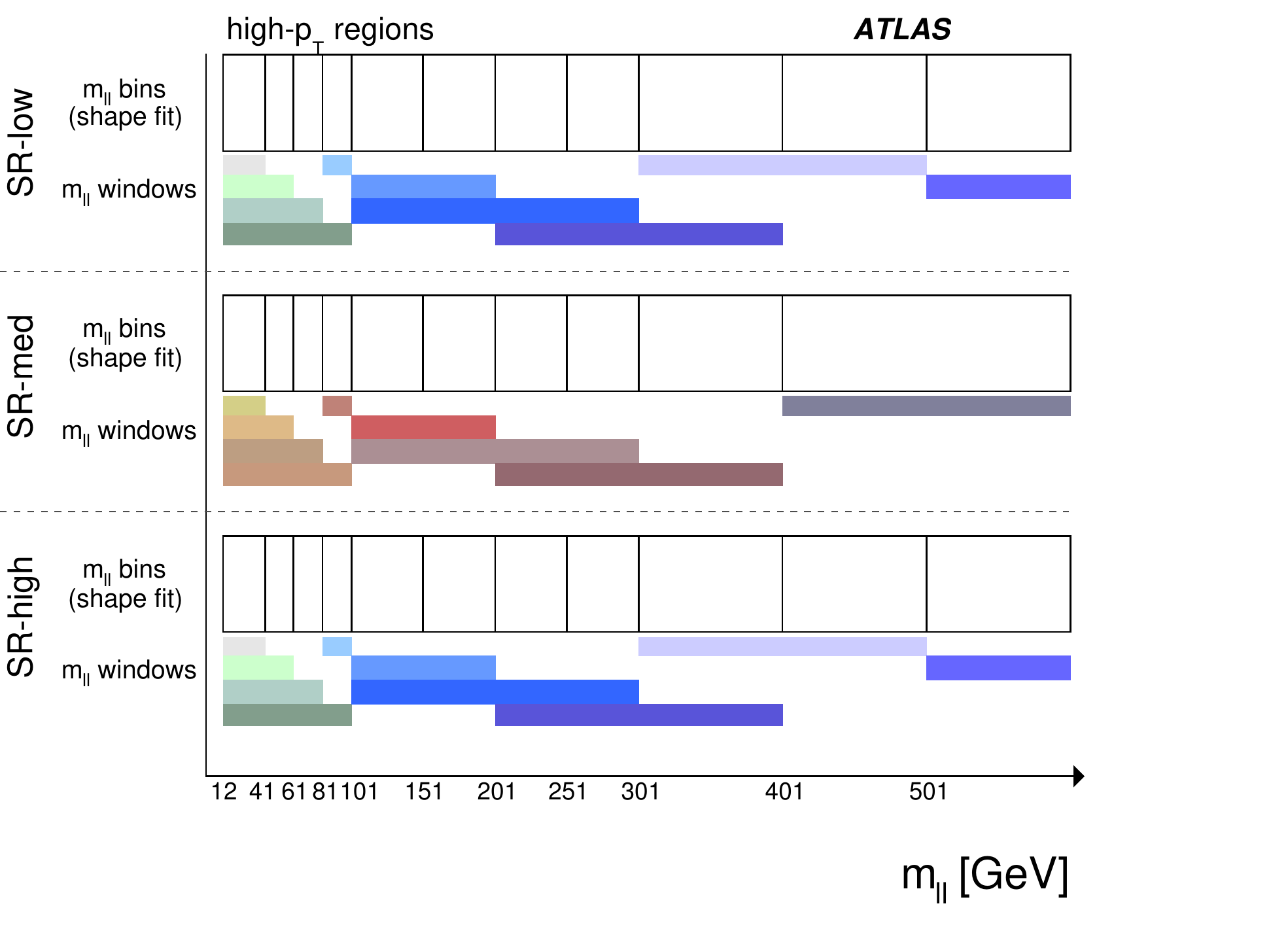}\\
\includegraphics[width=.8\textwidth]{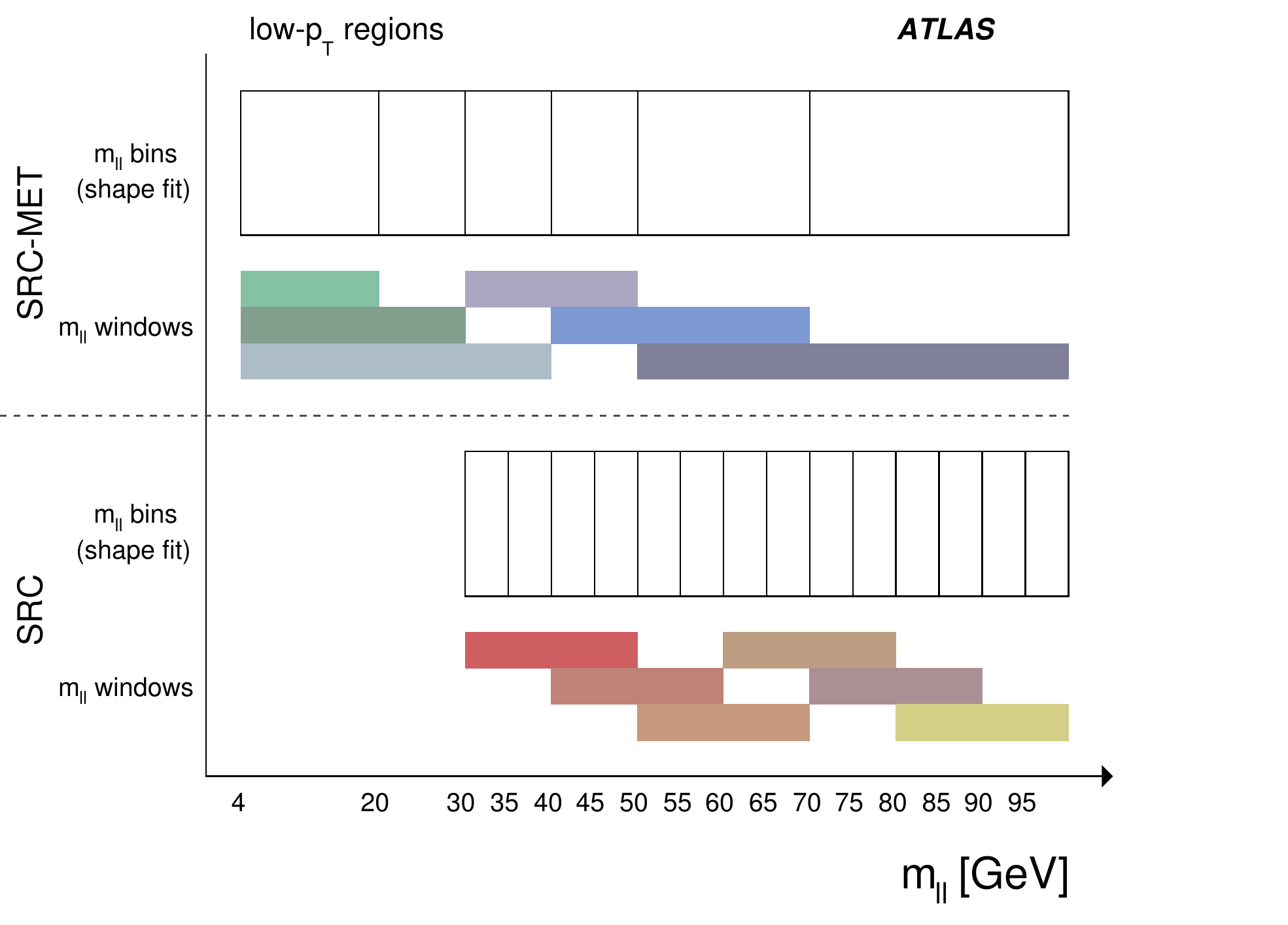}
\caption{
Schematic diagrams to show the \mll\ binning used in the various SRs alongside the overlapping \mll\ windows used for model-independent interpretations.
The unfilled boxes indicate the \mll\ bin edges for the shape fits used in the model-dependent interpretations. 
Each filled region underneath indicates one of the \mll\ windows, formed of one or more \mll\ bins, used to derive model-independent results for the given SR.
In each case, the last \mll\ bin includes the overflow.
\label{fig:bin_diagrams}
}
\end{figure}

\section{Background estimation}
\label{sec:bg}
In most SRs, the dominant background processes are ``flavour-symmetric'' (FS), 
where the ratio of $ee$, $\mu\mu$ and $e\mu$ dileptonic branching fractions is expected to be 1:1:2 because the two leptons originate from independent $W\to\ell\nu$ decays.
Dominated by \ttbar, this background, described in Section~\ref{sec:FS}, also includes $WW$, $Wt$, and $Z\to\tau\tau$ processes, and typically makes up 50--95\% of the total SM background in the SRs. 
The FS background is estimated using data control samples of different-flavour (DF) events for the high-\pt\ search, 
whereas the \lowpt\ search uses such samples to normalise the dominant top-quark (\ttbar\ and $Wt$) component of this background, with the shape taken from MC simulation.

As all the SRs have a high \met\ requirement, 
\dyjets\ events generally enter the SRs when there is large \met\ originating from instrumental effects or from neutrinos from the decays of hadrons produced in jet fragmentation. 
This background is always relatively small, contributing less than $10$\% of the total background in the SRs, 
but is difficult to model with MC simulation. 
A control sample of \gjets\ events in data, which have similar kinematic properties to those of \dyjets\ 
and similar sources of \met, is used to model this background for the high-\pt\ search by weighting the \gjets\ events to match \dyjets\ in another control sample, described in Section~\ref{sec:zjets}.  For the \lowpt\ analysis, where \dyjets\ processes make up at most $8$\% of the background in the SRs, MC simulation is used to estimate this background.

The contribution from events with fake or misidentified leptons in the \lowpt\ SRs is at most $20$\%, 
and is estimated using a data-driven matrix method, described in Section~\ref{sec:fakes}.  The contribution to the SRs from $WZ/ZZ$ production, described in Section~\ref{sec:diboson_other}, while small for the most part ($<5$\%), can be up to $70$\% in the on-$Z$ bins of the high-\pt\ analysis.
These backgrounds are estimated from MC simulation and validated in dedicated $3\ell$ ($WZ$) and $4\ell$ ($ZZ$) VRs.  ``Rare top'' backgrounds, also described in Section~\ref{sec:diboson_other}, which include $t\bar{t}W$, $t\bar{t}Z$ and $t\bar{t}WW$ processes, constitute $<10$\% of the SM expectation in all SRs
and are estimated from MC simulation.

\subsection{Flavour-symmetric backgrounds}
\label{sec:FS}

For the high-\pt\ analysis the so-called ``flavour-symmetry'' method is used to estimate the contribution of the background from flavour-symmetric processes to each SR\@.
This method makes use of three $e\mu$ control regions, 
CR-FS-low, CR-FS-medium or CR-FS-high, with the same \mll\ binning as their corresponding SR\@. 
For SR-low, SR-medium or SR-high the flavour-symmetric contribution to each $\mll$ bin of the signal regions is predicted using data 
from the corresponding bin from CR-FS-low, CR-FS-medium or CR-FS-high, respectively (precise region definitions can be found in Table~\ref{tab:regions-edge}). 
These CRs are $>95$\% pure in flavour-symmetric processes (estimated from MC simulation).
Each of these regions has the same kinematic requirements as their respective SR, with the exception of CR-FS-high, 
in which the $1200$~\GeV\ \HT\ and $200$~\GeV\ \met\ thresholds of SR-high are loosened to $1100$~\GeV\ and $100$~\GeV, respectively, 
in order to increase the number of $e\mu$ events available to model the FS background.  

The data events in these regions are subject to lepton $\pt$- and $\eta$-dependent correction factors determined in data. 
These factors are measured separately for 2015 and 2016 to take into account the differences between the triggers available in those years, 
and account for the different trigger efficiencies for the dielectron, 
dimuon and electron--muon selections, as well as the different identification and reconstruction efficiencies for electrons and muons.
The estimated numbers of events in the SF channels, $N^\text{est}$, are given by:

\begin{align}
\label{eq:fs_est}
N^\text{est} = \frac{f_{\text{SR}}}{2} \cdot 
\left[ \sum^{N_{e\mu}^\text{data}}_{i} 
\Bigl( k_{e}(\pT^{i,\mu}, \eta^{i,\mu}) + k_{\mu}(\pT^{i,e}, \eta^{i,e}) 
\Bigr) \cdot \alpha(\pT^{i,\ell_{1}}, \eta^{i,\ell_{1}}) \notag \right.\\
\left.- \sum^{N_{e\mu}^\text{MC}}_{i} 
\Bigl( k_{e}(\pT^{i,\mu}, \eta^{i,\mu}) + k_{\mu}(\pT^{i,e}, \eta^{i,e}) 
\Bigr) \cdot \alpha(\pT^{i,\ell_{1}}, \eta^{i,\ell_{1}}) \right],
\end{align}

\noindent where $N_{e\mu}^\text{data}$ is the number of data events observed in a given control region (CR-FS-low, CR-FS-medium or CR-FS-high). 
Events from non-FS processes are subtracted from the $e\mu$ data events using MC simulation, the second term in Eq.~\ref{eq:fs_est}, 
where $N_{e\mu}^\text{MC}$ is the number of events from non-FS processes in MC simulation in the respective CRs. 
The factor $\alpha(\pT^i, \eta^i)$ accounts for the different trigger efficiencies for SF and DF events, 
and $k_{e}(\pT^i, \eta^i)$ and $k_{\mu}(\pT^i, \eta^i)$ are the electron and muon selection efficiency factors for the kinematics of the lepton being replaced in event $i$.
The trigger and selection efficiency correction factors are derived from the events in an inclusive on-$Z$ selection ($81<\mll<101\GeV$, $\geq2$ signal jets), 
according to:

\begin{eqnarray*}
k_{e}(\pT, \eta) = \sqrt{\frac{N_{ee}^{\text{meas}(\pT, \eta)}}{N_{\mu\mu}^{\text{meas}(\pT, \eta)}}}, \\
k_{\mu}(\pT, \eta) = \sqrt{\frac{N_{\mu\mu}^{\text{meas}(\pT, \eta)}}{N_{ee}^{\text{meas}(\pT, \eta)}}}, \\
\alpha(\pT, \eta) = \frac{\sqrt{\epsilon^\text{trig}_{ee}(\pt^{\ell_1},\eta^{\ell_1})\times\epsilon^\text{trig}_{\mu\mu}(\pt^{\ell_1},\eta^{\ell_1})}}{\epsilon^\text{trig}_{e\mu}(\pt^{\ell_1},\eta^{\ell_1})},
\end{eqnarray*}

\noindent where $\epsilon^\text{trig}_{ee/\mu\mu/e\mu}$ is the trigger efficiency as a function of the leading-lepton ($\ell_1$) kinematics and $N_{ee}^{\text{meas}}$ $(N_{\mu\mu}^{\text{meas}})$ 
is the number of $ee$ $(\mu\mu)$ data events in the inclusive on-$Z$ region (or a DF selection in the same mass window in the case of $\epsilon^\text{trig}_{e\mu}$, for example) outlined above. 
Here $k_{e}(\pT, \eta)$ and $k_{\mu}(\pT, \eta)$ are calculated separately for leading and sub-leading leptons. 
The correction factors are typically within 10\% of unity, except in the region $|\eta|<0.1$ where, 
because of a lack of coverage of the muon spectrometer, they deviate by up to 50\% from unity.
To account for the extrapolation from $\HT>1100$~\GeV\ and $\met>100$~\GeV\ to $\HT>1200$~\GeV\ and $\met>200$~\GeV\ going from CR-FS-high to SR-high, 
an additional factor, $f_{\text{SR}}$, derived from simulation, is applied as given in Eq.~\ref{eq:extrapolationHTMET}.

\begin{equation}\label{eq:extrapolationHTMET}
f_{\text{SR}} = \frac{ N_{e\mu}^{\text{CR-FS-high}} (\met>200 \text{ GeV},\HT>1200 \text{ GeV}) }{ N_{e\mu}^{\text{CR-FS-high}} (\met>100 \text{ GeV}, \HT>1100 \text{ GeV})}
\end{equation}

\noindent In CR-FS-high this extrapolation factor is found to be constant over the full \mll\ range. 

The FS method is validated by performing a closure test using MC simulated events, 
with FS simulation in the $e\mu$ channel being scaled accordingly to predict the expected contribution in the SRs. 
The results of this closure test can be seen on the left of Figure~\ref{fig:fs}, 
where the \mll\ distribution is well modelled after applying the FS method to the $e\mu$ simulation. 
This is true in particular in SR-high, where the \met- and \HT-based extrapolation is applied.
The small differences between the predictions and the observed distributions are used to assign an MC non-closure uncertainty to the estimate. 
To further validate the FS method, the full procedure is applied to data in VR-low, VR-medium and VR-high (defined in Table~\ref{tab:regions-edge}) at lower \met, 
but otherwise with identical kinematic requirements. 
The FS contribution in these three VRs is estimated using three analogous $e\mu$ regions: VR-FS-low, VR-FS-med and VR-FS-high, also defined in Table~\ref{tab:regions-edge}. 
In the right of Figure~\ref{fig:fs}, the estimate taken from $e\mu$ data is shown to model the SF data well.

\begin{figure*}[!th]
\centering
\includegraphics[width=0.43\textwidth]{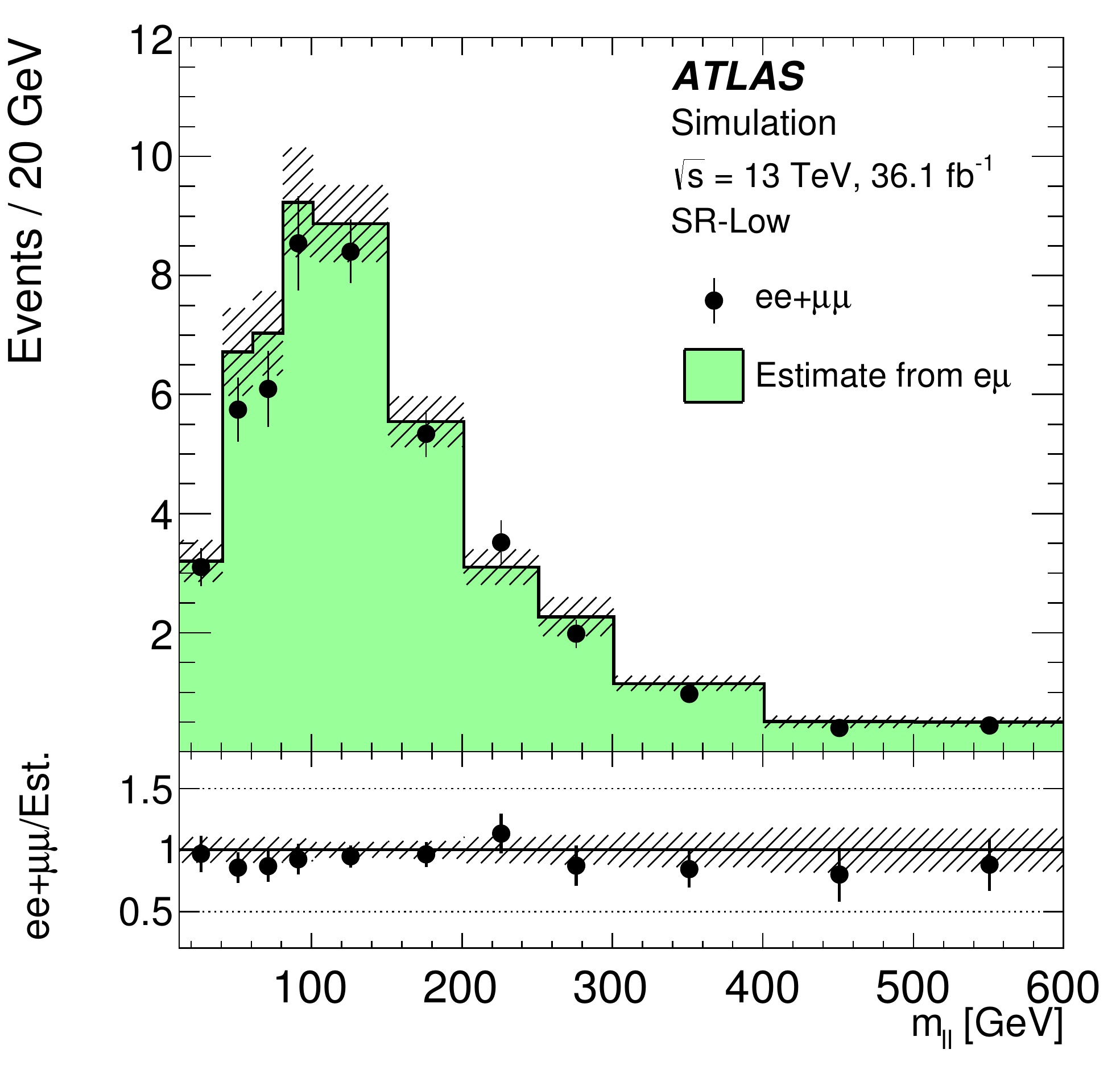}
\includegraphics[width=0.43\textwidth]{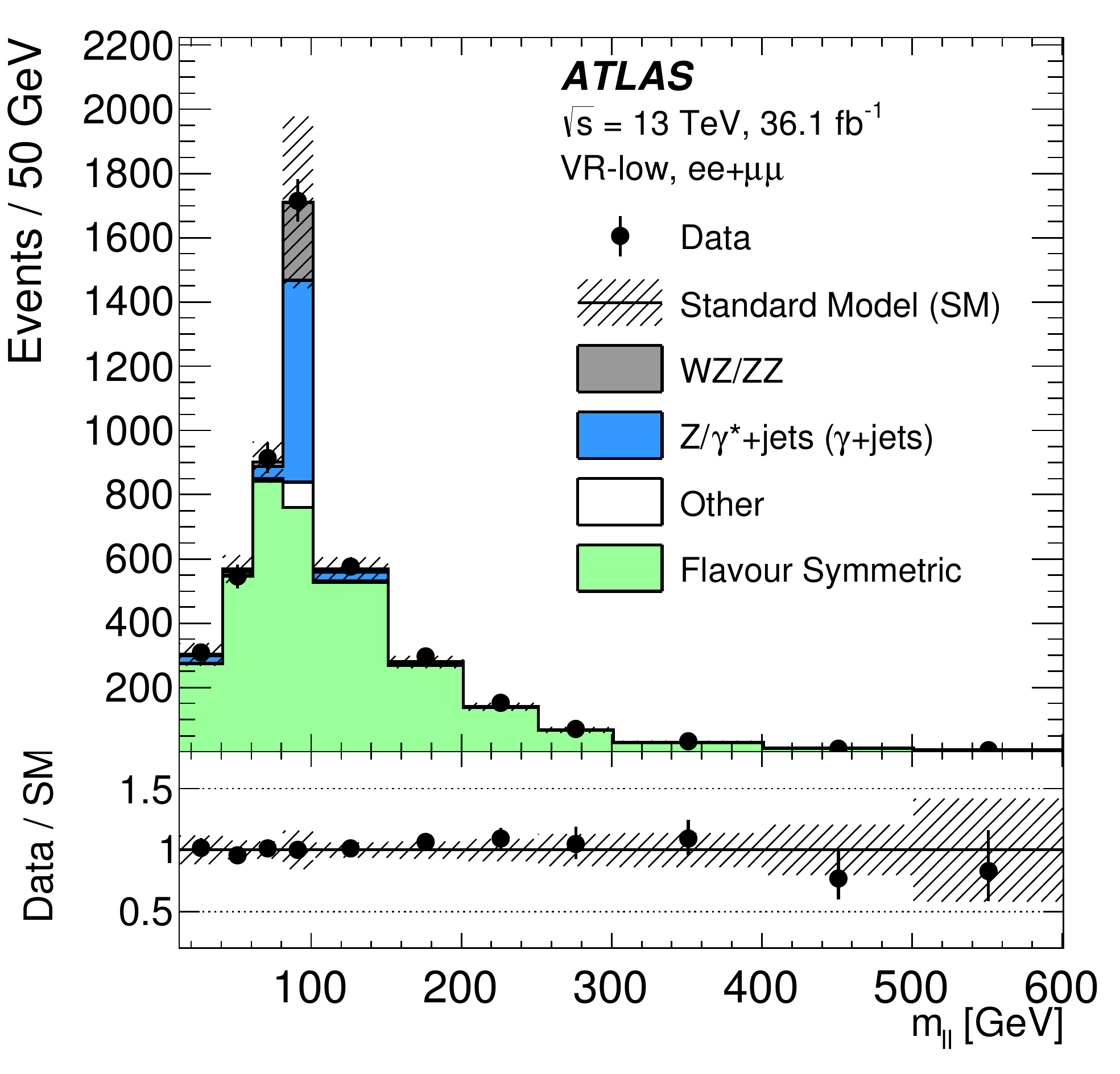}
\includegraphics[width=0.43\textwidth]{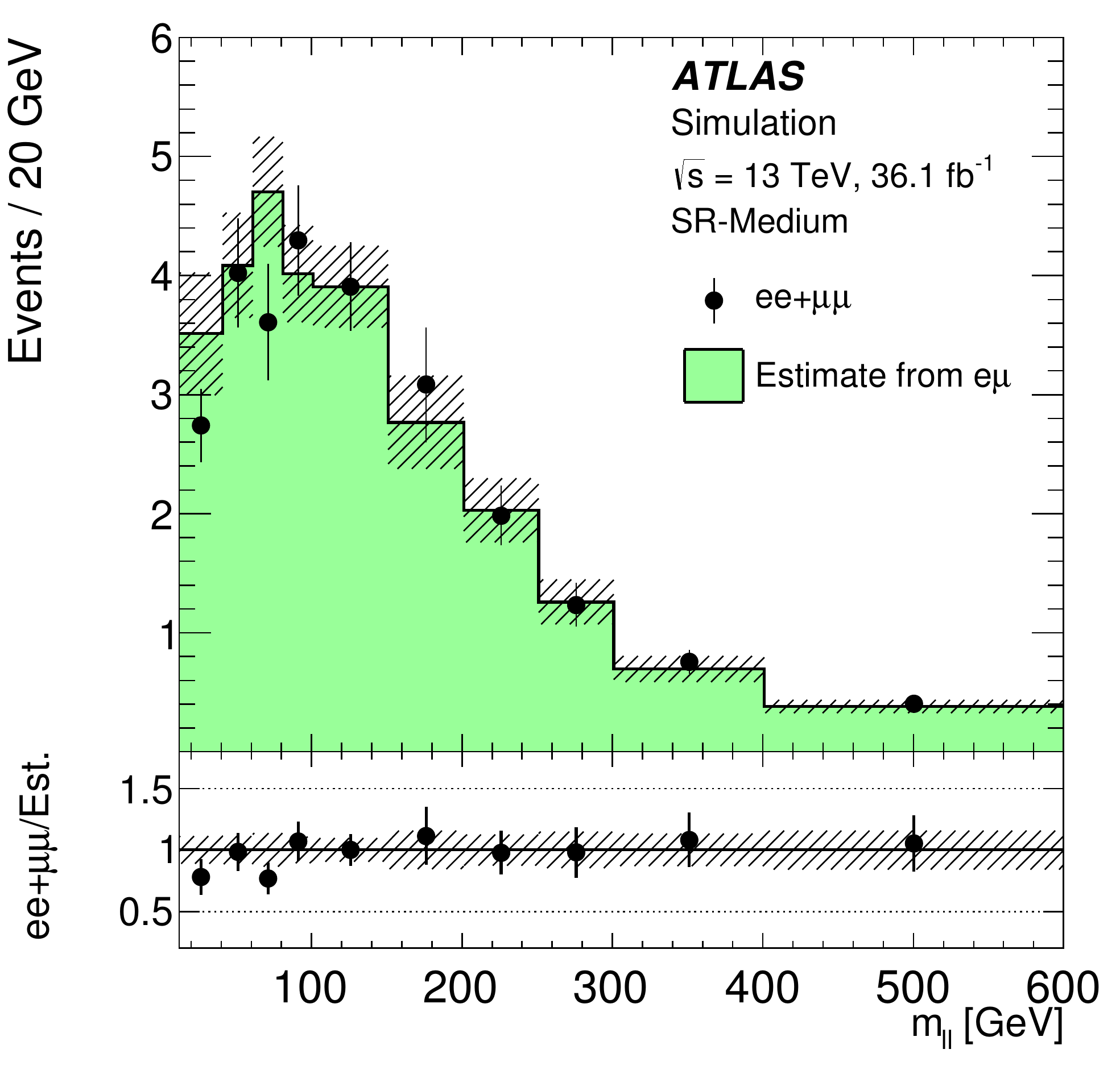}
\includegraphics[width=0.43\textwidth]{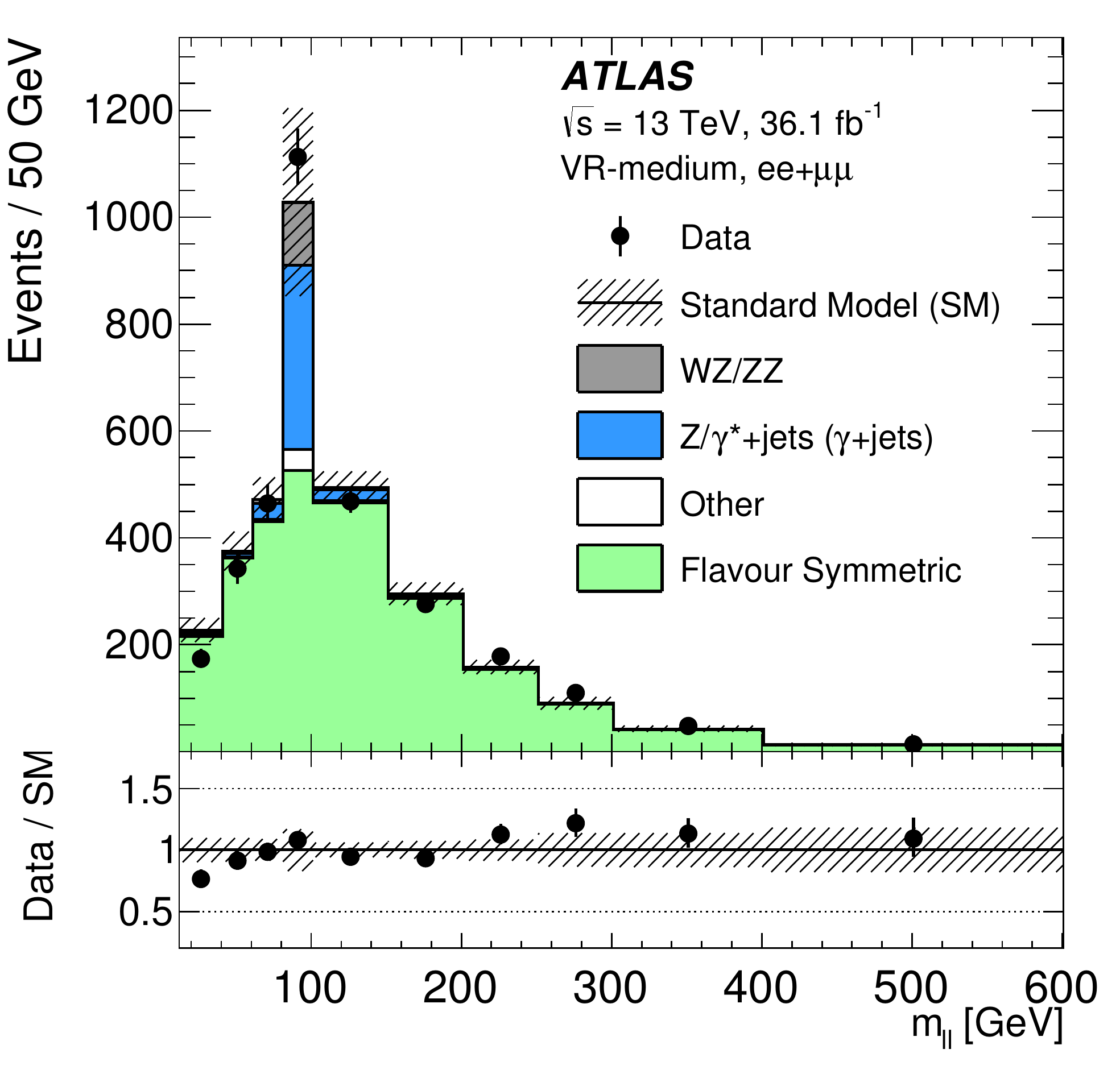}
\includegraphics[width=0.43\textwidth]{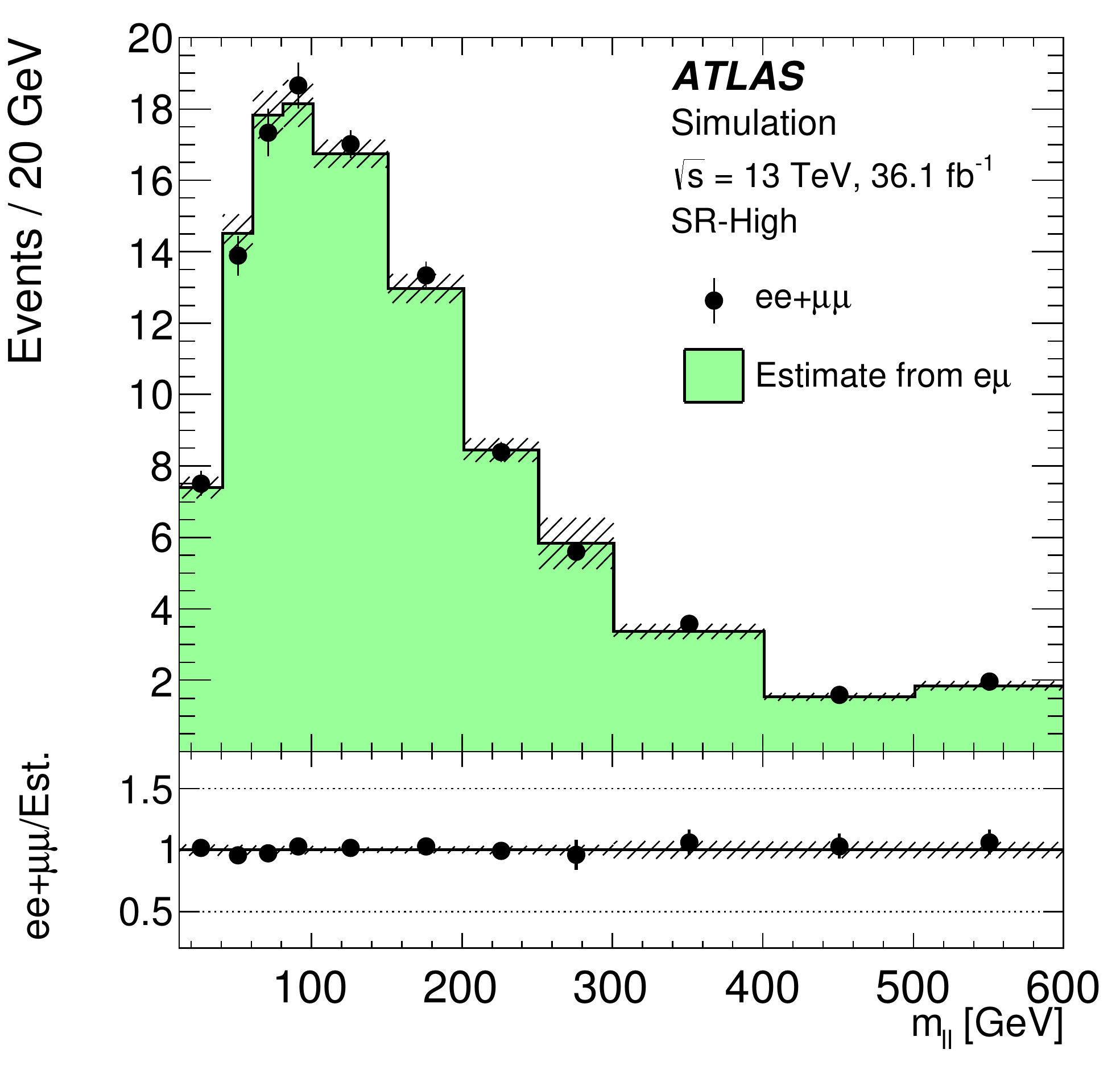}
\includegraphics[width=0.43\textwidth]{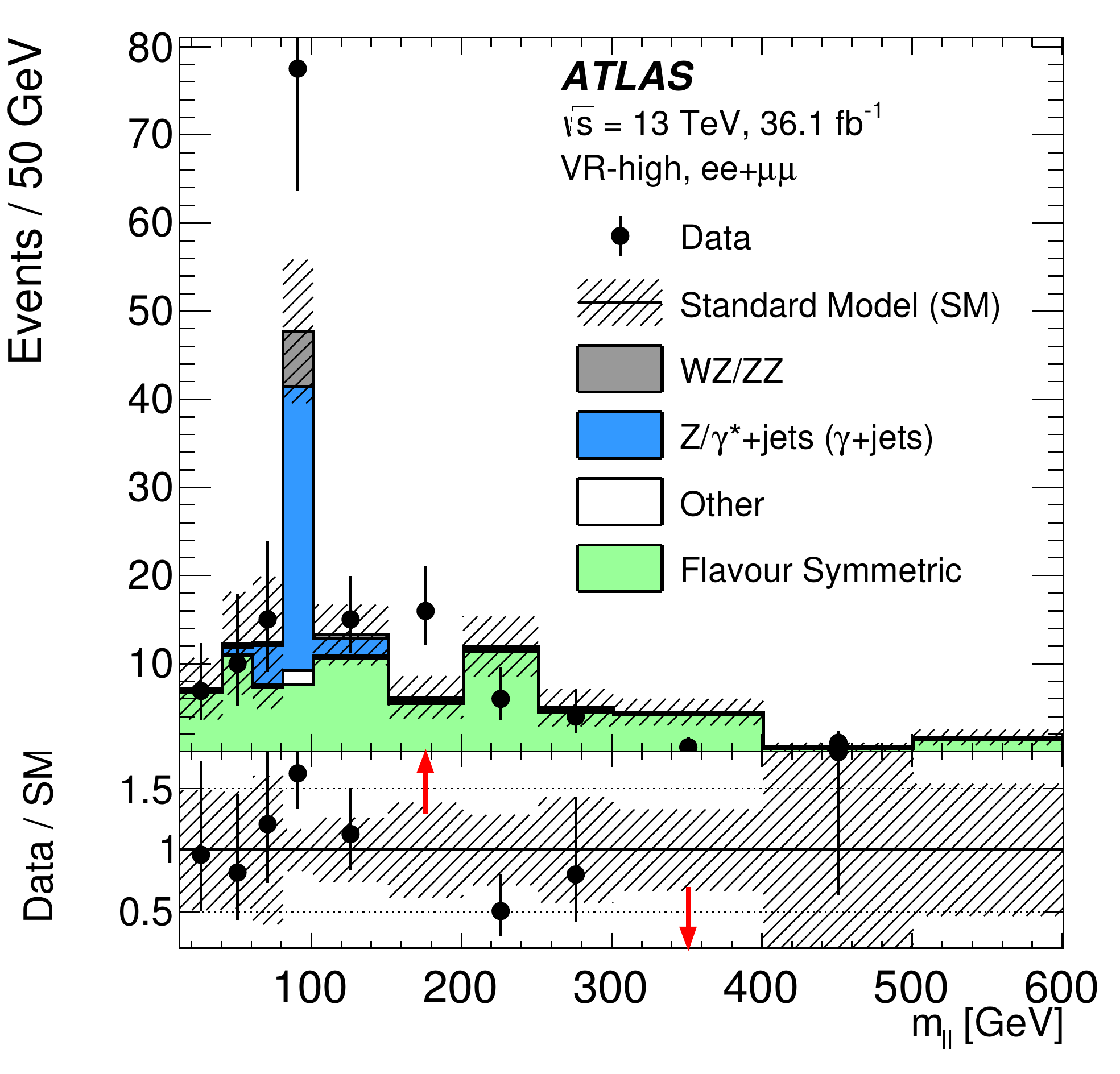}
\caption{Validation of the flavour-symmetry method using MC simulation (left) and data (right), 
in SR-low and VR-low (top), SR-medium and VR-medium (middle), and SR-high and VR-high (bottom).
On the left the flavour-symmetry estimate from \ttbar, $Wt$, $WW$ and $Z\rightarrow \tau\tau$ MC samples in the $e\mu$ channel 
is compared with the SF distribution from these MC samples. The MC statistical uncertainty is indicated by the hatched band. 
In the data plots, all uncertainties in the background expectation are included in the hatched band.  
The bottom panel of each figure shows the ratio of the observation to the prediction.
In cases where the data point is not accommodated by the scale of this panel, 
an arrow indicates the direction in which the point is out of range.
The last bin always contains the overflow. 
\label{fig:fs}}
\end{figure*}

For the low-\pt\ search, FS processes constitute the dominant background in SRC, 
comprising $>90$\% \ttbar, $\sim 8$\% $Wt$, with a very small contribution from $WW$ and $Z\rightarrow \tau\tau$. 
These backgrounds are modelled using MC simulation, with the dominant \ttbar\ and $Wt$ components being normalised to data in dedicated $e\mu$ CRs. 
The top-quark background normalisation in SRC is taken from CRC, while CRC-MET is used to extract the top-quark background normalisation for SRC-MET. 
The modelling of these backgrounds is tested in four VRs: VRA, VRA2, VRB and VRC,
where the normalisation for \ttbar\ and $Wt$ is $1.00\pm0.22$, $1.01\pm0.13$, $1.00\pm0.21$ and $0.86\pm0.13$, respectively, calculated from identical regions in the $e\mu$ channel.  
Figure~\ref{fig:VRs_lowpt} shows a comparison between data and prediction in these four VRs.
VRA probes low \ptll\ in the range equivalent to that in SRC, but at lower \met, 
while VRB and VRC are used to check the background modelling at \ptll$>20$~\GeV, but with \met\ between 250 and 500~\GeV. 
Owing to poor background modelling at very low \mll\ and \ptll, the \mll\ range in VRA and SRC does not go below $30$~\GeV.

\begin{figure*}[hb!t]
\centering
\includegraphics[width=.45\textwidth]{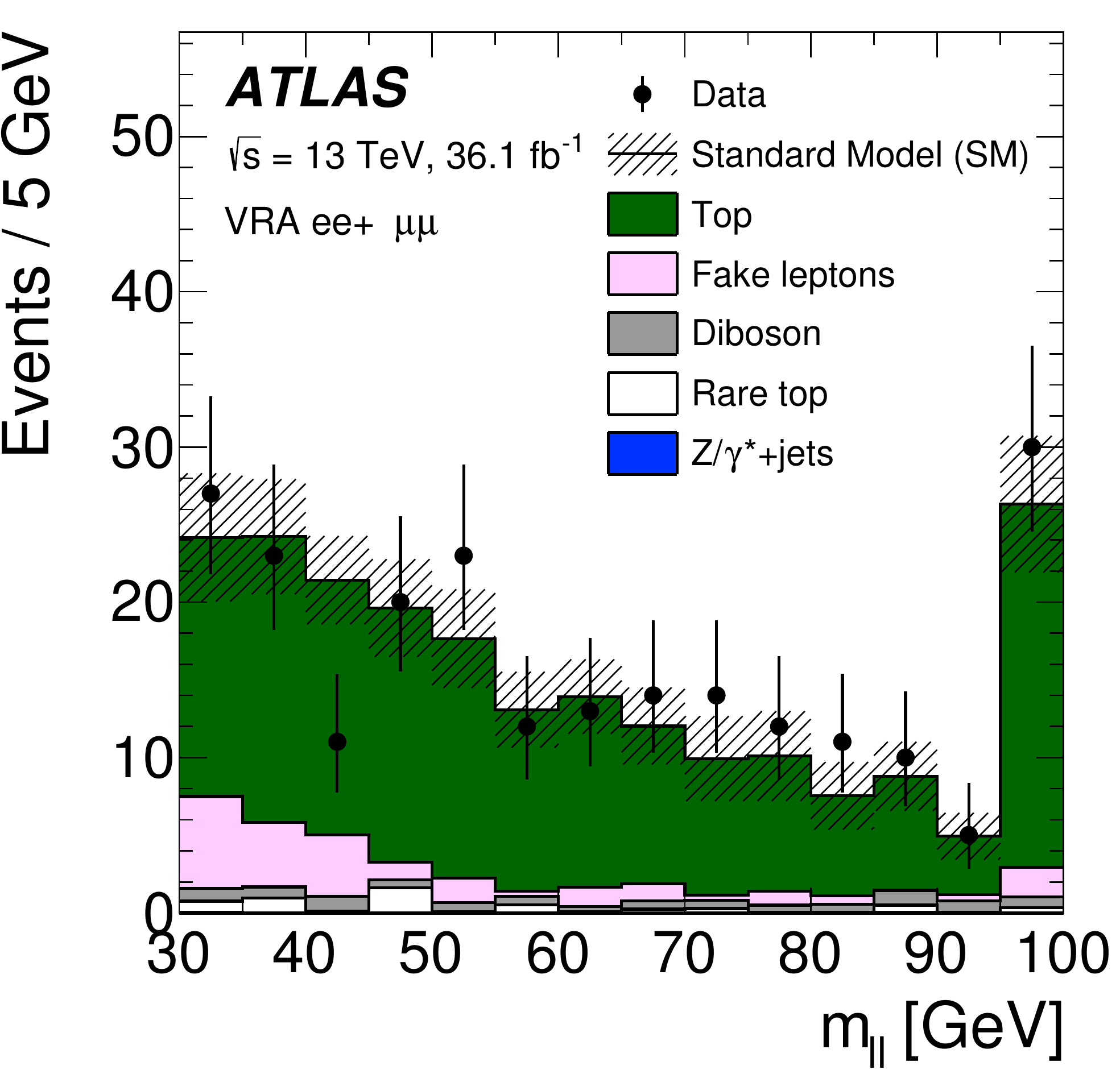}
\includegraphics[width=.45\textwidth]{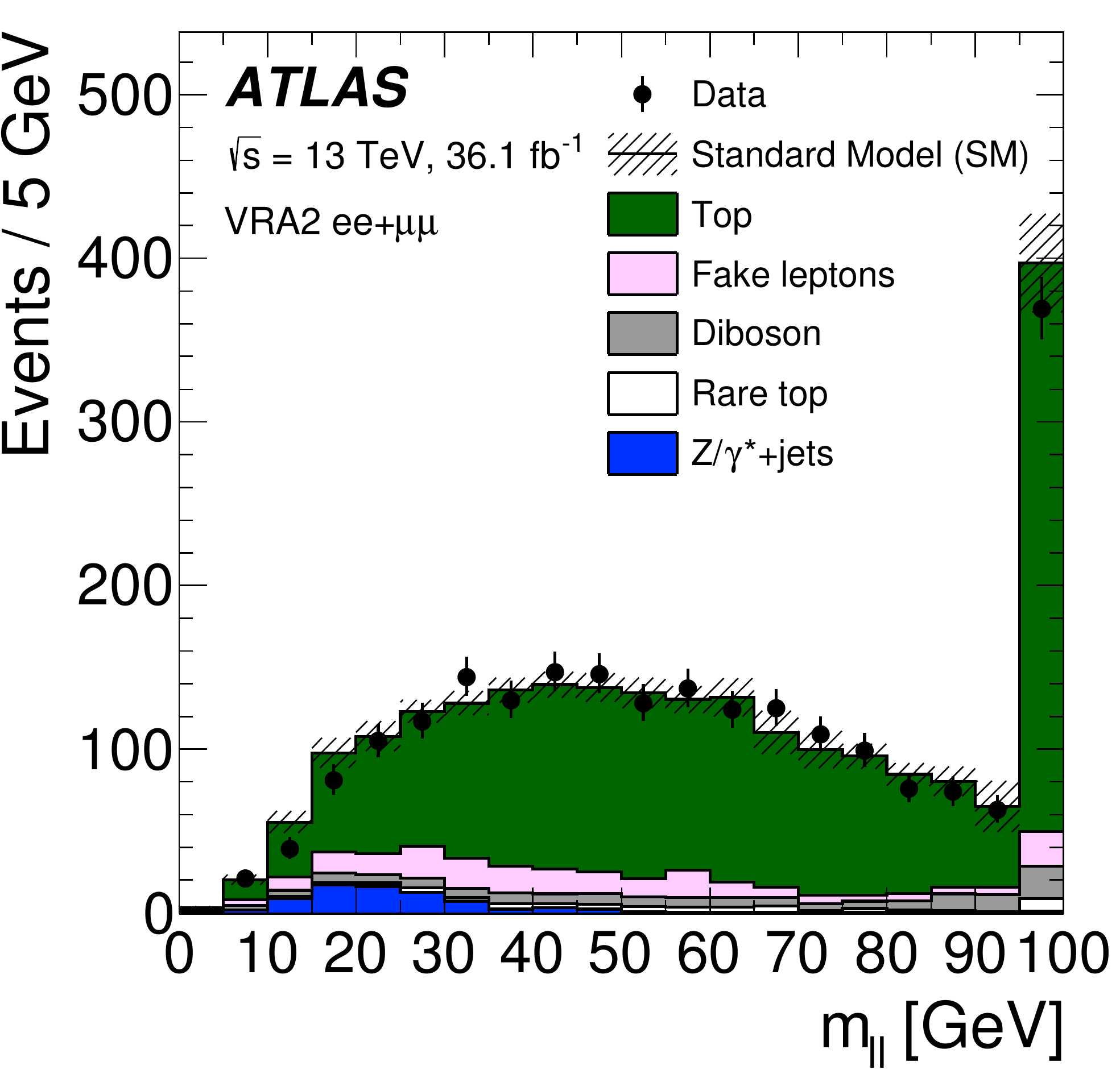}
\includegraphics[width=.45\textwidth]{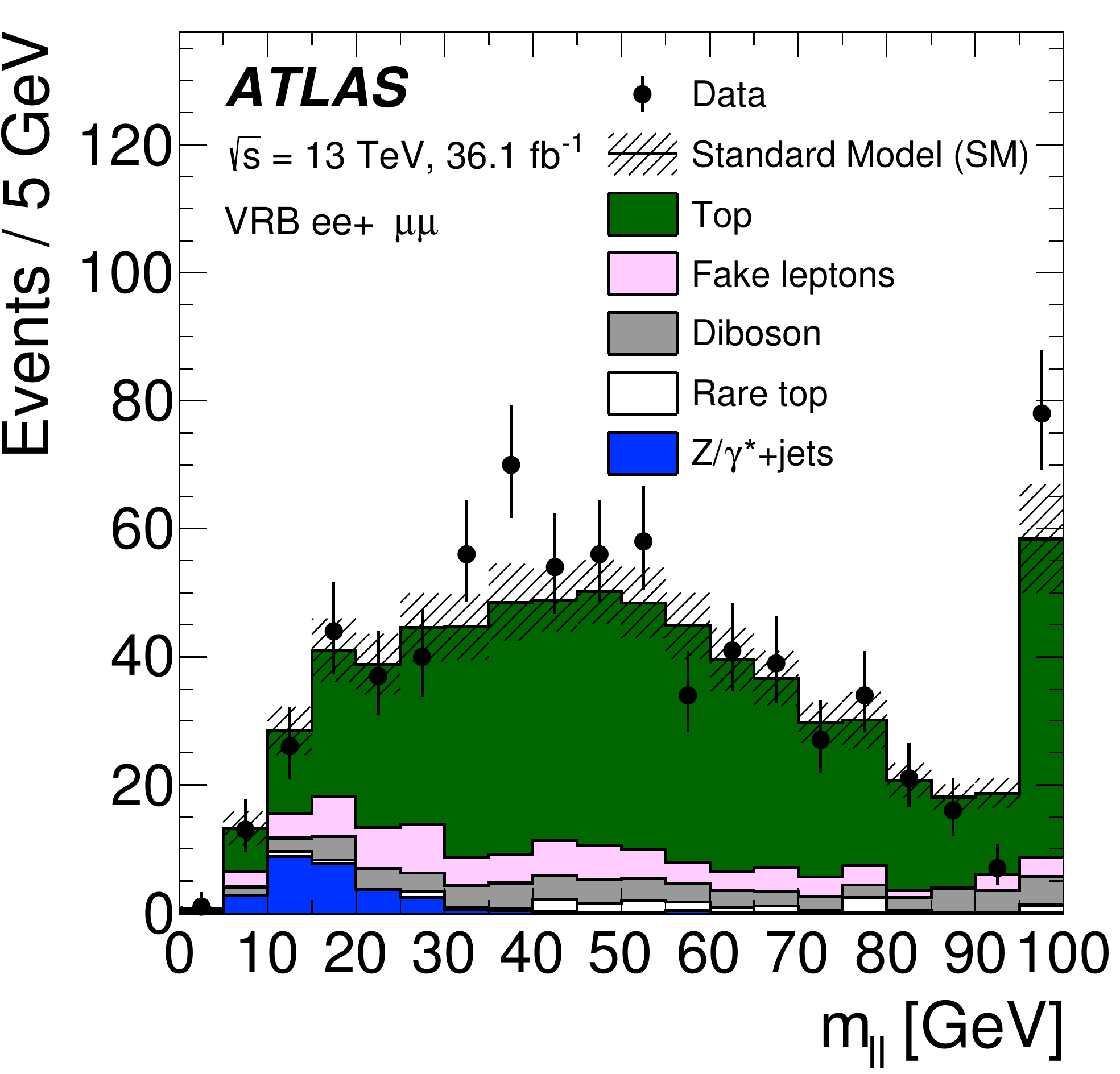}
\includegraphics[width=.45\textwidth]{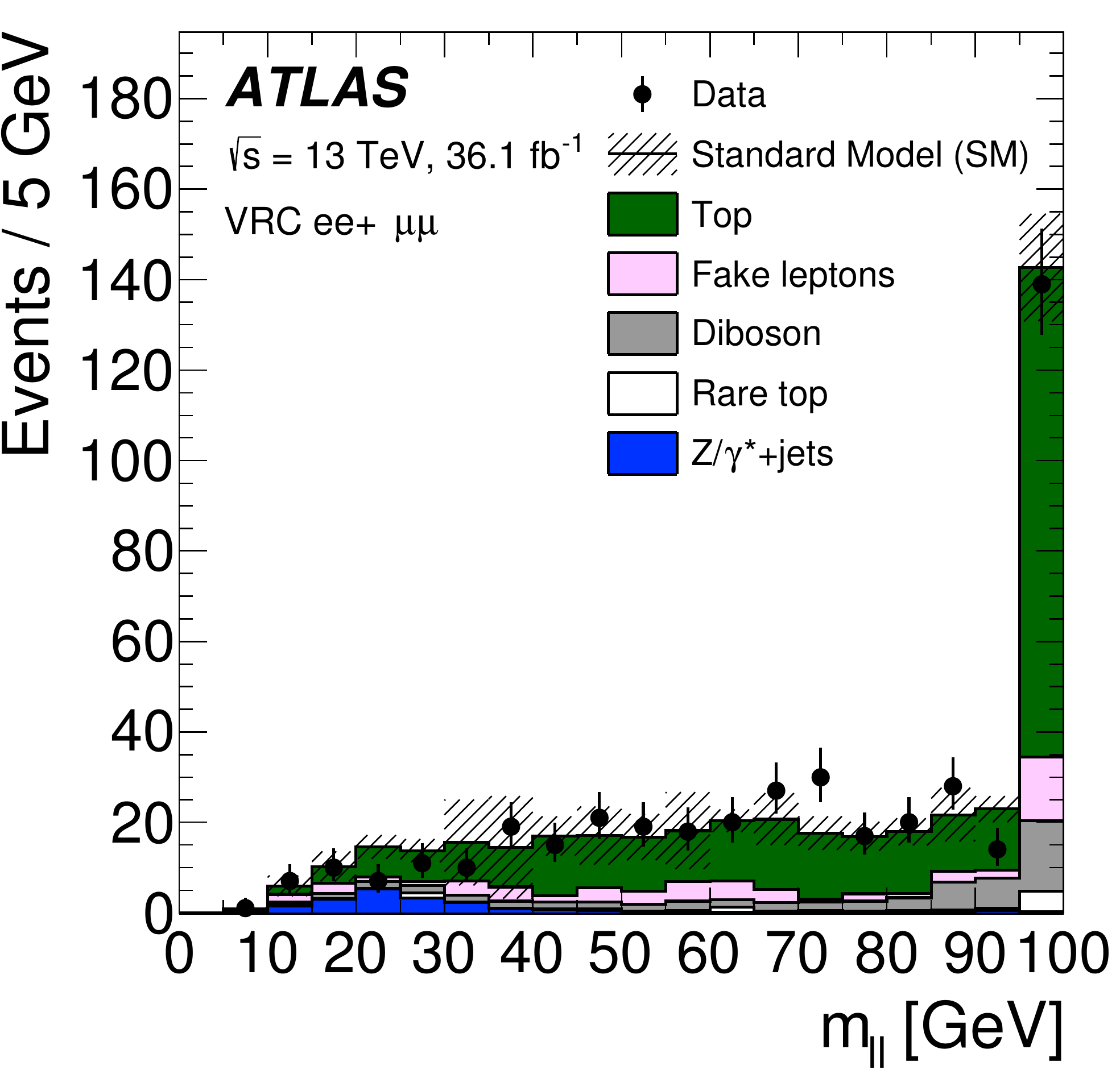}
\caption{
Validation of the background modelling for the \lowpt\ analysis in VRA (top left), 
VRA2 (top right), VRB (bottom left) and VRC (bottom right) in the SF channels. 
The \ttbar\ and $Wt$ backgrounds are normalised in $e\mu$ data samples for which the requirements are otherwise the same as in the VR in question.
All uncertainties in the background expectation are included in the hatched band.
The last bin always contains the overflow. }
\label{fig:VRs_lowpt}
\end{figure*}

\FloatBarrier

\subsection{\dyjets\ background}
\label{sec:zjets}

The \dyjets\ processes make up to $10$\% of the background in the on-$Z$ \mll\ bins in SR-low, SR-medium and SR-high.
For the high-\pt\ analysis this background is estimated using a data-driven method that takes \gjets\ events in data to model the \met\ distribution of \dyjets.
These two processes have similar event topologies, with a well-measured object recoiling against a hadronic system, 
and both tend to have \met\ that stems from jet mismeasurements and neutrinos in hadron decays. 
In this method, different control regions (CR$\gamma$-low, CR$\gamma$-medium, CR$\gamma$-high) are constructed, 
which contain at least one photon and no leptons.  They have the same kinematic selection as their corresponding SRs, 
with the exception of \met\ and $\Delta\phi(\text{jet}_{12},{\boldsymbol p}_{\mathrm{T}}^\mathrm{miss})$ requirements. 
Detailed definitions of these regions are given in Table~\ref{tab:regions-edge}.

The \gjets\ events in CR$\gamma$-low, CR$\gamma$-medium and CR$\gamma$-high are reweighted such that the photon \pt\ distribution matches that of the \dyjets\ dilepton $\pt$ distribution of events in CRZ-low, 
CRZ-medium and CRZ-high, respectively.
This procedure accounts for small differences in event-level kinematics between the \gjets\ events and \dyjets\ events,
which arise mainly from the mass of the $Z$ boson. 
Following this, to account for the difference in resolution between photons, electrons, and muons, which can be particularly significant at high boson \pt, 
the photon \pt\ is smeared according to a $Z\rightarrow ee$ or $Z\rightarrow\mu\mu$ resolution function. 
The smearing function is derived by comparing the ${\boldsymbol p}_{\mathrm{T}}^\mathrm{miss}$-projection along the boson momentum in \dyjets\ and \gjets\ MC events in a 1-jet control region with no other event-level kinematic requirements.
A deconvolution procedure is used to avoid including the photon resolution in the $Z$ bosons's \pT\ resolution function.
For each event, a photon \pt\ smearing $\Delta\pt$ is obtained by sampling the smearing function.
The photon \pt\ is shifted by $\Delta\pt$, with the parallel component of the ${\boldsymbol p}_{\mathrm{T}}^\mathrm{miss}$ vector being correspondingly adjusted by $-\Delta\pt$.

Following this smearing and reweighting procedure, the \met\ of each \gjets\ event is recalculated, 
and the final \met\ distribution is obtained after applying the $\Delta\phi(\text{jet}_{12},{\boldsymbol p}_{\mathrm{T}}^\mathrm{miss} ) > 0.4$ requirement.
For each SR, the resulting \met\ distribution is normalised to data in the corresponding CRZ before the SR \met\ selection is applied.
The \mll\ distribution is modelled by binning the \mll\ in \dyjets\ MC events as a function of the ${\boldsymbol p}_{\mathrm{T}}^\mathrm{miss}$-projection along the boson momentum, 
with this being used to assign an \mll\ value to each \gjets\ event via a random sampling of the corresponding distribution.
The \mttwo\ distribution is modelled by assigning leptons to the event, with the direction of the leptons drawn from a flat distribution in the $Z$ boson rest frame.
The process is repeated until both leptons fall into the detector acceptance after boosting to the lab frame.

The full smearing, reweighting, and \mll\ assignment procedure is applied to both the $V\gamma$ MC and the \gjets\ data events. 
After applying all corrections to both samples, the $V\gamma$ contribution to the \gjets\ data sample is subtracted to remove contamination from the main backgrounds with real \met\ from neutrinos.
Contamination by events with fake photons in these \gjets\ data samples is small, and as such this contribution is neglected.
 
The procedure is validated using \gjets\ and \dyjets\ MC events. 
For this validation, the \gjets\ MC simulation is reweighted according to the \pt\ distribution given by the \dyjets\ MC simulation.
The \dyjets\ \met\ distribution in MC events can be seen on the left of Figure~\ref{fig:VRs_zjets} and is found to be well reproduced by \gjets\ MC events. 
In addition to this, three VRs, VR-$\Delta\phi$-low, VR-$\Delta\phi$-medium and VR-$\Delta\phi$-high, 
which are orthogonal to SR-low SR-medium and SR-high due to the inverted $\Delta\phi(\text{jet}_{12},{\boldsymbol p}_{\mathrm{T}}^\mathrm{miss})$ requirement, 
are used to validate the method with data.
Here too, as shown on the right of Figure~\ref{fig:VRs_zjets}, good agreement is seen between the \dyjets\ prediction from \gjets\ data and the data in the three VRs.
The systematic uncertainties associated with this method are described in Section~\ref{sec:syst}.

\begin{figure*}[!htb]
\centering
\includegraphics[width=.42\textwidth]{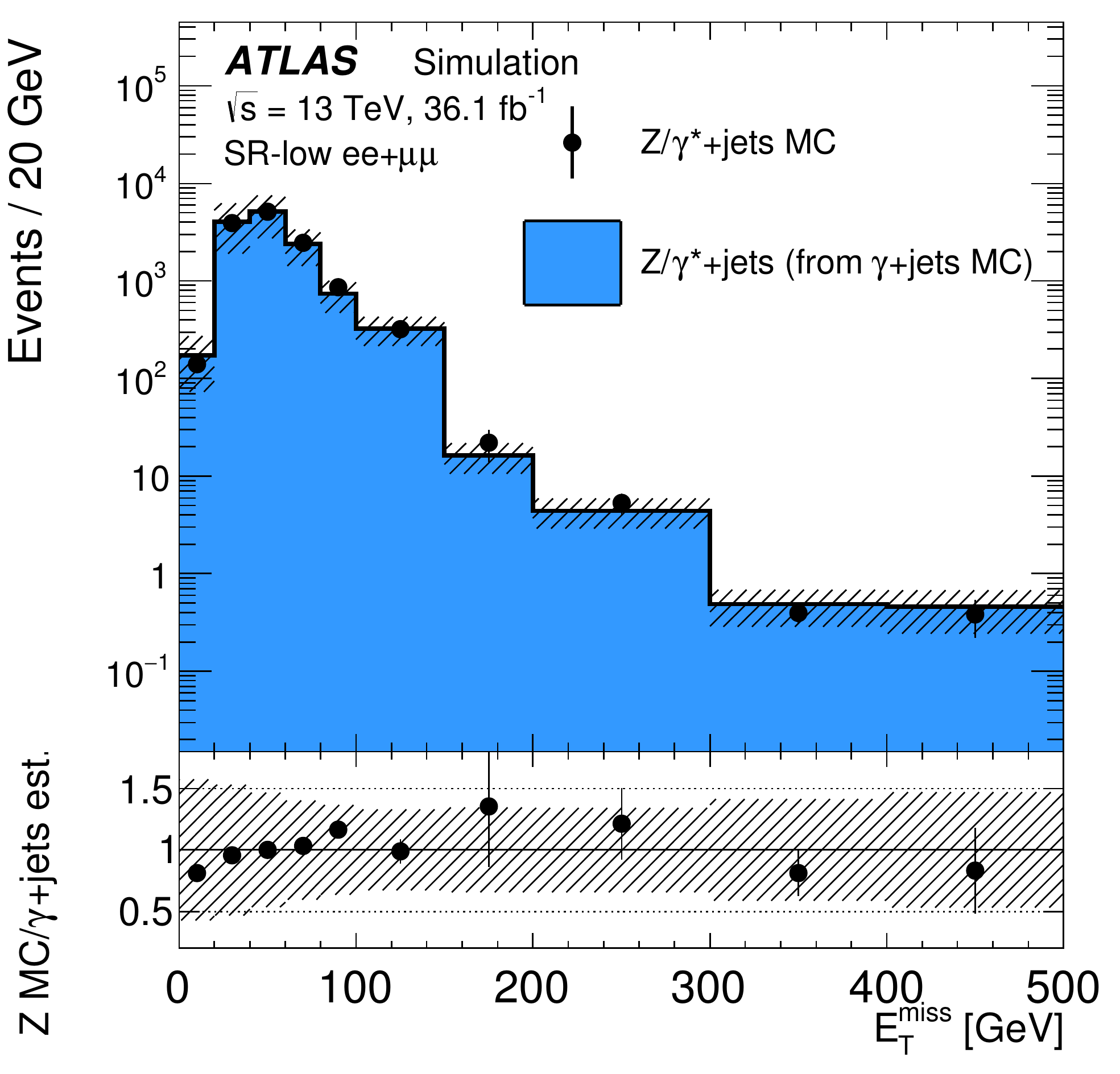}
\includegraphics[width=.42\textwidth]{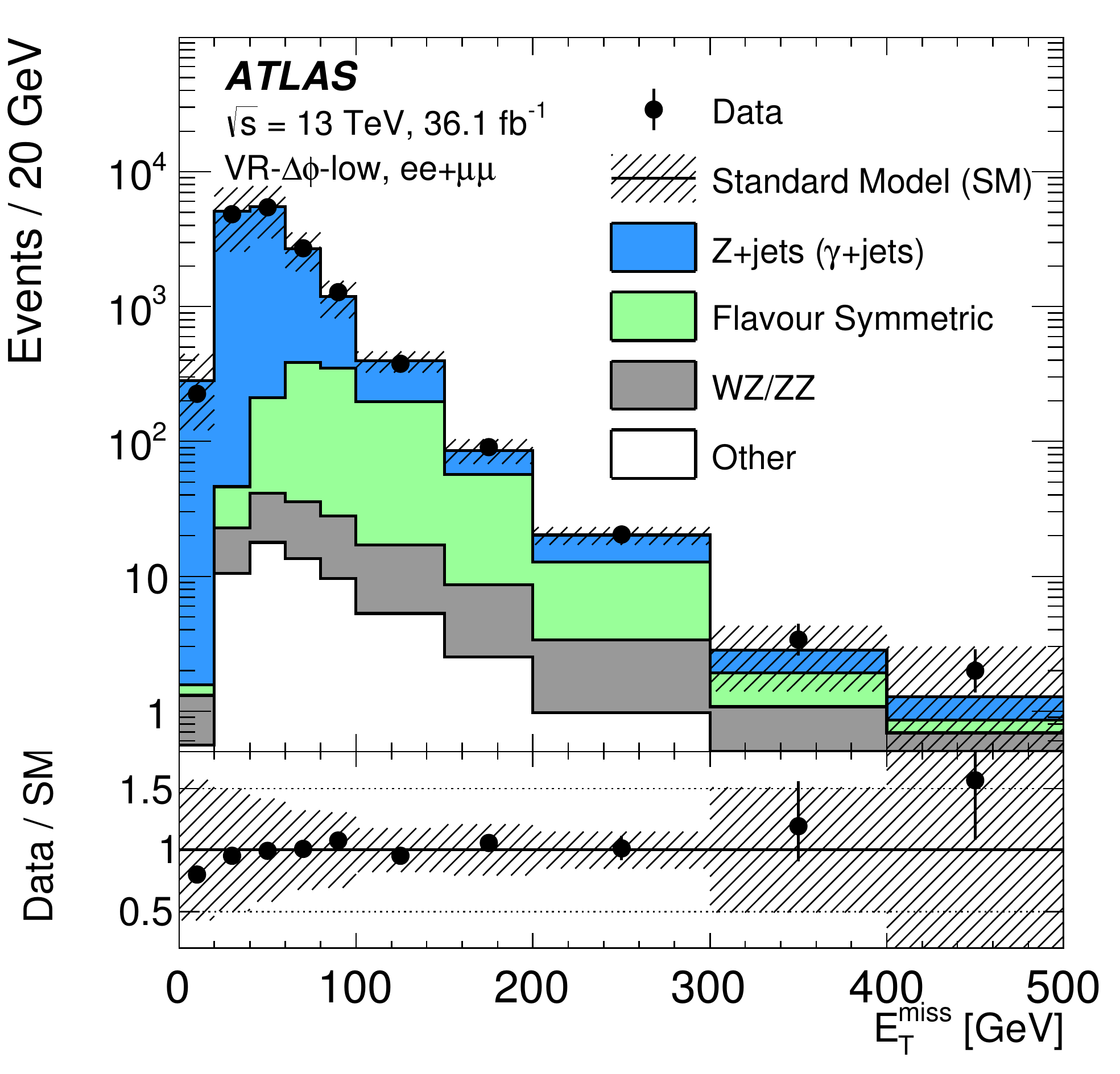}
\includegraphics[width=.42\textwidth]{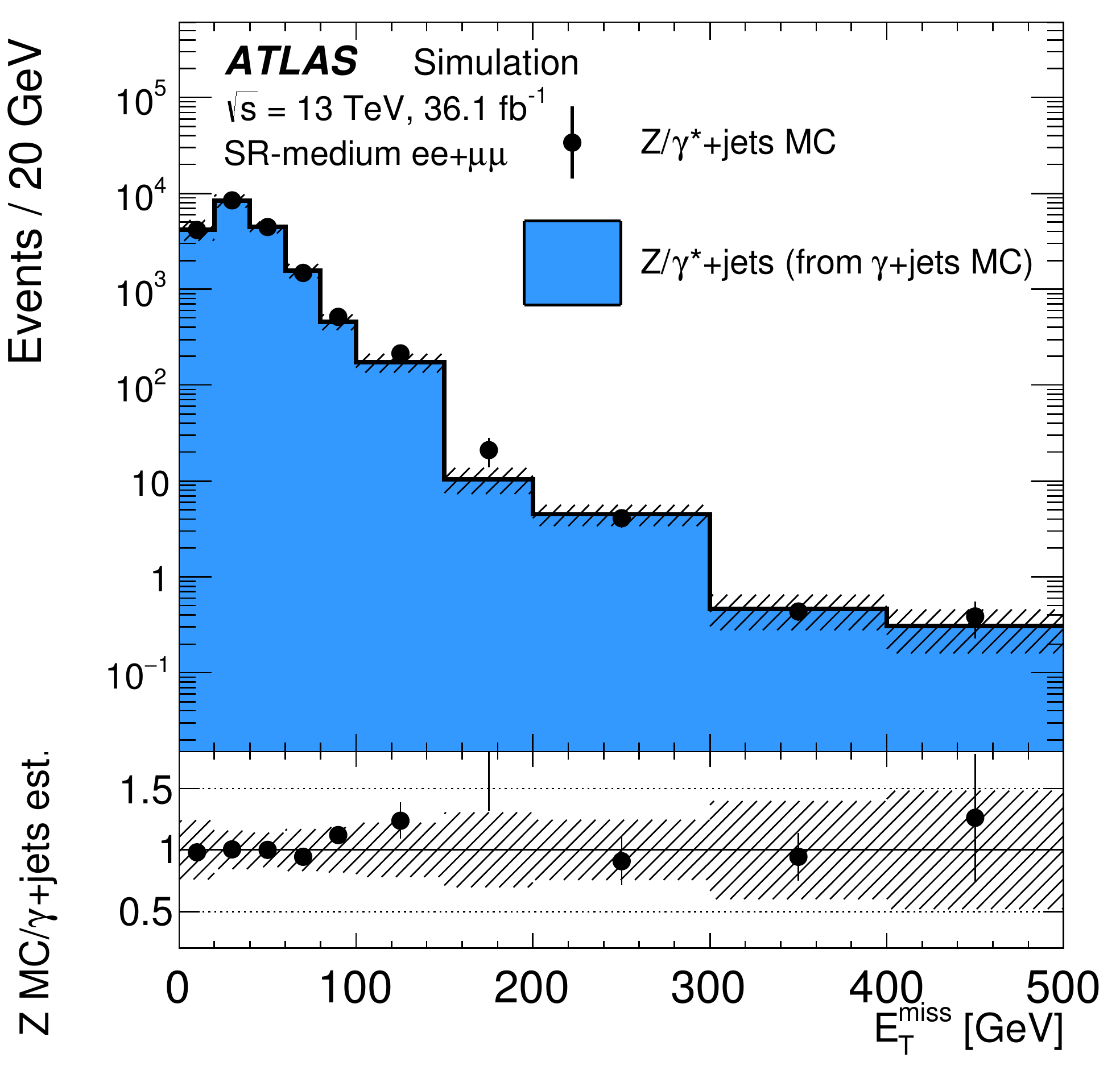}
\includegraphics[width=.42\textwidth]{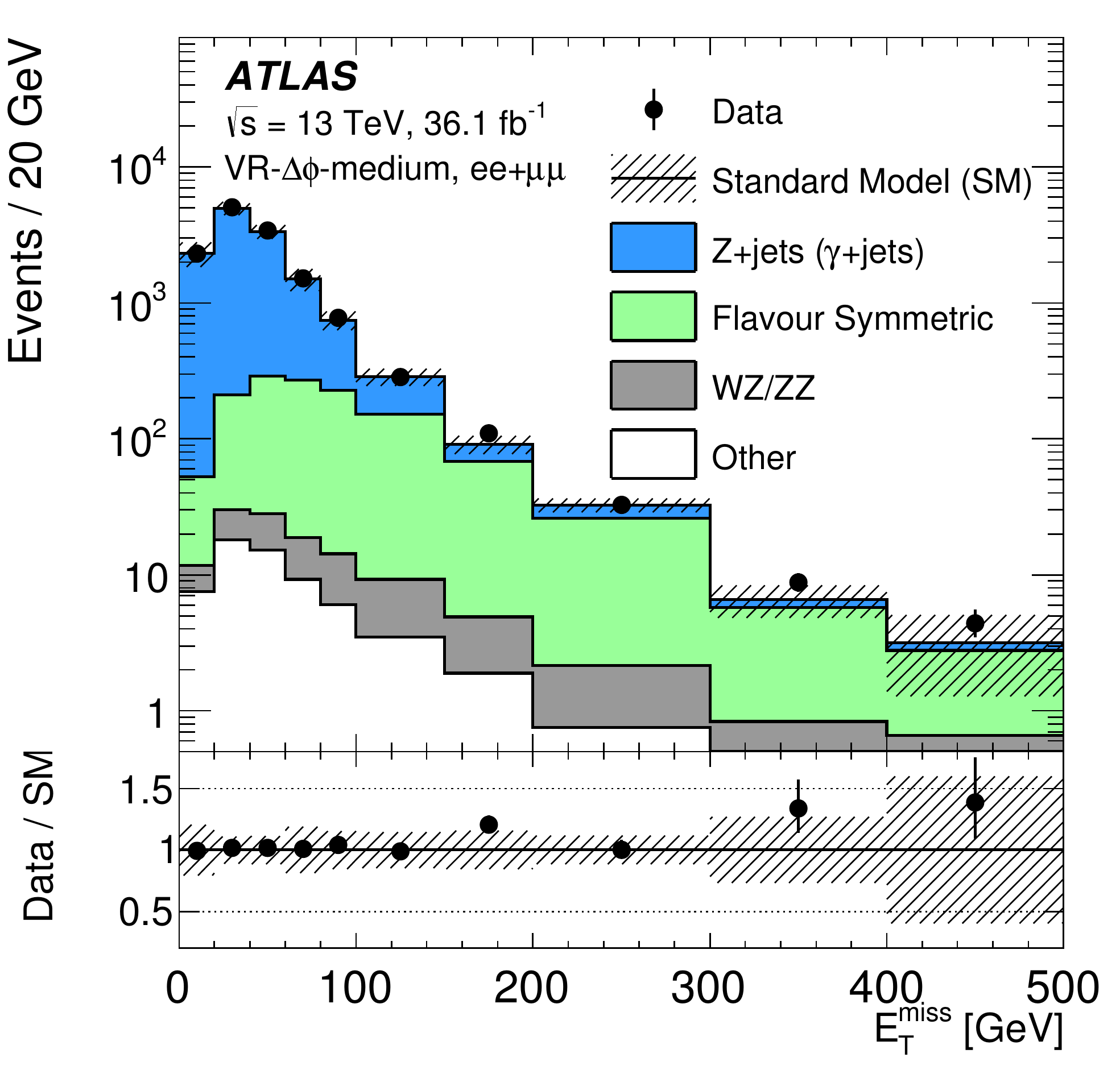}
\includegraphics[width=.42\textwidth]{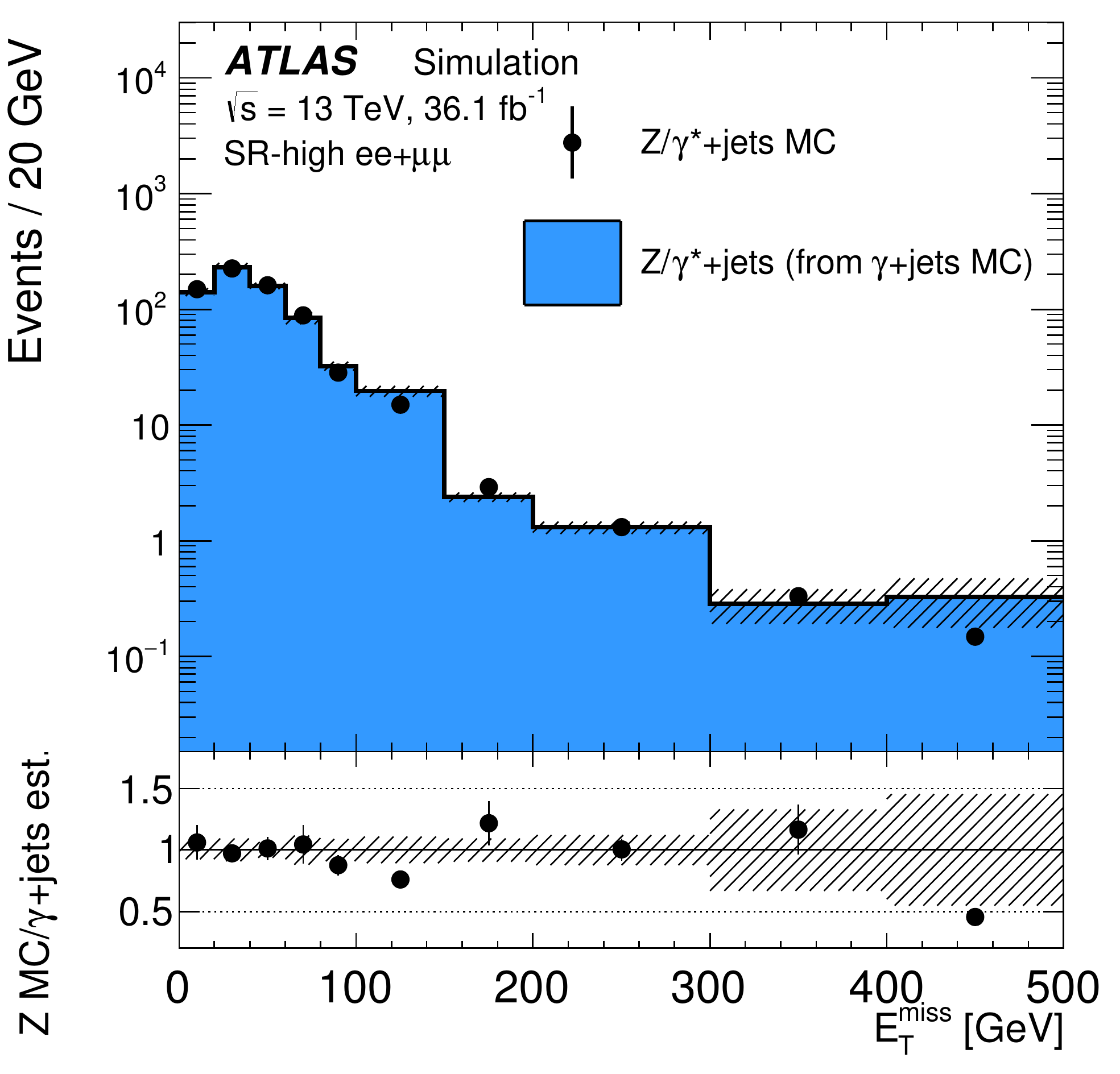}
\includegraphics[width=.42\textwidth]{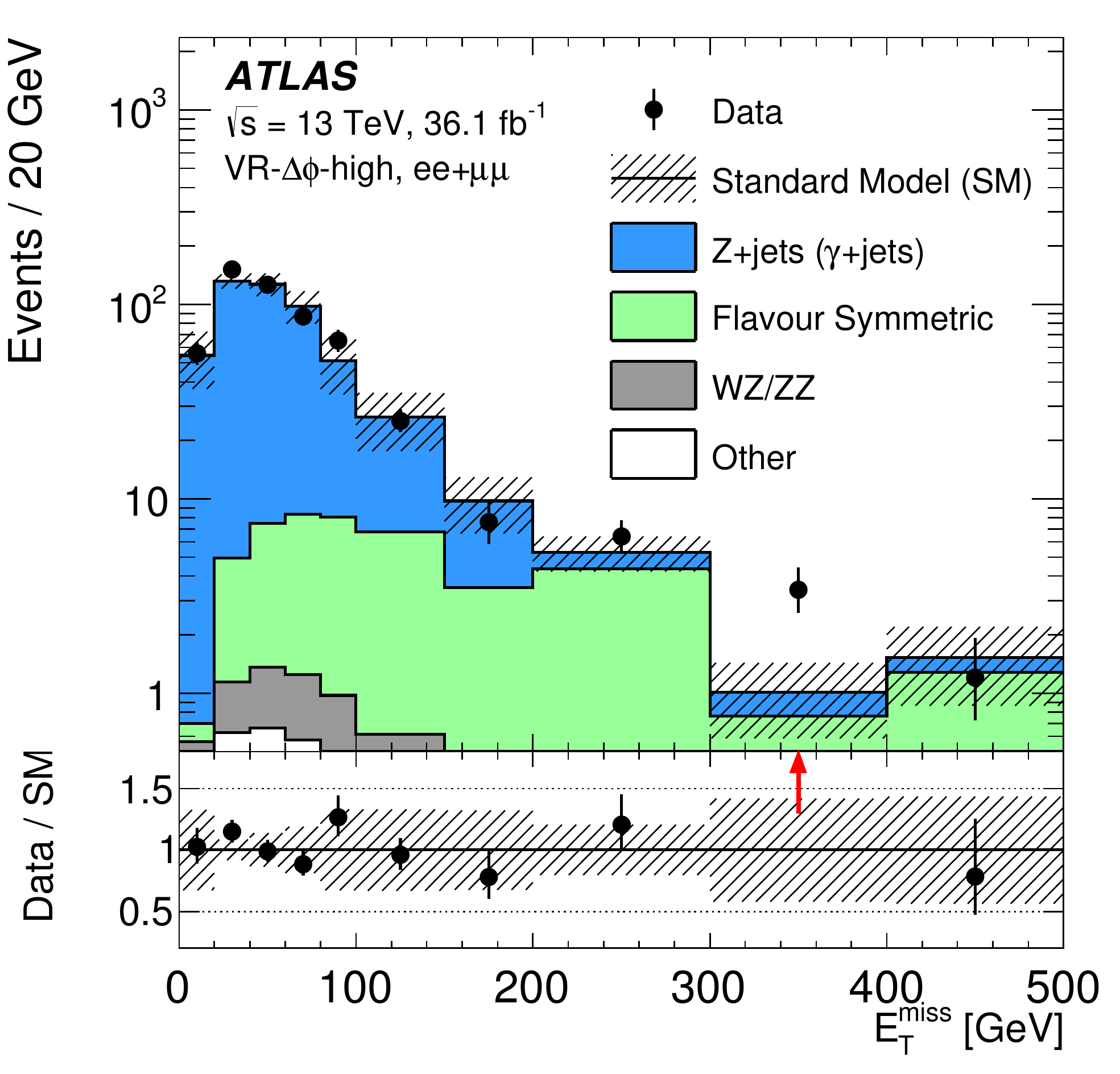}
\caption{
Left, the \met\ spectrum in \dyjets\ MC simulation compared to that of the \gjets\ method applied to 
\gjets\ MC simulation in SR-low (top), SR-medium (middle) and SR-high (bottom).  No selection on \met\ is applied.
The error bars on the points indicate the statistical uncertainty of the \dyjets\ MC simulation,
and the hashed uncertainty bands indicate the statistical and reweighting systematic uncertainties of the $\gamma+$jet background method.
Right, the \met\ spectrum when the method is applied to data in VR-$\Delta\phi$-low (top), VR-$\Delta\phi$-medium (middle) and VR-$\Delta\phi$-high (bottom).
The bottom panel of each figure shows the ratio of observation (left, in MC simulation; right, in data) to prediction.
In cases where the data point is not accommodated by the scale of this panel, 
an arrow indicates the direction in which the point is out of range.
The last bin always contains the overflow. 
}
\label{fig:VRs_zjets}
\end{figure*}

\FloatBarrier

While the \gjets\ method is used in the high-\pt\ analysis, 
\sherpa\ \dyjets\ simulation is used to model this background in the \lowpt\ analysis. 
This background is negligible in the very low \ptll\ SRC, and while it can contribute up to $\sim 30$\% in some \mll\ bins in SRC-MET, 
this is in general only a fraction of a small total number of expected events. 
In order to validate the \dyjets\ estimate in this \lowpt\ region, the data are compared to the MC prediction in VR-$\Delta\phi$, 
where the addition of a $b$-tagged-jet veto is used to increase the \dyjets\ event fraction. 
The resulting background prediction in this region is consistent with the data.

\subsection{Fake-lepton background}
\label{sec:fakes}

Events from semileptonic \ttbar, $W\rightarrow \ell\nu$ and single top (s- and t-channel) decays enter the dilepton channels via lepton ``fakes.''
These can include misidentified hadrons, converted photons or non-prompt leptons from heavy-flavour decays. 
In the high-\pt\ SRs the contribution from fake leptons is negligible, but fakes can contribute up to $\sim 12$\% in SRC and SRC-MET. 
In the \lowpt\ analysis this background is estimated using the matrix method, detailed in Ref.~\cite{SUSY-2013-20}.
In this method a control sample is constructed using baseline leptons, 
thereby enhancing the probability of selecting a fake lepton compared to the signal-lepton selection. 
For each relevant CR, VR or SR, the region-specific kinematic requirements are placed upon this sample of baseline leptons. 
The events in this sample in which the selected leptons subsequently pass ($N_{\text{pass}}$) or fail ($N_{\text{fail}}$) 
the signal lepton requirements of Section~\ref{sec:objects} are then counted. 
In the case of a one-lepton selection, the number of fake-lepton events ($N_{\text{pass}}^{\text{fake}}$) in a given region is then estimated according to:

\begin{equation*}
N_{\text{pass}}^{\text{fake}} = \frac{N_{\text{fail}} - (1/\epsilon^{\text{real}} - 1) \times N_{\text{pass}} }{1/\epsilon^{\text{fake}} - 1/\epsilon^{\text{real}}}.
\end{equation*}

\noindent Here $\epsilon^{\text{real}}$ is the relative identification efficiency (from baseline to signal) for 
genuine, prompt (``real'') leptons and $\epsilon^{\text{fake}}$ is the relative identification efficiency (again from baseline to signal) with which non-prompt leptons or jets might be misidentified as prompt leptons. 
This principle is then expanded to a dilepton selection by using a four-by-four matrix to account for the various possible real--fake combinations for the two leading leptons in an event.

The real-lepton efficiency, $\epsilon^{\text{real}}$, is measured in $Z\rightarrow\ell\ell$ data events using a tag-and-probe method in CR-real, 
defined in Table~\ref{tab:regions-lowpt}. 
In this region the \pt\ of the leading lepton is required to be $>40$~\GeV, and only events with exactly two SFOS leptons are selected. 
The efficiency for fake leptons, $\epsilon^{\text{fake}}$, is measured in CR-fake, a region enriched with fake leptons by requiring same-sign lepton pairs. 
The lepton \pt\ requirements are the same as those in CR-real, with the leading lepton being tagged as the ``real'' lepton and the fake-lepton efficiency being evaluated using the sub-leading lepton in the event. 
A requirement of $\met<125~\GeV$ is used to reduce possible contamination from non-SM processes (e.g. SUSY).
In this region, the background due to prompt-lepton production, estimated from MC simulation, is subtracted from the total data contribution. 
Prompt-lepton production makes up $7$\% ($10$\%) of the baseline electron (muon) sample and $10$\% ($60$\%) of the signal electron (muon) sample in CR-fake. 
From the resulting data sample the fraction of events in which the baseline leptons pass the signal selection requirements yields the fake-lepton efficiency.  
The \pt\ and $\eta$ dependence of both fake- and real-lepton efficiencies is taken into account.

This method is validated in an OS VR, VR-fakes, which covers a region of phase space similar to that of the \lowpt\ SRs, but with a DF selection. 
The left panel of Figure~\ref{fig:fakesVR} shows the level of agreement between data and prediction in this region.
In the SF channels, an SS selection is used to obtain a VR, VR-SS in Table~\ref{tab:regions-lowpt}, dominated by fake leptons. 
The data-driven prediction is close to the data in this region, as shown on the right of Figure~\ref{fig:fakesVR}.
The large systematic uncertainty in this region is mainly from the flavour composition, as described in Section~\ref{sec:syst}.

\begin{figure}[htbp]
\centering
\includegraphics[width=0.49\textwidth]{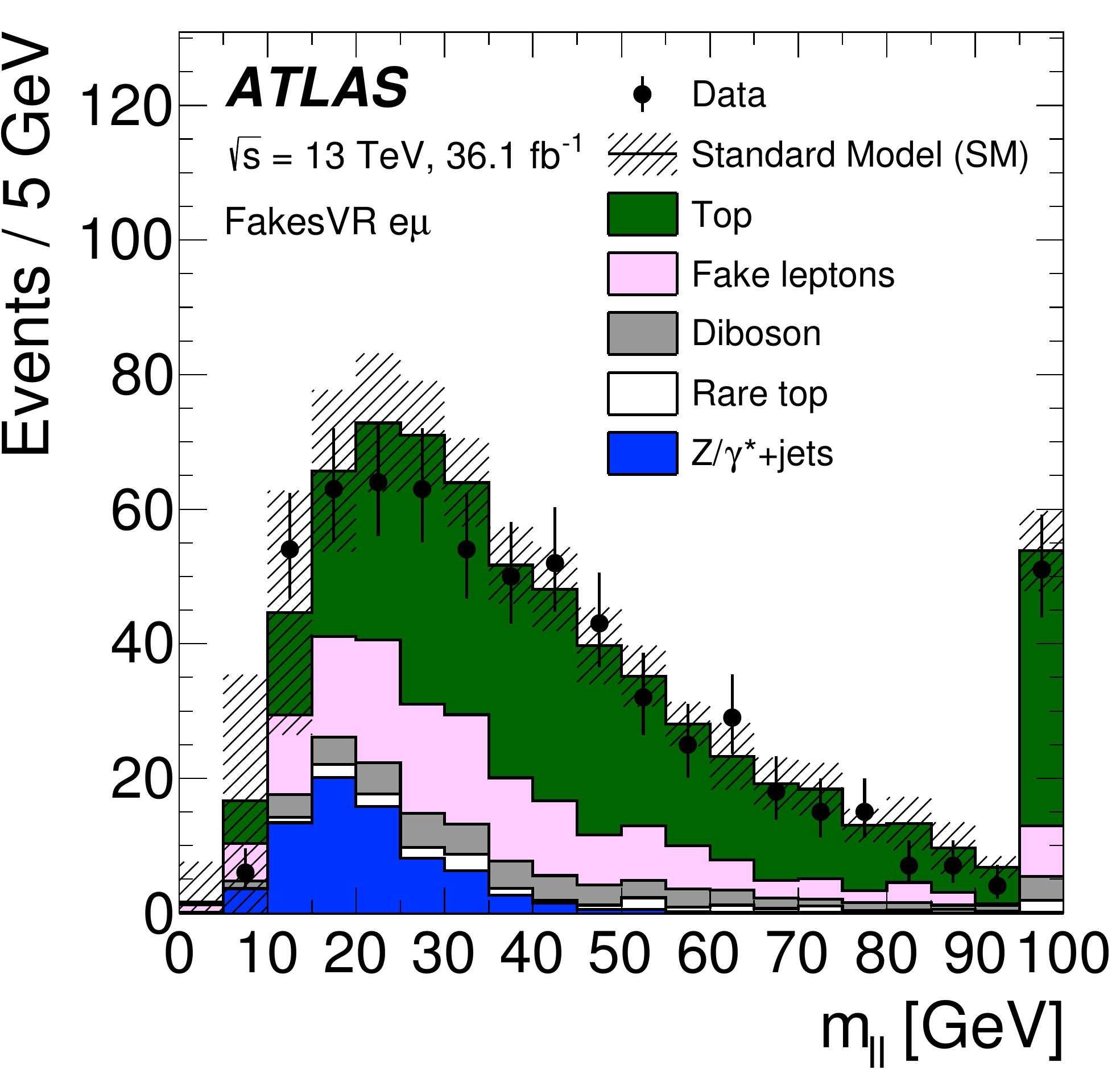}
\includegraphics[width=0.49\textwidth]{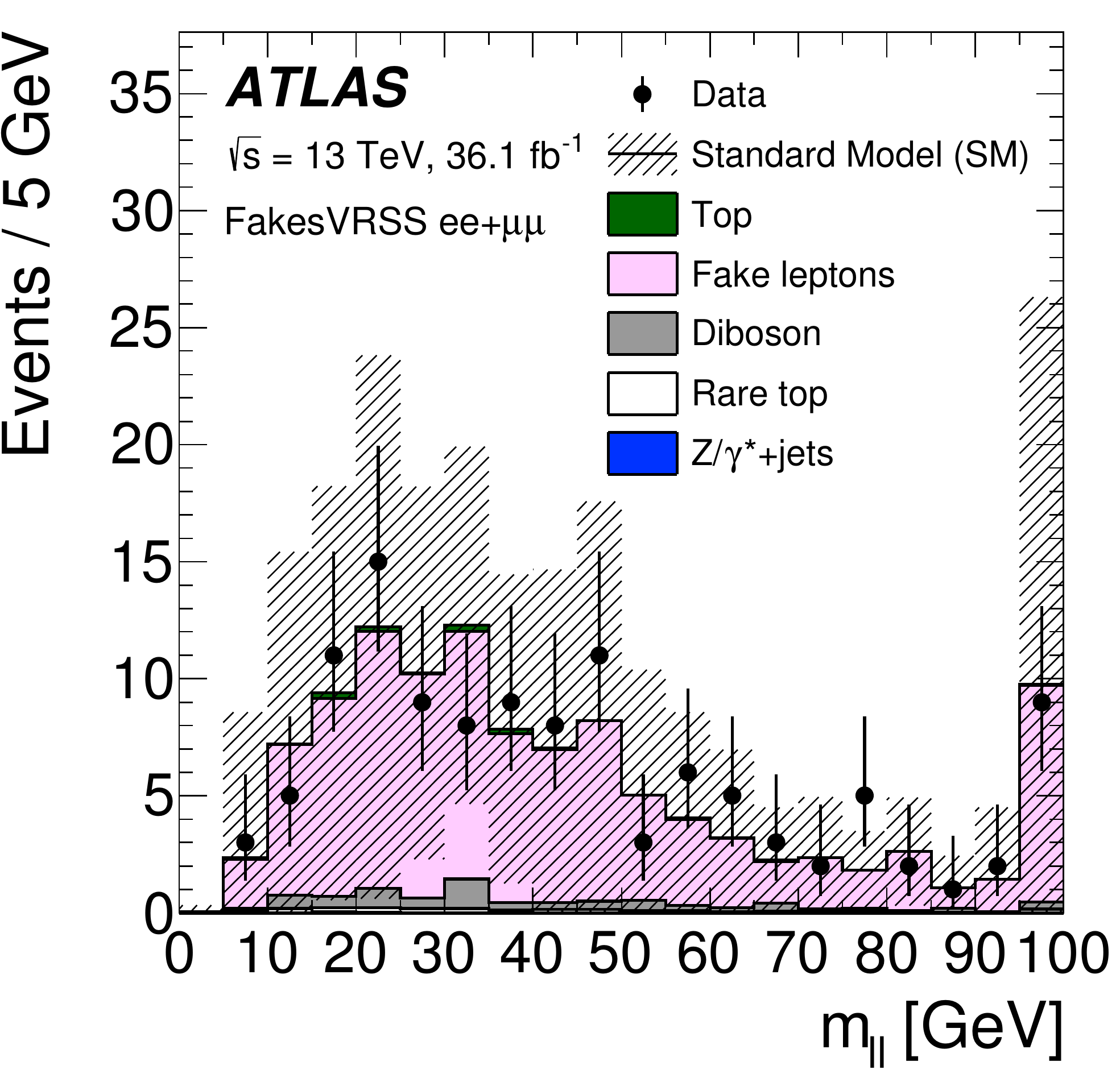}
\caption{Validation of the data-driven fake-lepton background for the \lowpt\ analysis. 
The \mll\ distribution in VR-fakes (left) and VR-SS (right). 
Processes with two prompt leptons are modelled using MC simulation. 
The hatched band indicates the total systematic and statistical uncertainty of the background prediction.
The last bin always contains the overflow.  }
\label{fig:fakesVR}
\end{figure}

\FloatBarrier

\subsection{Diboson and rare top processes}
\label{sec:diboson_other}
The remaining SM background contribution in the SRs is due to $WZ/ZZ$ diboson production and rare top processes 
($\antibar{t}Z$, $\antibar{t}W$ and $\antibar{t}WW$). 
The rare top processes contribute $<10\%$ of the SM expectation in the SRs and are taken directly from MC simulation.

The contribution from the production of $WZ/ZZ$ dibosons is generally small in the SRs, but in the on-$Z$ bins in the high-\pt\ SRs it is up to $70$\% of the expected background, whereas in SRC-MET it is up to $40\%$ of the expected background. 
These backgrounds are estimated from MC simulation, 
and are validated in VRs with three-lepton (VR-WZ) and four-lepton (VR-ZZ) requirements, as defined in Table~\ref{tab:regions-edge}.
VR-$WZ$, with $\HT>200$~\GeV, forms a $WZ$-enriched region in a kinematic phase space as close as possible to the high-\pt\ SRs. 
In VR-ZZ an $\met<100$~\GeV\ requirement is used to suppress $WZ$ and top processes to form a region with high purity in $ZZ$ production.
The yields and kinematic distributions observed in these regions are well-modelled by MC simulation.
In particular, the \met, $H_\text{T}$, jet multiplicity, and dilepton $\pt$ distributions show good agreement. 
For the \lowpt\ analysis, VR-WZ-\lowpt\ and VR-ZZ-\lowpt, defined in Table~\ref{tab:regions-lowpt}, are used to check the modelling of these processes at low lepton \pt, and good modelling is also observed.
Figure~\ref{fig:db_VR_summary} shows the level of agreement between data and prediction in these validation regions.

\begin{figure}[htbp]
\centering
\includegraphics[width=0.8\textwidth]{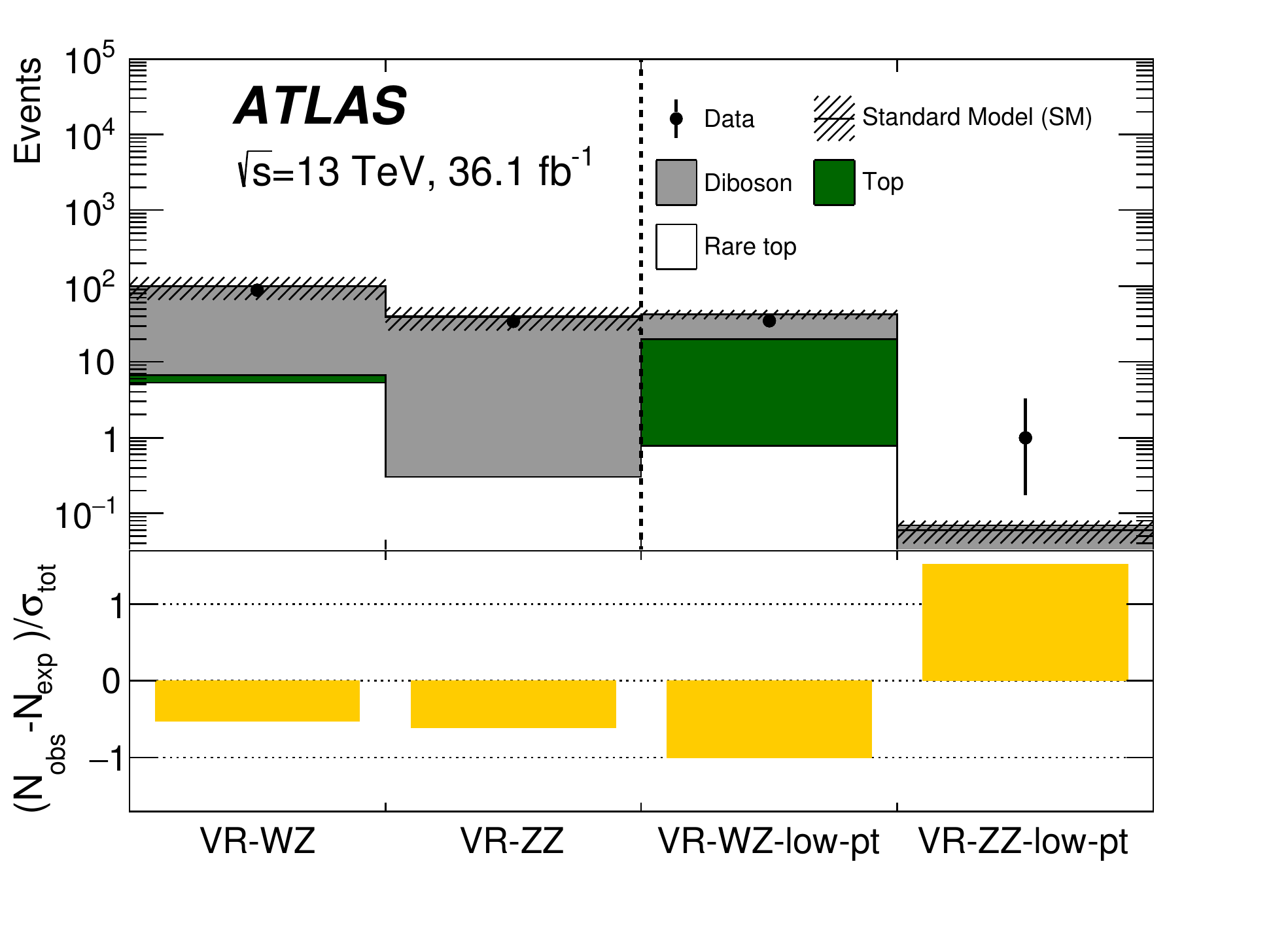}
\caption{
The observed and expected yields in the diboson VRs. 
The data are compared to the sum of the expected backgrounds. 
The observed deviation from the expected yield normalised to the total uncertainty is shown in the bottom panel.
The hatched uncertainty band includes the statistical and systematic uncertainties of the background prediction.
\label{fig:db_VR_summary}}
\end{figure}

\section{Systematic uncertainties}
\label{sec:syst}
The data-driven background estimates are subject to uncertainties associated with the methods employed and the limited 
number of events used in their estimation. 
The dominant source of uncertainty for the flavour-symmetry-based background estimate in the high-\pt\ SRs is due to the limited statistics in the corresponding DF CRs, 
yielding an uncertainty of between $10\%$ and $90$\% depending on the \mll\ range in question.  
Other systematic uncertainties assigned to this background estimate include those due to 
MC closure, the measurement of the efficiency correction factors and the extrapolation in \met\ and \HT\ in the case of SR-high.

Several sources of systematic uncertainty are associated with the data-driven \dyjets\ background prediction for the high-\pt\ analysis. 
The boson $\pt$ reweighting procedure is assigned an uncertainty based on a comparison of the nominal results with those obtained by reweighting events using
the \HT\ distribution instead. 
For the smearing function 
an uncertainty is derived by comparing the results obtained using the nominal smearing function derived from MC simulation with those obtained using a smearing function derived from data in a 1-jet control region. 
The full reweighting and smearing procedure is carried out using \gjets\ MC events such that an MC non-closure uncertainty can be derived 
by comparing the resulting \gjets\ MC \met\ distribution to that in \dyjets\ MC events. 
An uncertainty of 10\% is obtained for the $V\gamma$ backgrounds, 
based on a data-to-MC comparison in a $V\gamma$-enriched control region where events are required to have a photon and one lepton. 
This uncertainty is propagated to the final \dyjets\ estimate following the subtraction of the $V\gamma$ background. 
Finally, the statistical precision of the estimate also enters as a systematic uncertainty in the final background estimate. 
Depending on the \mll\ range in question, the uncertainties in the \dyjets\ prediction can vary from $\sim 10$\% to $>100$\%. 

For the \lowpt\ analysis the uncertainties in the fake-lepton background stem from the number of events in the regions used to measure the real- and fake-lepton efficiencies, 
the limited sample size of the inclusive loose-lepton sample,
varying the prompt-lepton contamination in the region used to measure the fake-lepton efficiency,
and from varying the region used to measure the fake-lepton efficiency.  
The nominal fake-lepton efficiency is compared with those measured in regions where the presence of $b$-tagged jets is either required or explicitly vetoed. 
Varying the sample composition via $b$-jet tagging makes up the largest uncertainty.  

Theoretical and experimental uncertainties are taken into account for the signal models, as well as background processes that rely on MC simulation.
A $2.1\%$ uncertainty is applied to the luminosity measurement~\cite{DAPR-2013-01}. 
The jet energy scale is subject to uncertainties associated with the jet flavour composition, 
the pile-up and the jet and event kinematics~\cite{Aaboud:2017jcu}. 
Uncertainties in the jet energy resolution are included to account for differences between data and MC simulation~\cite{Aaboud:2017jcu}.
An uncertainty in the \met\ soft-term resolution and scale is taken into account~\cite{Aaboud:2018tkc}, 
and uncertainties due to the lepton energy scales and resolutions, as well as trigger, reconstruction, and identification efficiencies, are also considered.
The experimental uncertainties are generally $<1$\% in the SRs, with the exception of those associated with the jet energy scale, which can be up to $14$\% in the \lowpt\ SRs.

In the \lowpt\ analysis, theoretical uncertainties are assigned to the \mll-shape of the \ttbar\ and $Wt$ backgrounds, which are taken from MC simulation. 
For these backgrounds an uncertainty in the parton shower modelling is derived from comparisons between samples generated with {\sc Powheg+Pythia6} and {\sc Powheg+Herwig++} \cite{Bahr:2008pv,Bellm:2015jjp}.
For \ttbar\ an uncertainty in the hard-scatter process generation is assessed using samples generated using {\sc Powheg+Pythia8} to compare with {\sc MG5\_aMC@NLO+Pythia8}. 
Samples using either the diagram subtraction scheme or the diagram removal scheme to estimate interference effects in the single-top production diagrams are used to assess an interference uncertainty for the $Wt$ background~\cite{Re:2010bp}. 
Variations of the renormalisation and factorisation scales are taken into account for both \ttbar\ and $Wt$.

Again in the \lowpt\ analysis, theoretical uncertainties are assigned to the \dyjets\ background, which is also taken from MC simulation.
Variations of the renormalisation, resummation and factorisation scales are taken into account, as are parton shower matching scale uncertainties. 
Since the \dyjets\ background is not normalised to data, a total cross-section uncertainty of 5\% is assigned~\cite{STDM-2015-03}.

The $WZ/ZZ$ processes are assigned a cross-section uncertainty of $6\%$~\cite{ATL-PHYS-PUB-2017-005} and an additional uncertainty of up to $30\%$ in the SRs, which is based on comparisons between {\sc Sherpa} and {\sc Powheg} MC samples.
Uncertainties due to the choice of factorisation, resummation and renormalisation scales are calculated by varying the nominal values up and down by a factor of two. 
The parton shower scheme is assigned an uncertainty from a comparison of samples generated using the schemes proposed in Ref.~\cite{Schumann:2007mg} and Ref.~\cite{Hoeche:2009xc}.
These scale and parton shower uncertainties are generally $<20$\%. 
For rare top processes, a total uncertainty of 26\% is assigned to the cross-section ~\cite{Alwall:2014hca,Campbell:2012,Lazopoulos:2008,Garzelli:2012bn}.

For signal models, 
the nominal cross-section and its uncertainty are taken from an envelope of cross-section predictions using different PDF sets and factorisation and renormalisation scales, as described in Ref.~\cite{Borschensky:2014cia}. 

The uncertainties that have the largest impact in each SR vary from SR-to-SR. 
For most of the high-\pt\ SRs the dominant uncertainty is that due to the limited numbers of events in the $e\mu$ CRs used for the flavour-symmetric prediction.
Other important uncertainties include the systematic uncertainties associated with this method and uncertainties in the \gjets\ method for the \dyjets\ background prediction. 
In SRs that include the on-$Z$ \mll\ bin, diboson theory uncertainties also become important. 
The total uncertainty in the high-\pt\ SRs ranges from 12\% in the most highly populated SRs to $>100$\% 
in regions where less than one background event is expected.
The \lowpt\ SRs are generally impacted by uncertainties due to the limited size of the MC samples used in the background estimation, 
with these being dominant in SRC-MET. 
In SRC the theoretical uncertainties in the \ttbar\ background dominate, with these also being important in SRC-MET.  
The total background uncertainty in the \lowpt\ SRs is typically 10--20\% in SRC and 25--35\% in SRC-MET.

\section{Results}
\label{sec:result}
\label{sec:results}

The integrated yields in the high- and \lowpt\ signal regions are compared to the expected background in Tables~\ref{tab:EdgeIntegratedYields} and~\ref{tab:results:lowptIntegratedYields}, respectively.  
The full \mll\ distributions in each of these regions are compared to the expected background in Figures~\ref{fig:edgemll} and~\ref{fig:lowptmll}. 

\begin{table}[htbp!]
\begin{center}
\caption{
Breakdown of the expected background and observed data yields for SR-low, SR-medium and SR-high, integrated over the \mll\ spectrum.
The quoted uncertainties include statistical and systematic contributions, and due to anti-correlations with the CR, the total uncertainty may be less than the sum of individual parts.
\label{tab:EdgeIntegratedYields}}
\setlength{\tabcolsep}{0.0pc}
\begin{tabular*}{\textwidth}{@{\extracolsep{\fill}}lP{-1}P{-1}P{-1}}
\noalign{\smallskip}\hline\noalign{\smallskip}
                                                                            & \multicolumn{1}{c}{SR-low}                     & \multicolumn{1}{c}{SR-medium}            & \multicolumn{1}{c}{SR-high}  \\[-0.05cm]
\noalign{\smallskip}\hline\noalign{\smallskip}
Observed events                                                             & 134                      & 40                     &   72 \\
\noalign{\smallskip}\hline\noalign{\smallskip}
Total expected background events                                             & 144, 22 & 40, 10 & 83,9\\
\noalign{\smallskip}\hline\noalign{\smallskip}
  Flavour-symmetric (\ttbar, $Wt$, $WW$ and $Z\rightarrow\tau\tau$) events   & 86, 12 & 29, 9 & 75,8\\
  \dyjets\ events                                                            & 9\rlap{$^{+13}_{-9}$}  & 0.2\rlap{$^{+0.8}_{-0.2}$}  & 2.0,1.2\\
  $WZ/ZZ$ events                                                             & 43, 12 & 9.8, 3.2 & 4.1,1.2\\
  Rare top events                                                            & 6.7, 1.8 & 1.20, 0.35 & 1.8,0.5\\
\noalign{\smallskip}\hline\noalign{\smallskip}
\end{tabular*}
\end{center}
\end{table}

As signal models may produce kinematic endpoints at any value of \mll, any excess must be searched for across the \mll\ distribution. 
To do this a ``sliding window'' approach is used, as described in Section~\ref{sec:selection}.  
The 41 \mll\ windows (10 for SR-low, 9 for SR-medium, 10 for SR-high, 6 for SRC and 6 for SRC-MET) 
are chosen to make model-independent statements about the possible presence of new physics. 
The results in these \mll\ windows are summarised in Figure~\ref{fig:edge_summary}, 
with the observed and expected yields in the combined $ee+\mu\mu$ channel for all 41 \mll\ windows.
In general the data are consistent with the expected background across the full \mll\ range. 
The largest excess is observed in SR-medium with $101<\mll<201$~\GeV, 
where a total of 18 events are observed in data, compared to an expected $7.6 \pm 3.2$ events, 
corresponding to a local significance of $2\sigma$.

\begin{table}
\begin{center}
\caption{Breakdown of the expected and observed data yields for the \lowpt\ signal regions and their corresponding control regions.  
The quoted uncertainties include the statistical and systematic contributions, 
and due to anti-correlations with the CRs, the total uncertainty may be less than the sum of individual parts. }
\label{tab:results:lowptIntegratedYields}
\setlength{\tabcolsep}{0.0pc}
{\small
\begin{tabular*}{\textwidth}{@{\extracolsep{\fill}}lP{-1}P{-1}P{-1}P{-1}}
\noalign{\smallskip}\hline\noalign{\smallskip}
           & \multicolumn{1}{c}{SRC} & \multicolumn{1}{c}{CRC}   & \multicolumn{1}{c}{SRC-MET}  & \multicolumn{1}{c}{CRC-MET}      \\[-0.05cm]
\noalign{\smallskip}\hline\noalign{\smallskip}
Observed events          & 93 & 98 & 17  & 10 \\
\noalign{\smallskip}\hline\noalign{\smallskip}
Total expected background events          & 104, 17 & 98, 10 & 10, 4 & 10.0, 2.6\\
\noalign{\smallskip}\hline\noalign{\smallskip}
        Top-quark events             & 85, 17 & 81, 14 & 3\rlap{$^{+4}_{-3}$}  & 2.5\rlap{$^{+3.0}_{-2.5}$} \\
        Fake-lepton events        & 8.3, 1.5 & 10, 10  & 2.00, 0.35 & 3.6, 1.2\\
        Diboson events         & 7.6, 1.3 & 5.7, 1.6 & 4.4, 1.3 & 3.1, 1.2\\    
        Rare top events        & 3.26, 0.95 & 1.8, 0.7 & 0.53, 0.15 & 0.59, 0.18\\
        \dyjets\ events           & 0.050, 0.010 & 0.0, 0.0 & 0.52, 0.12 & 0.18, 0.05\\
\noalign{\smallskip}\hline\noalign{\smallskip}
\end{tabular*}
}
\end{center}
\end{table}

\begin{figure}[h]
\centering
\includegraphics[width=.45\textwidth]{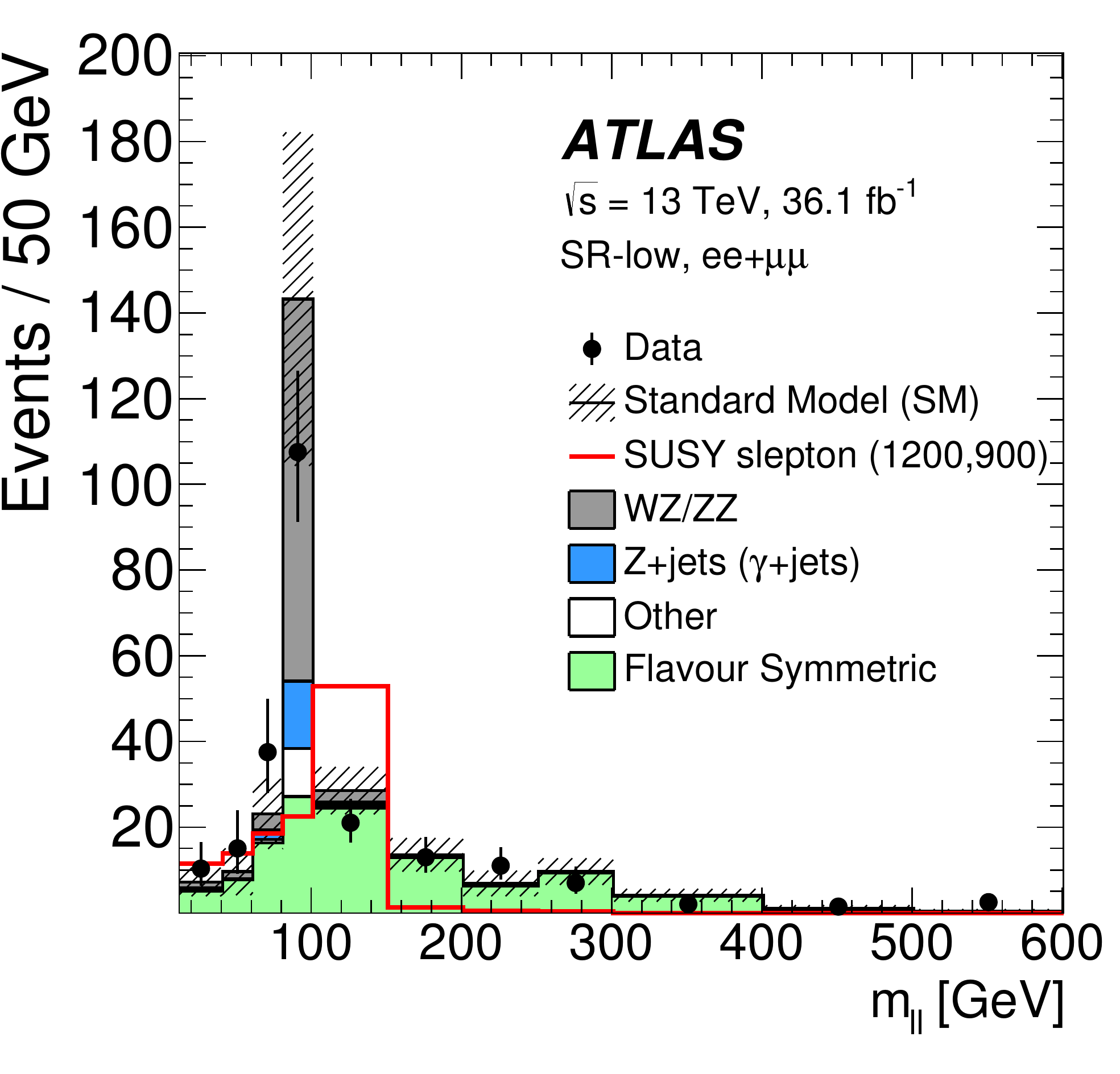}
\includegraphics[width=.45\textwidth]{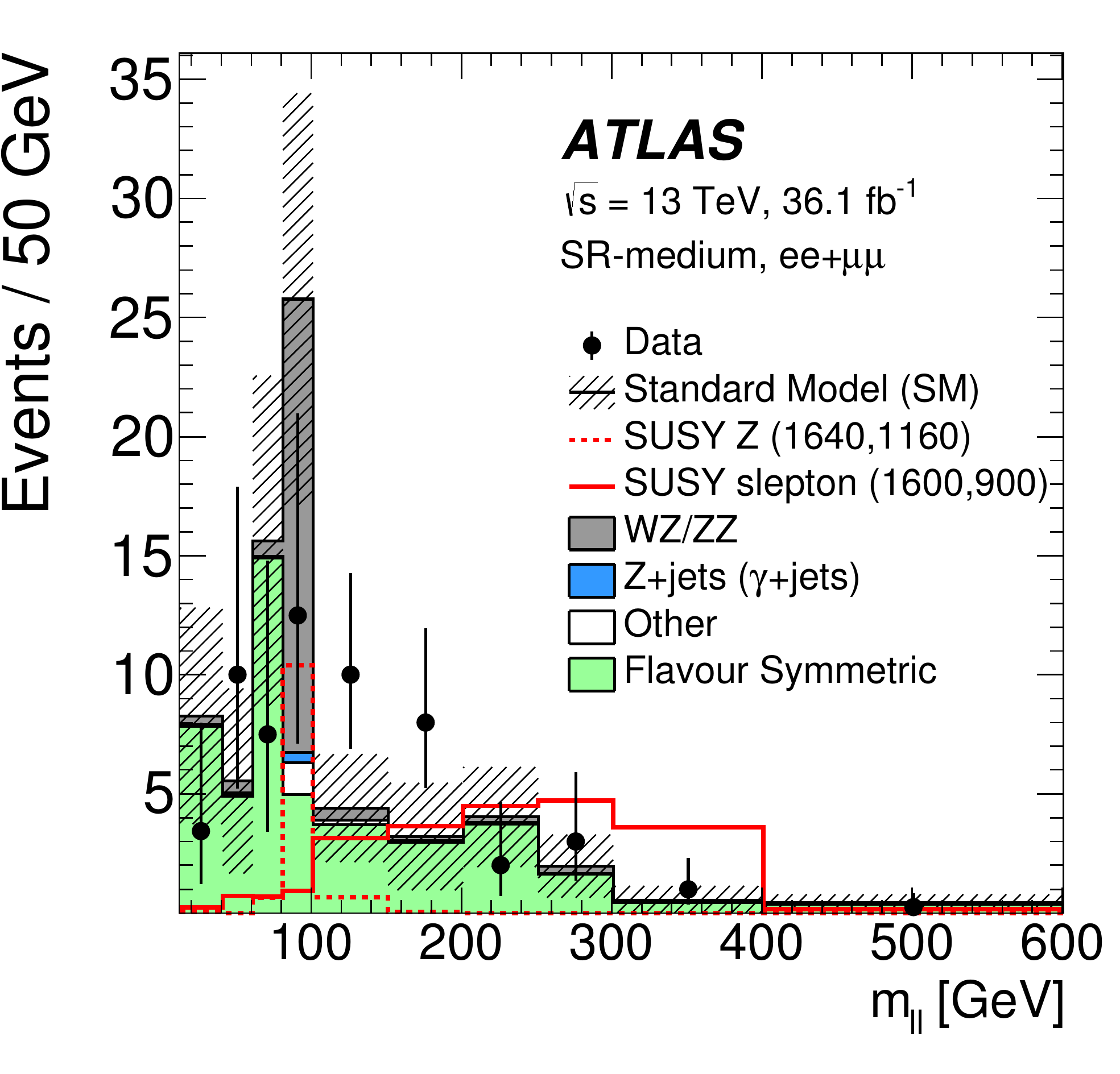}\\
\includegraphics[width=.45\textwidth]{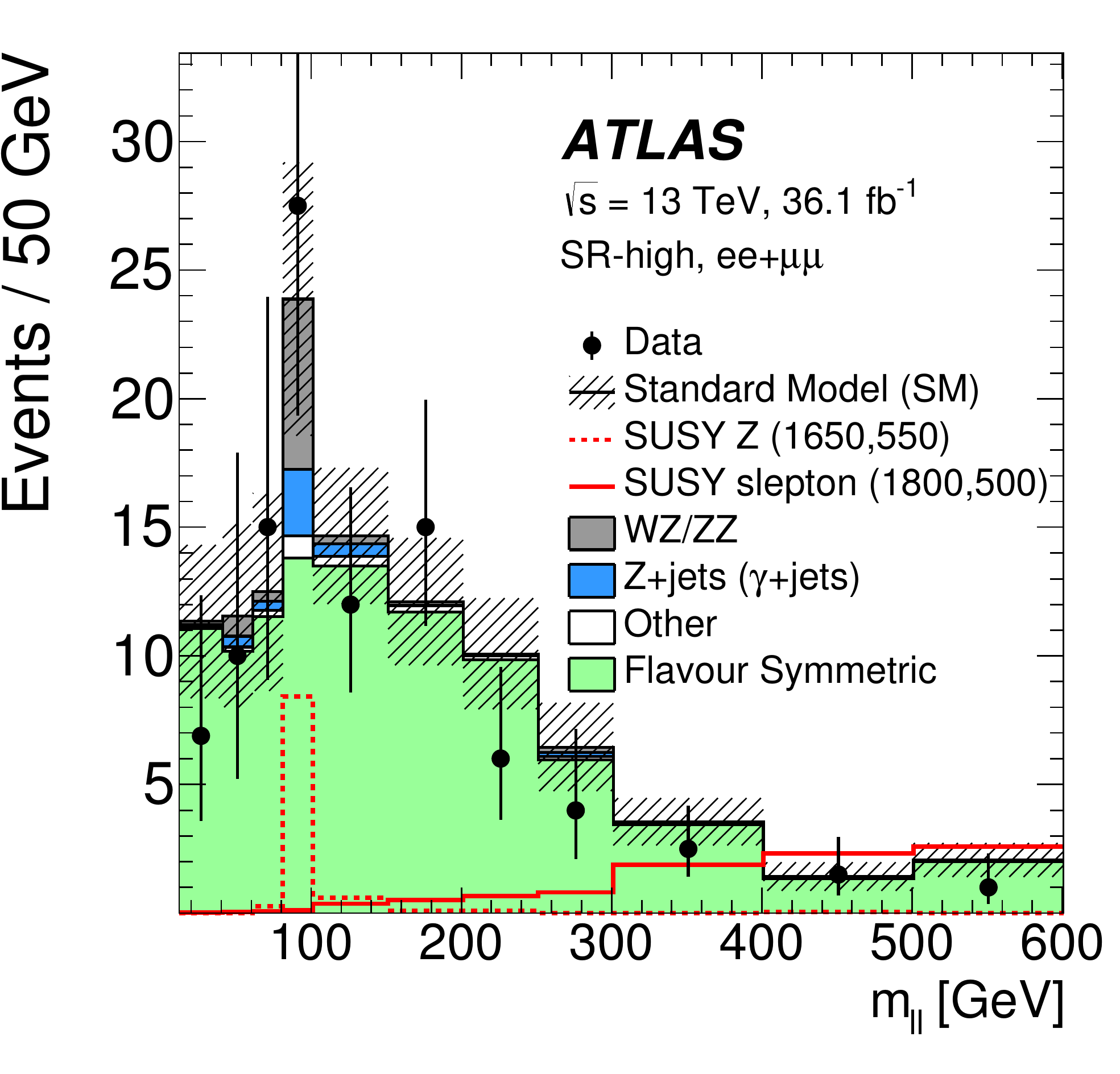}
\caption{ Observed and expected dilepton mass distributions, with the bin boundaries considered for the interpretation, 
in (top left) SR-low, (top-right) SR-medium, and (bottom) SR-high of the edge search. 
All statistical and systematic uncertainties of the expected background are included in the hatched band. 
The last bin contains the overflow.
One (two) example signal model(s) are overlaid on the top left (top right, bottom). 
For the slepton model, the numbers in parentheses in the legend indicate the gluino and $\tilde{\chi}_1^0$ masses of the example model point.
In the case of the $Z$ model illustrated, the numbers in parentheses indicate the gluino and $\tilde{\chi}_2^0$ masses, 
with the $\tilde{\chi}_1^0$ mass being fixed at 1~\GeV\ in this model.
}
\label{fig:edgemll}
\end{figure}

\begin{figure}[htbp]
\centering
\includegraphics[width=.45\textwidth]{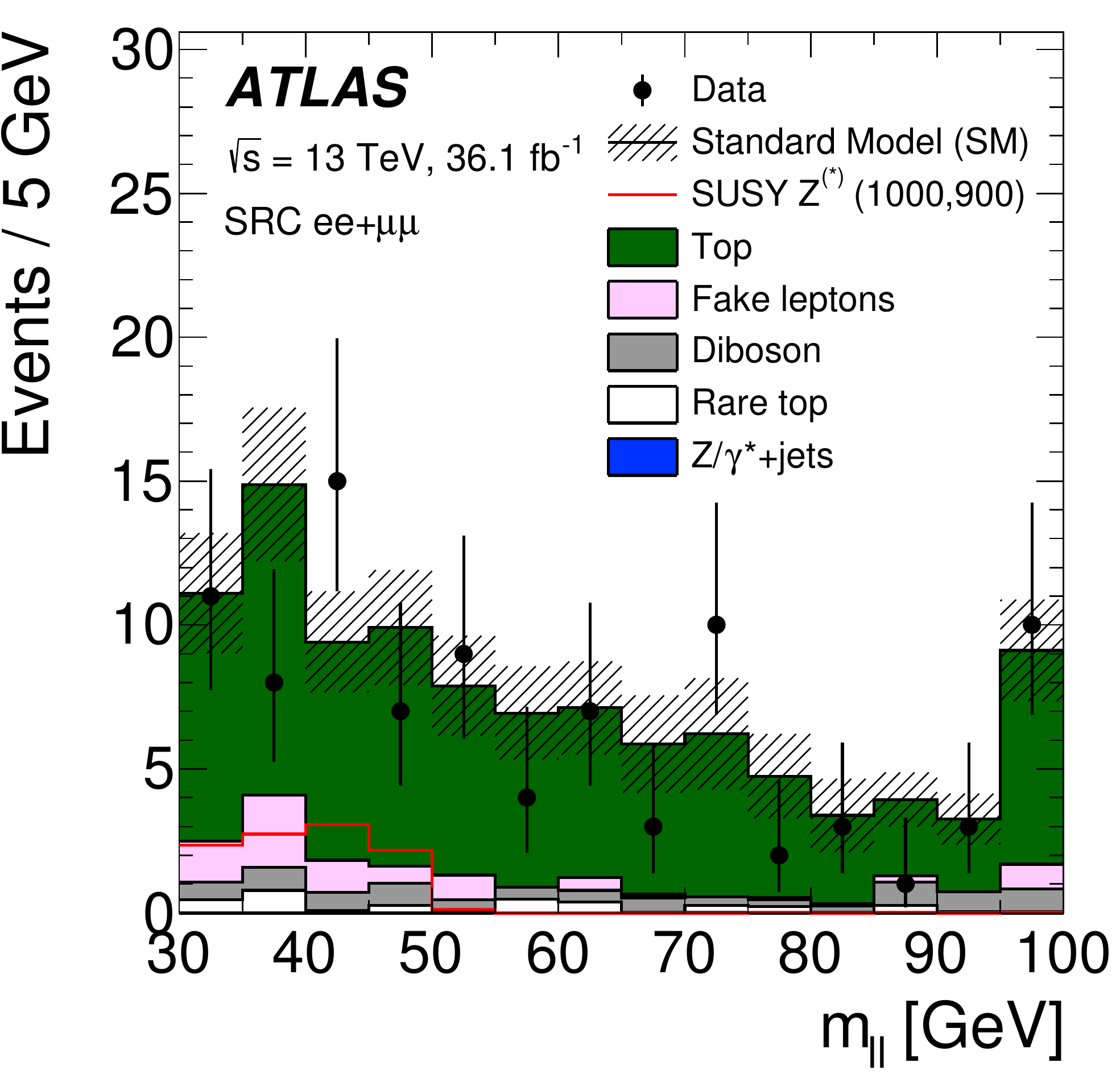}
\includegraphics[width=.45\textwidth]{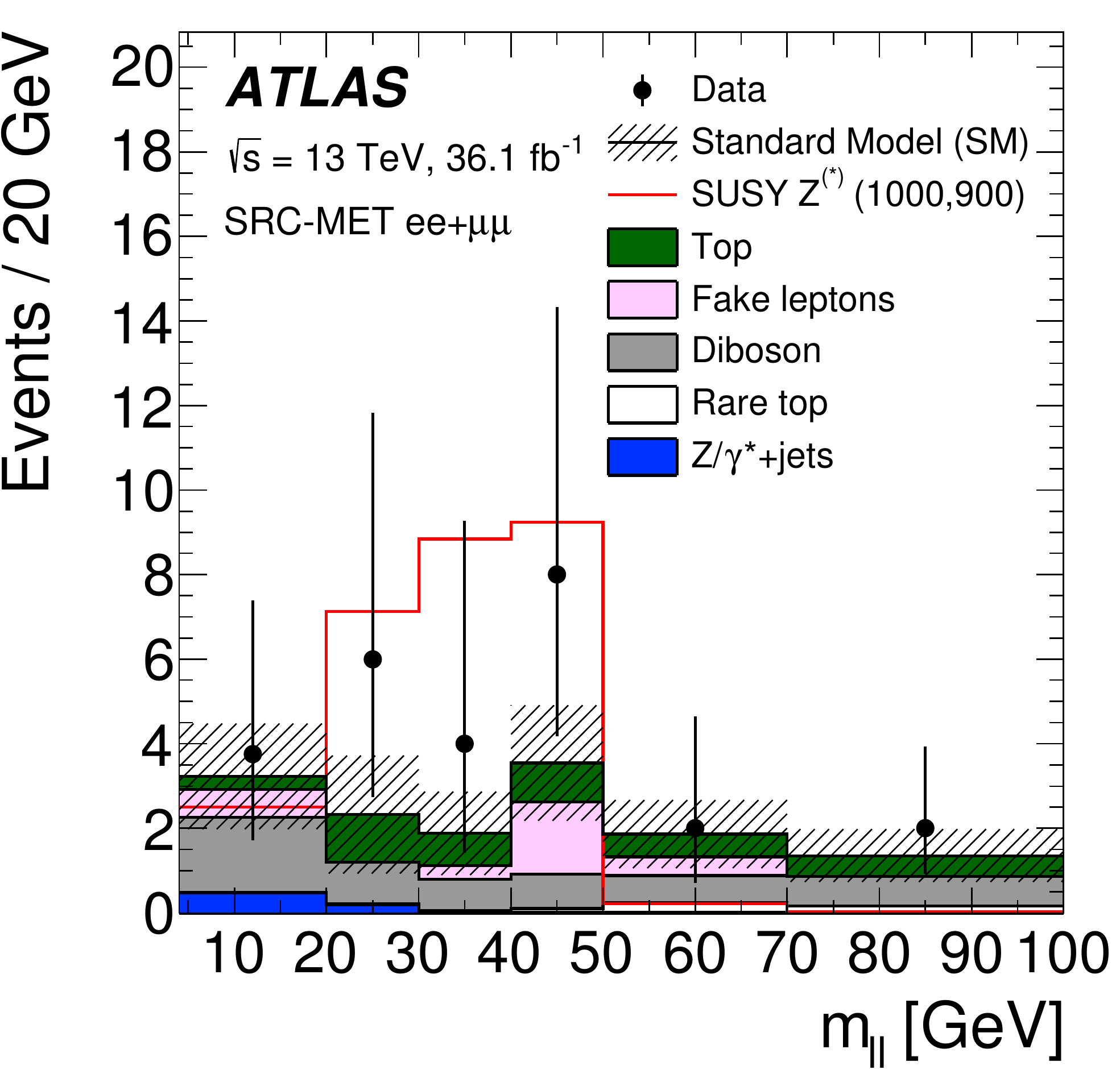} \\
\caption{ Observed and expected dilepton mass distributions, with the bin boundaries considered for the interpretation, in (left) SRC and (right) SRC-MET of the \lowpt\ edge search. 
All statistical and systematic uncertainties of the expected background are included in the hatched band. 
An example signal from the $Z^{(*)}$ model with $m(\tilde g)=1000~\GeV$ and $m(\tilde{\chi}_1^0)=900~\GeV$ is overlaid.}
\label{fig:lowptmll}
\end{figure}

\begin{figure}[htbp]
\centering

\includegraphics[width=1\textwidth]{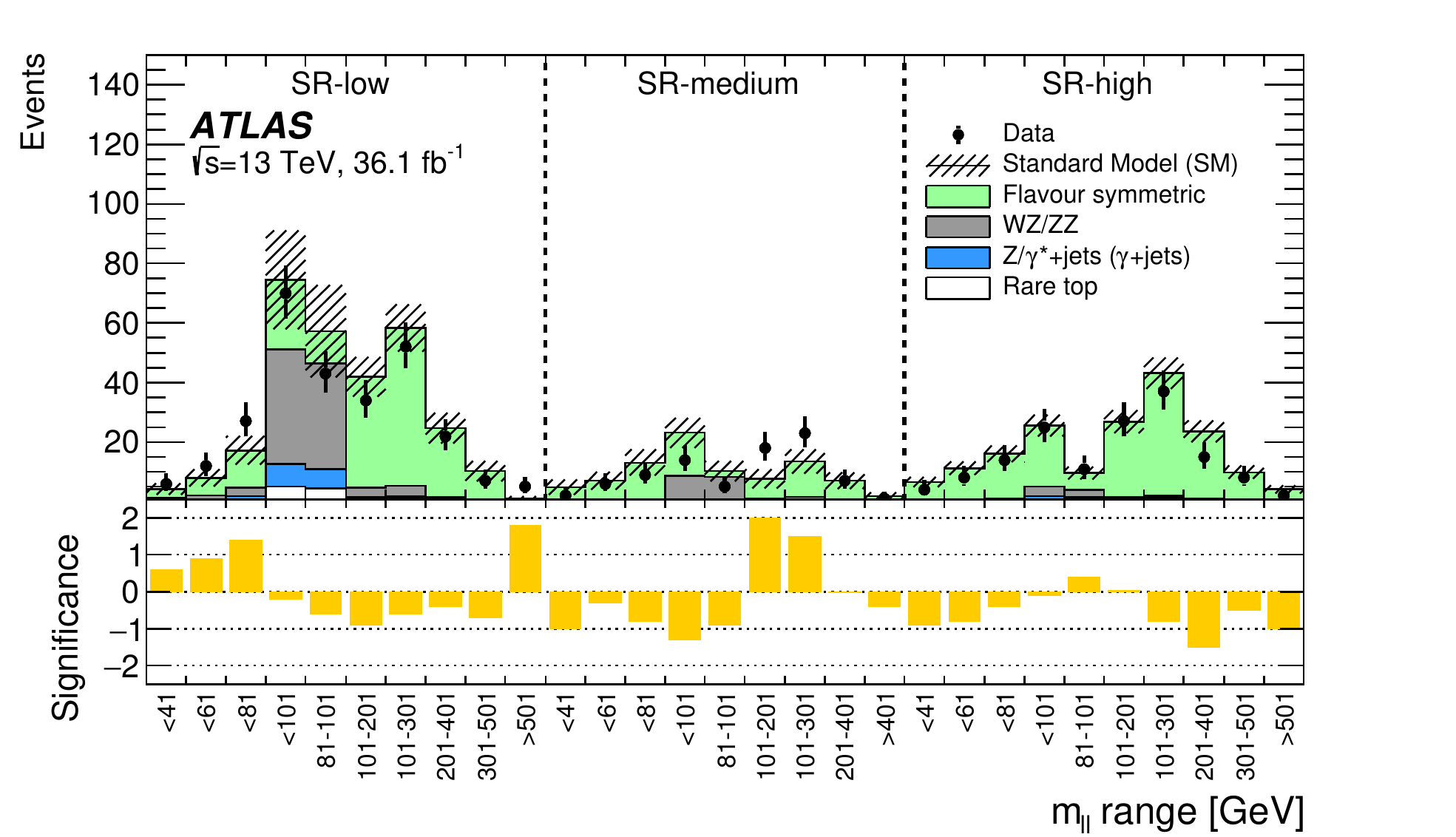} \\
\includegraphics[width=0.8\textwidth]{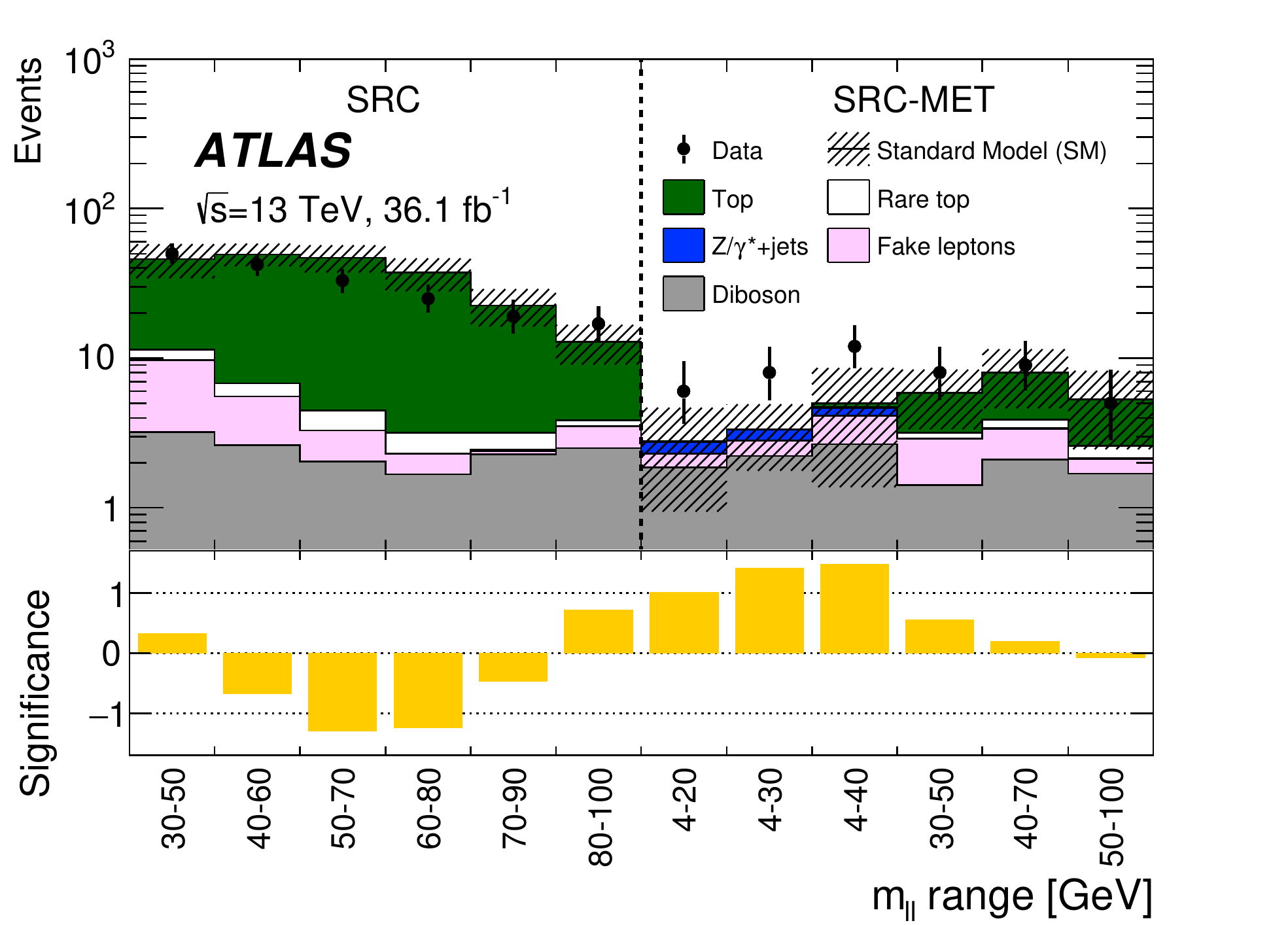} \\
\caption{ The observed and expected yields in the (overlapping) \mll\ windows of SR-low, SR-medium, SR-high, SRC and SRC-MET.
These are shown for the 29 \mll\ windows for the high-\pt\ SRs (top) and the 12 \mll\ windows for the \lowpt\ SRs (bottom). 
The data are compared to the sum of the expected backgrounds. 
The significance of the difference between the observed and expected yields is shown in the bottom plots.  
For cases where the $p$-value is less than 0.5 a negative significance is shown. 
The hatched uncertainty band includes the statistical and systematic uncertainties of the background prediction.
\label{fig:edge_summary}}
\end{figure}

Model-independent upper limits at 95\% confidence level (CL) on the number of events ($S^{95}$) that could be attributed to non-SM sources are derived using the
$\text{CL}_{\text{S}}$ prescription~\cite{clsread}, implemented in the HistFitter program~\cite{Baak:2014wma}. 
A Gaussian model for nuisance parameters is used for all but two of the uncertainties. 
The exceptions are the statistical uncertainties in the flavour-symmetry method and MC-based backgrounds, which are treated as Poissonian nuisance parameters.
This procedure is carried out using the \mll\ windows from the high-\pt\ and \lowpt\ analyses, neglecting possible signal contamination in the CRs.
For these upper limits, pseudo-experiments are used. 
Upper limits on the visible BSM cross-section $\langle A\epsilon{\rm \sigma}\rangle_{\rm obs}^{95}$ are obtained by dividing the observed upper limits on the number of BSM events by the integrated luminosity.
Expected and observed upper limits are given in Tables~\ref{tab:edge_results}~and~\ref{tab:interp:lowpt_UL} for the high-\pt\ and \lowpt\ SRs, respectively. 
The $p$-values, which represent the probability of the SM background alone to fluctuate to the observed number of events or higher, 
are also provided using the asymptotic approximation~\cite{statforumlimits}. 

\FloatBarrier
\begin{table}
\begin{center}
\caption{Breakdown of the expected background and observed data yields in the high-\pt\ signal regions.
The results are given for SR-low, SR-medium and SR-high in all 29 \mll\ windows.
The \mll\ range is indicated in the left-most column of the table.
Left to right: the total expected background, with combined statistical and systematic uncertainties, observed data, 95\% CL upper limits on the visible cross section
($\langle A\epsilon\sigma\rangle_{\rm obs}^{95}$) and on the number of
signal events ($S_{\rm obs}^{95}$).  The sixth column
($S_{\rm exp}^{95}$) shows the expected 95\% CL upper limit on the number of
signal events, given the expected number (and $\pm 1\sigma$
excursions on the expectation) of background events.
The last two columns
indicate the discovery $p$-value ($p(s = 0)$), and the Gaussian significance ($Z(s=0)$). 
For cases where $p(s=0)<0.5$ a negative significance is shown.
\label{tab:edge_results}}
\resizebox{\textwidth}{!}{
\setlength{\tabcolsep}{0.0pc}
\small
\begin{tabular*}{\textwidth}{@{\extracolsep{\fill}}lP{-1}d{-1}d{-1}d{-1}d{-1}d{-1}d{-1}}
\noalign{\smallskip}\hline\noalign{\smallskip}
{\bf Signal Region}    &  \multicolumn{1}{c}{Total Bkg.}  &  \multicolumn{1}{c}{Data}  &  \multicolumn{1}{c}{$\langle A\epsilon{\rm \sigma}\rangle_{\rm obs}^{95}$[fb]}   &   \multicolumn{1}{c}{$S_{\rm obs}^{95}$}   &   \multicolumn{1}{c}{$S_{\rm exp}^{95}$}  &  \multicolumn{1}{c}{$p(s=0)$}  &  \multicolumn{1}{c}{$Z(s=0)$} \\
\mll\ range [GeV] & & & & & && \\
\noalign{\smallskip}\hline\noalign{\smallskip}
\noalign{\smallskip}\hline\noalign{\smallskip}
SR-low & & & & & &&\\
\noalign{\smallskip}\hline\noalign{\smallskip}
12$-$41   & 4.2, 2.0 &  6      &  0.28  &  10.2   & 6.9\rlap{$^{+3.3}_{-1.3}$}  &    0.27   &  0.6 \\[.5ex]
12$-$61   & 8.0, 3.0 &  12     &  0.44  &  15.8   & 9.9\rlap{$^{+4}_{-2.5}$}  &    0.19   &  0.9 \\[.5ex] 
12$-$81   & 17, 5 &  27    &  0.73  &  26.3       & 15\rlap{$^{+6}_{-4}$}  &    0.086   &  1.4 \\[.5ex] 
12$-$101  & 75, 17 &  70   &  1.56  &  56.2       & 60\rlap{$^{+7}_{-5}$}  &    0.6   &  -0.2 \\[.5ex] 
81$-$101  & 57, 16 &  43   &  1.13  &  40.6       & 47\rlap{$^{+6}_{-6}$}  &    0.73   &  -0.6 \\[.5ex]
101$-$201 & 42, 7 &  34    &  0.38  &  13.8       & 19\rlap{$^{+9}_{-5}$}  &    0.81   &  -0.9 \\[.5ex]
101$-$301 & 58, 8 &  52    &  0.46  &  16.5       & 23\rlap{$^{+9}_{-8}$}  &    0.72   &  -0.6 \\[.5ex]
201$-$401 & 25, 5 &  22    &  0.37  &  13.4       & 15\rlap{$^{+11}_{-4}$}  &    0.65   &  -0.4 \\[.5ex]
301$-$501 & 10.2, 3.5 &  7     &  0.20  &  7.1    & 9.4\rlap{$^{+4}_{-2.8}$}  &    0.77   &  -0.7 \\[.5ex]
501$-$    & 0.9\rlap{$^{+0.95}_{-0.9}$}  &  5      &  0.27  &  9.9   & 6.0\rlap{$^{+2.3}_{-1.0}$}  &    0.039   &  1.8 \\[.5ex]
\noalign{\smallskip}\hline\noalign{\smallskip}
SR-medium & & & & & &&\\[.5ex]%
\noalign{\smallskip}\hline\noalign{\smallskip}
12$-$41   & 4.8, 2.6 &  2      &  0.16  &   5.7   & 6.9\rlap{$^{+3.2}_{-1.3}$}  &     0.83   &  -1.0 \\[.5ex]
12$-$61   & 7.0, 3.0 &  6      &  0.20  &   7.4   & 8.2\rlap{$^{+4}_{-2.1}$}  &     0.6   &  -0.3 \\[.5ex]
12$-$81   & 13, 4 &  9     &  0.22  &   7.8   & 11.0\rlap{$^{+4}_{-3.3}$}  &     0.78   &  -0.8 \\[.5ex]
12$-$101  & 23, 5 &  14    &  0.25  &   9.1   & 13.5\rlap{$^{+5}_{-3.5}$}  &     0.91   &  -1.3 \\[.5ex]
81$-$101  & 10.3, 3.4 &  5     &  0.22  &   8.0   & 10.0\rlap{$^{+2.8}_{-2.5}$}  &     0.82   &  -0.9 \\[.5ex]
101$-$201 & 7.6, 3.2 &  18     &  0.53  &   19.1   & 11.1\rlap{$^{+4}_{-2.7}$}  &     0.024   &  2.0 \\[.5ex]
101$-$301 & 14, 4 &  23    &  0.68  &   24.5   & 14\rlap{$^{+6}_{-4}$}  &     0.063   &  1.5 \\[.5ex]
201$-$401 & 7.1, 2.8 &  7      &  0.27  &   9.8   & 8.6\rlap{$^{+4}_{-2.4}$}  &     0.51   &  -0.0 \\[.5ex]
401$-$    & 1.8, 1.4 &  1      &  0.12  &   4.3   & 4.8\rlap{$^{+2.5}_{-1.0}$}  &      0.67   &  -0.4 \\[.5ex]
\noalign{\smallskip}\hline\noalign{\smallskip}
SR-high & & & & & && \\[.5ex]%
\noalign{\smallskip}\hline\noalign{\smallskip}
12$-$41   & 6.6, 1.7 &  4      &  0.14  &   5.0   & 7.0\rlap{$^{+2.7}_{-2.1}$}  &   0.82   &  -0.9 \\[.5ex]
12$-$61   & 11.2, 2.3 &  8     &  0.18  &   6.5   & 8.6\rlap{$^{+4}_{-2.5}$}  &   0.8   &  -0.8 \\[.5ex]
12$-$81   & 16.1, 2.9 &  14    &  0.25  &   9.1   & 10.7\rlap{$^{+4}_{-2.5}$}  &   0.67   &  -0.4 \\[.5ex]
12$-$101  & 26, 4 &  25    &  0.37  &   13.4  & 14\rlap{$^{+5}_{-4}$}  &   0.54   &  -0.1 \\[.5ex] 
81$-$101  & 9.6, 2.1 &  11     &  0.30  &   11.0   & 10.8\rlap{$^{+3.4}_{-2.2}$}  &   0.35   &  0.4 \\[.5ex]
101$-$201 & 27, 4 &  27    &  0.35  &   12.8   & 12.9\rlap{$^{+7}_{-3.1}$}  &   0.49   &  0.0 \\[.5ex]
101$-$301 & 43, 5 &  37    &  0.35  &   12.7   & 17\rlap{$^{+6}_{-5}$}  &   0.77   &  -0.8 \\[.5ex]
201$-$401 & 24, 4 &  15    &  0.19  &   6.8   & 12\rlap{$^{+5}_{-4}$}  &   0.94   &  -1.5 \\[.5ex] 
301$-$501 & 9.9, 2.2 &  8      &  0.21  &   7.5   & 8.6\rlap{$^{+4}_{-2.7}$}  &   0.7   &  -0.5 \\[.5ex]
501$-$    & 4.1, 1.3 &  2      &  0.12  &   4.3   & 5.6\rlap{$^{+2.3}_{-1.5}$}  &   0.84   &  -1.0 \\[.5ex]
\noalign{\smallskip}\hline\noalign{\smallskip}
\end{tabular*}
}
\end{center}
\end{table}

\begin{table}[htbp]
\centering
\caption{Breakdown of the expected background and observed data yields in the \lowpt\ signal regions.
The results are given for SRC and SRC-MET in all 12 \mll\ windows.
The \mll\ range in units of \GeV\ is indicated in the left-most column of the table.
Left to right: the total expected background, with combined statistical and systematic uncertainties, observed data, 95\% CL upper limits on the visible cross section
($\langle A\epsilon\sigma\rangle_{\rm obs}^{95}$) and on the number of
signal events ($S_{\rm obs}^{95}$).  The sixth column
($S_{\rm exp}^{95}$) shows the expected 95\% CL upper limit on the number of
signal events, given the expected number (and $\pm 1\sigma$
excursions on the expectation) of background events.
The last two columns
indicate the discovery $p$-value ($p(s = 0)$), and the Gaussian significance ($Z(s=0)$). 
\label{tab:interp:lowpt_UL}}
\setlength{\tabcolsep}{0.0pc}
\begin{tabular*}{\textwidth}{@{\extracolsep{\fill}}lP{-1}d{-1}d{-1}d{-1}d{-1}d{-1}d{-1}}
\noalign{\smallskip}\hline\noalign{\smallskip}
{\bf Signal Region}   &  \multicolumn{1}{c}{Total Bkg.}  &  \multicolumn{1}{c}{Data}  &  \multicolumn{1}{c}{$\langle A\epsilon{\rm \sigma}\rangle_{\rm obs}^{95}$[fb]}   &   \multicolumn{1}{c}{$S_{\rm obs}^{95}$}   &  \multicolumn{1}{c}{$S_{\rm exp}^{95}$}  &  \multicolumn{1}{c}{$p(s=0)$}  &  \multicolumn{1}{c}{$Z(s=0)$} \\
{\mll\ range [GeV]}  & & & & & & & \\
\noalign{\smallskip}\hline\noalign{\smallskip}
\noalign{\smallskip}\hline\noalign{\smallskip}
\multicolumn{8}{l}{SRC} \\
\noalign{\smallskip}\hline\noalign{\smallskip}
30-50  &  46, 12  &  50  &   1.29   &    46.4   &  42\rlap{$^{+10}_{-8}$}   &    0.38   &   0.3  \\[.5ex]
40-60  &  50, 9  &  42  &   0.54   &    19.5   &  25\rlap{$^{+9}_{-8}$}   &    0.75   &   -0.7  \\[.5ex]
50-70  &  47, 10  &  33  &  0.43   &    15.6   &  24\rlap{$^{+9}_{-7}$}   &     0.90   &   -1.3  \\[.5ex]
60-80  &  37, 9  &  25 &   0.37   &    13.3   &  28\rlap{$^{+4}_{-12}$}   &     0.89   &   -1.3  \\[.5ex]
70-90  &  23, 6  &  19 &   0.31   &    11.1   &  16\rlap{$^{+6}_{-4}$}   &     0.68   &   -0.5  \\[.5ex]
80-    &  13, 4  &  17  &   0.42   &    15.3   &  12.8\rlap{$^{+5}_{-4}$}   &    0.24   &   0.7  \\[.5ex]
\noalign{\smallskip}\hline\noalign{\smallskip}
\multicolumn{8}{l}{SRC-MET} \\
\noalign{\smallskip}\hline\noalign{\smallskip}
4-20    &  2.8, 1.9  &  6  &  0.31   &    11.0   &  8.4\rlap{$^{+5}_{-2.2}$}   &    0.15   &   1.0  \\[.5ex]
4-30    &  3.3, 1.6  &  8 &   0.35   &    12.5   &  8.6\rlap{$^{+4}_{-2.0}$}   &     0.078   &   1.4  \\[.5ex]
4-40    &  5, 4  &  12 &   0.45   &    16.3   &  10.2\rlap{$^{+5}_{-1.9}$}   &      0.069   &   1.5  \\[.5ex]
30-50   &  5.9, 2.5  &  8  &   0.30   &    10.7   &  8.8\rlap{$^{+4}_{-2.2}$}   &     0.29   &   0.6 \\[.5ex]
40-70   &  8.0, 3.4  &  9  &   0.32   &    11.5   &  10.6\rlap{$^{+4}_{-2.8}$}   &    0.42   &   0.2  \\[.5ex]
50-     &  5.3, 2.9  &  5 &   0.24   &    8.8   &  8.8\rlap{$^{+3.4}_{-1.9}$}   &    0.53   &   -0.1  \\[.5ex]
\noalign{\smallskip}\hline\noalign{\smallskip}
\end{tabular*}
\end{table}

\clearpage
\section{Interpretation}
\label{sec:interpretation}
In this section, exclusion limits are shown for the SUSY models detailed in Section~\ref{sec:susy}. 
For these model-dependent exclusion limits a shape fit is performed on each of the binned \mll\ distributions in Figures~\ref{fig:edgemll} and~\ref{fig:lowptmll}. 
The $CL_{\text{S}}$ prescription in the asymptotic approximation is used.
Experimental uncertainties are treated as correlated between signal and background events.
The theoretical uncertainty of the signal cross-section is not accounted for in the limit-setting procedure.
Instead, following the initial limit determination, 
the impact of varying the signal cross-section within its uncertainty is evaluated separately and indicated in the exclusion results. 
For the high-\pt\ analysis, possible signal contamination in the CRs is neglected in the limit-setting procedure;
the contamination is found to be negligible for signal points near the exclusion boundaries. 
Signal contamination in the CRs is taken into account in the limit-setting procedure for the \lowpt\ analysis.

The top panel of Figure~\ref{fig:limit_slepton} shows the exclusion contours in the $m(\tilde{g})$--$m(\tilde{\chi}^{0}_{1})$ plane for a simplified model with gluino pair production, 
where the gluinos decay via sleptons. 
The exclusion contour shown is derived using a combination of results from the three high-\pt\ and two \lowpt\ SRs based on the best-expected sensitivity.
The \lowpt\ SRs drive the limits close to the diagonal, with the high-\pt\ SRs taking over at high gluino masses.
In SR-low there is good sensitivity at high gluino and high LSP masses.
Around gluino mass of 1.8~\TeV, the observed limit drops below the expected limit by 200~\GeV, 
where the dilepton kinematic edge is expected to occur around 800~\GeV. 
Here the highest \mll\ bin in SR-low (\mll>$501$~\GeV), which is the bin driving the limit in this region, has a mild excess in data, 
explaining this effect.
The region where the \lowpt\ search becomes the most sensitive can be seen close to the diagonal,
where there is a kink in the contour at $m(\gluino) \sim 1400$~\GeV. 
A zoomed-in view of the compressed region of phase space, the region close to the diagonal for this model, 
is provided in the $m(\tilde{g})-(m(\tilde{g})$--$m(\tilde{\chi}^{0}_{1}))$ plane in the bottom panel of Figure~\ref{fig:limit_slepton}.  
Here the exclusion contour includes only the \lowpt\ regions. 
SRC-MET has the best sensitivity almost everywhere, except at low values of LSP mass (at the top-left of the bottom panel of Figure~\ref{fig:limit_slepton}), where SRC drives the limit. 
An exclusion contour derived using a combination of results from the three high-\pt\ SRs alone is overlaid, 
demonstrating the increased sensitivity brought by the \lowpt\ analysis.

\begin{figure}[htbp]
\centering
\includegraphics[width=0.8\textwidth]{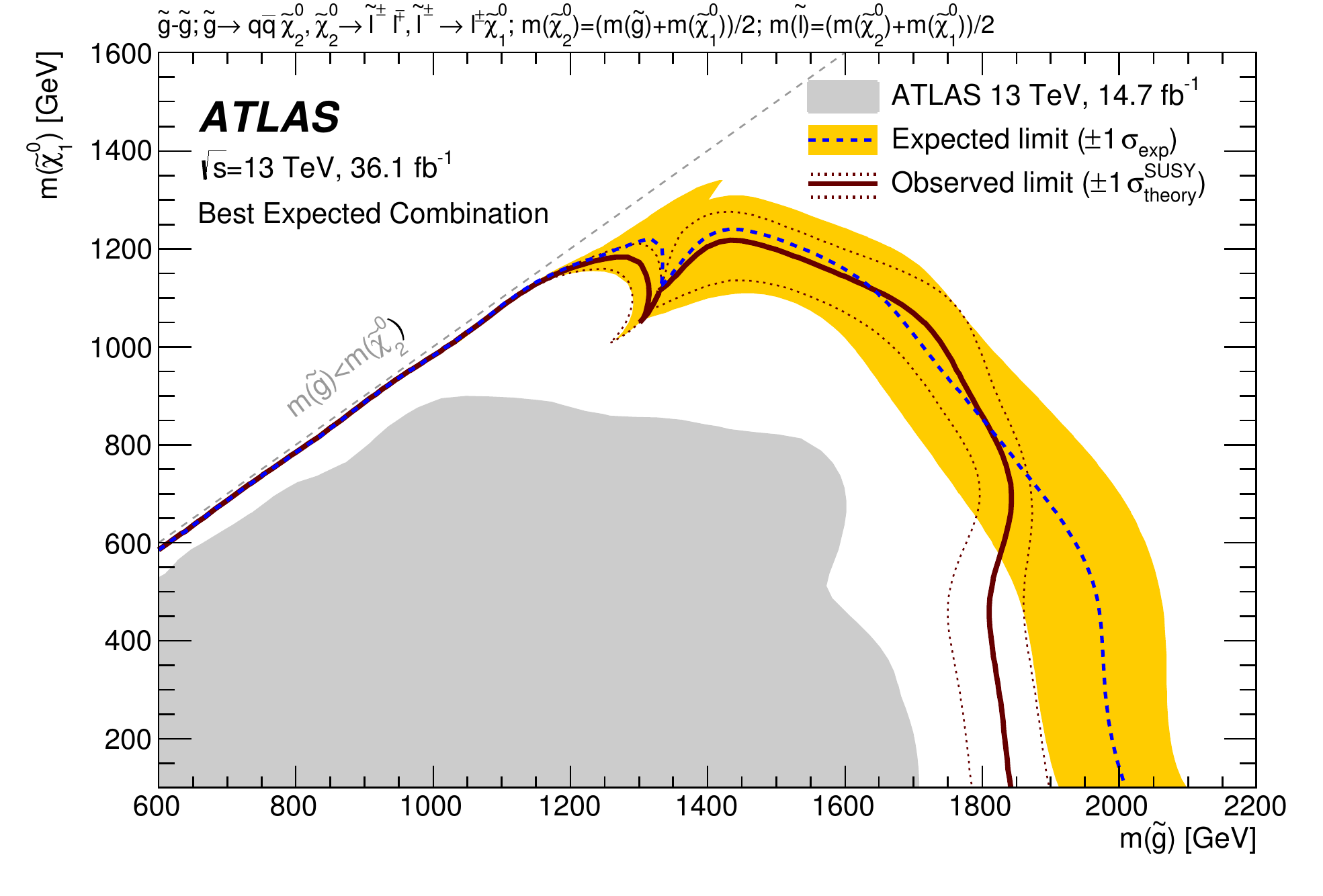}
\includegraphics[width=0.8\textwidth]{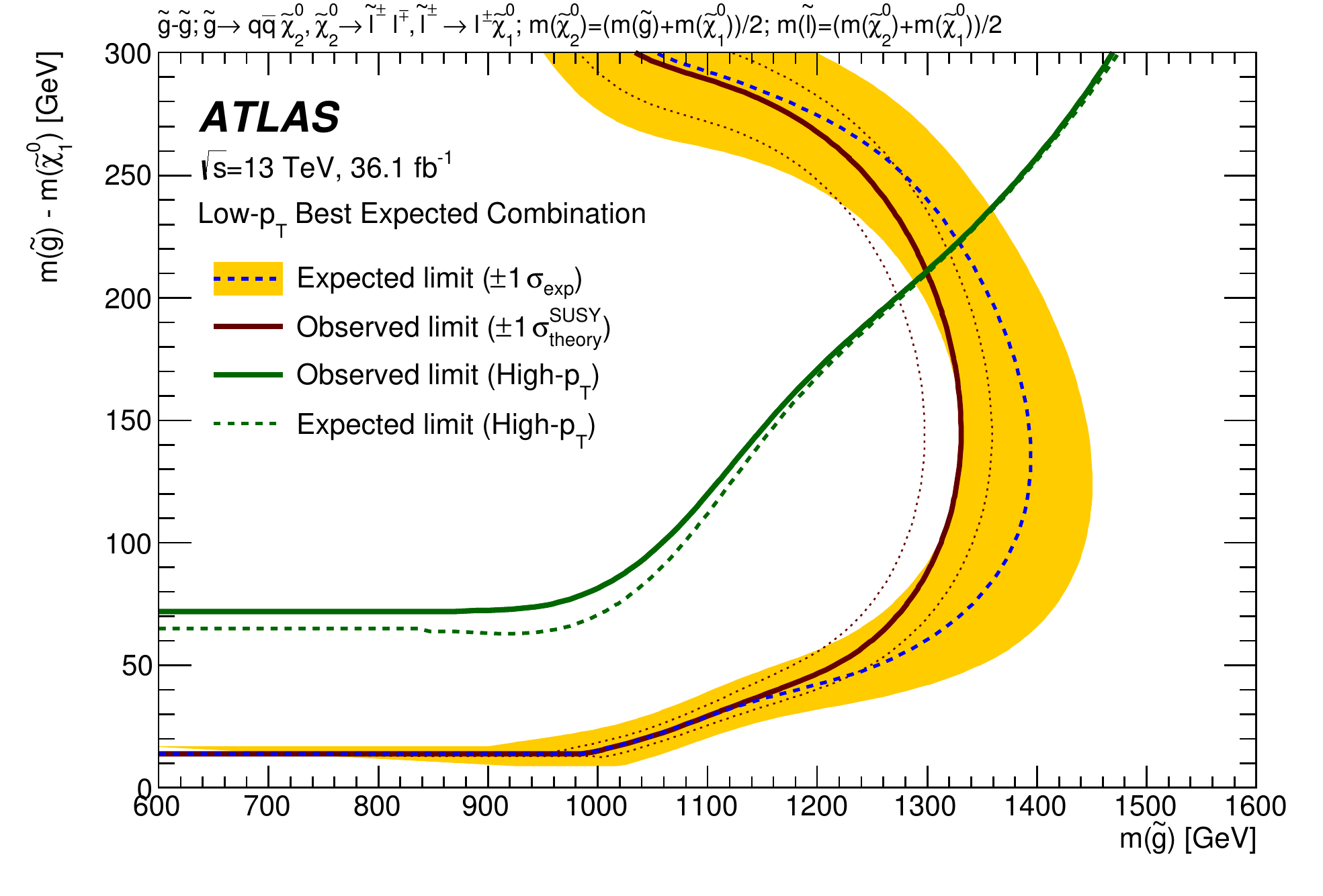}
\caption{
Expected and observed exclusion contours derived from the combination of the results in the high-\pt\ and \lowpt\ edge SRs based on the best-expected sensitivity (top) and zoomed-in view of the \lowpt\ only (bottom) for the slepton signal model.
The dashed line indicates the expected limits at $95\%$ CL and the surrounding band shows the $1\sigma$ variation of the expected limit as a consequence of the uncertainties in the background prediction and the experimental uncertainties in the signal ($\pm1\sigma_\text{exp}$).
The dotted lines surrounding the observed limit contours indicate the variation resulting from changing the signal cross-section within its uncertainty ($\pm1\sigma^\text{SUSY}_\text{theory}$). 
The shaded area on the upper plot indicates the observed limit on this model from Ref.~\cite{SUSY-2016-05}.
In the lower plot the observed and expected contours derived from the high-\pt\ SRs alone are overlaid, illustrating the added sensitivity from the low-\pt\ SRs.
Small differences between the contours in the compressed region are due to differences in interpolation between the top and bottom plot.
\label{fig:limit_slepton}}
\end{figure}

The top panel of Figure~\ref{fig:comb_limit_zstar} shows the exclusion contours for the \zstar\ simplified model in the $m(\tilde{g})$--$m(\tilde{\chi}^{0}_{1})$ plane, 
where on- or off-shell $Z$ bosons are expected in the final state.
Again, the \lowpt\ SRs have good coverage near the diagonal. 
SR-med drives the limits at high gluino mass, reaching beyond 1.6~\TeV.
For this interpretation the contour is mostly dominated by the on-$Z$ bin of the three edge SRs.
The kink in the exclusion contour at $m(\gluino)=1200$~\GeV\ occurs where the \lowpt\ SRs begin to dominate the sensitivity. 
A zoomed-in view of the compressed region of phase space where the \lowpt\ SRs dominate the sensitivity
is provided in the $m(\tilde{g})-(m(\tilde{g})$--$m(\tilde{\chi}^{0}_{1}))$ plane in the bottom panel of Figure~\ref{fig:comb_limit_zstar}.
Here the exclusion contour includes only the \lowpt\ regions, with the exclusion contour derived using a combination of results from the three high-\pt\ SRs alone overlaid. 

\begin{figure}[htbp]
\centering
\includegraphics[width=0.8\textwidth]{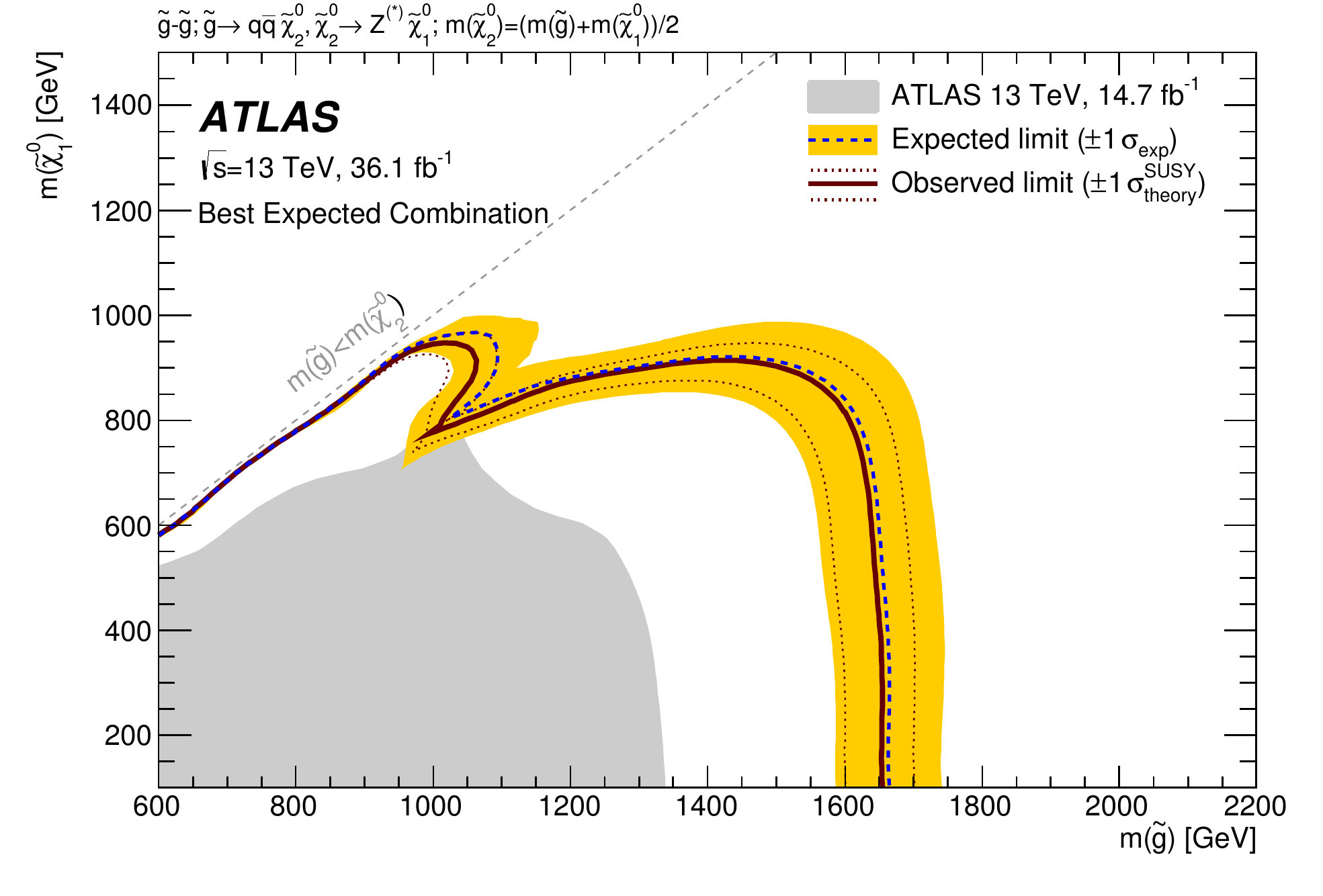}
\includegraphics[width=0.8\textwidth]{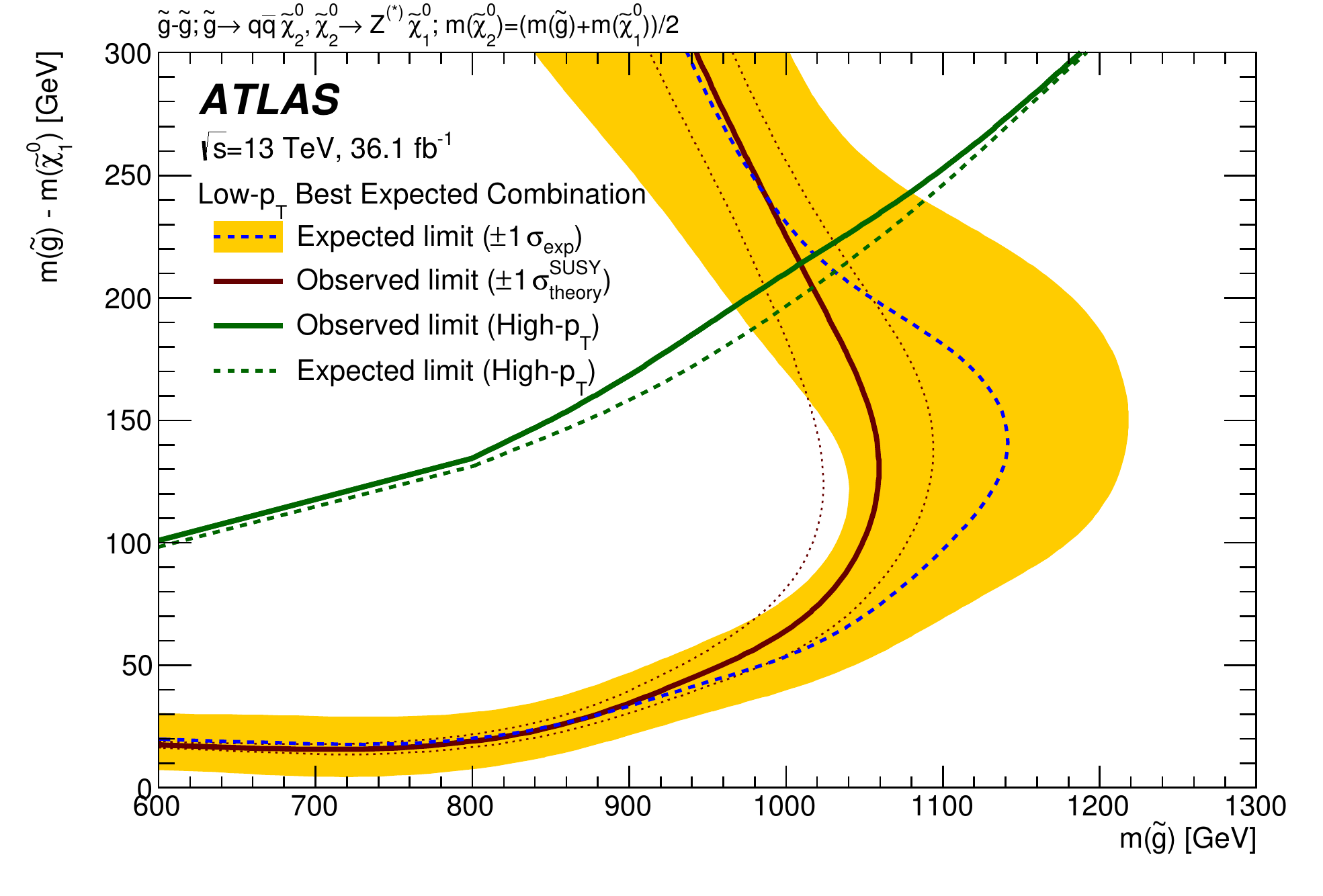}
\caption{
Expected and observed exclusion contours derived from the combination of the results in the high-\pt\ and \lowpt\ edge SRs based on the best-expected sensitivity (top) and zoomed-in view for the \lowpt\ only (bottom) for the \zstar\ model.
The dashed line indicates the expected limits at $95\%$ CL and the surrounding band shows the $1\sigma$ variation of the expected limit as a consequence of the uncertainties in the background prediction and the experimental uncertainties in the signal ($\pm1\sigma_\text{exp}$). 
The dotted lines surrounding the observed limit contours indicate the variation resulting from changing the signal cross-section within its uncertainty ($\pm1\sigma^\text{SUSY}_\text{theory}$). 
The shaded area on the upper plot indicates the observed limit on this model from Ref.~\cite{SUSY-2016-05}.
In the lower plot the observed and expected contours derived from the high-\pt\ SRs alone are overlaid, illustrating the added sensitivity from the low-\pt\ SRs.
Small differences in the contours in the compressed region are due to differences in interpolation between the top and bottom plot.
\label{fig:comb_limit_zstar}}
\end{figure}

The on-$Z$ windows ($81<\mll<101$~\GeV) of SR-medium and SR-high have good sensitivity to the on-shell $Z$ models discussed in Section~\ref{sec:susy}. 
These two \mll\ windows alone are used for the following three simplified model interpretations, where a best-expected-sensitivity combination of the results from the two windows is used. 
In Figure~\ref{fig:excl_SMGGN2_1}, these results are interpreted in a simplified model with gluino-pair production,
where each gluino decays as 
$\tilde{g} \rightarrow q\bar{q} \tilde{\chi}^{0}_{2}, \tilde{\chi}^{0}_{2} \rightarrow Z \tilde{\chi}^{0}_{1}$ and the $\tilde{\chi}^{0}_{1}$ mass is set to 1~\GeV. 
The expected and observed exclusion contours for this $\tilde{g}$--$\chitwozero$ on-shell grid are shown in the $m(\tilde{g})$--$m(\tilde{\chi}^{0}_{2})$ plane in Figure~\ref{fig:excl_SMGGN2_1}.
The expected (observed) lower limit on the gluino mass is about 1.60~\TeV\ (1.65~\TeV) for a $\tilde{\chi}_2^0$ with a mass of 1.2~\TeV\ in this model.
Here, the on-$Z$ window of SR-medium drives the limit close to the diagonal, while SR-high takes over at high $m(\tilde{g})$ and lower $m(\tilde{\chi}^{0}_{2})$. 
A kink can be seen in the observed limit contour at the point at which the SR with the best-expected sensitivity changes from SR-medium to SR-high.
Figure~\ref{fig:excl_SMGGN2_1} also shows the expected and observed exclusion limits for the $\tilde{q}$--$\chitwozero$ on-shell model 
in the $m(\tilde{q})$--$m(\tilde{\chi}^{0}_{2})$ plane. 
This is a simplified model with squark-pair production, where each squark decays into a quark and a neutralino, 
with the neutralino subsequently decaying into a $Z$ boson and an LSP with a mass of 1~\GeV.
In this model, exclusion is observed (expected) for squarks with masses below 1.3~\TeV\ (1.26~\TeV) for a $\tilde{\chi}^{0}_{2}$ mass of 900~\GeV.

Figure~\ref{fig:excl_SMGGN2} shows the expected and observed exclusion contours for the $\tilde{g}$--$\chionezero$ on-shell model in the $m(\tilde{g})$--$m(\tilde{\chi}^{0}_{1})$ plane, 
in which the produced gluinos follow the same decay chain as in the model above.
In this case the mass difference $\Delta m = m(\tilde{\chi}^{0}_{2})-m(\tilde{\chi}^{0}_{1})$ is set to 100~\GeV.
Overlaid on the figure is the observed limit from the previous analysis \cite{SUSY-2016-05}.
The sensitivity in the small $m(\tilde{g})-m(\tilde{\chi}^{0}_{1})$ difference regime is improved due to an optimisation of SRs including a change to define \HT\ only using jets, rather than also including leptons.

\begin{figure}[htbp]
\centering
\includegraphics[width=.8\textwidth]{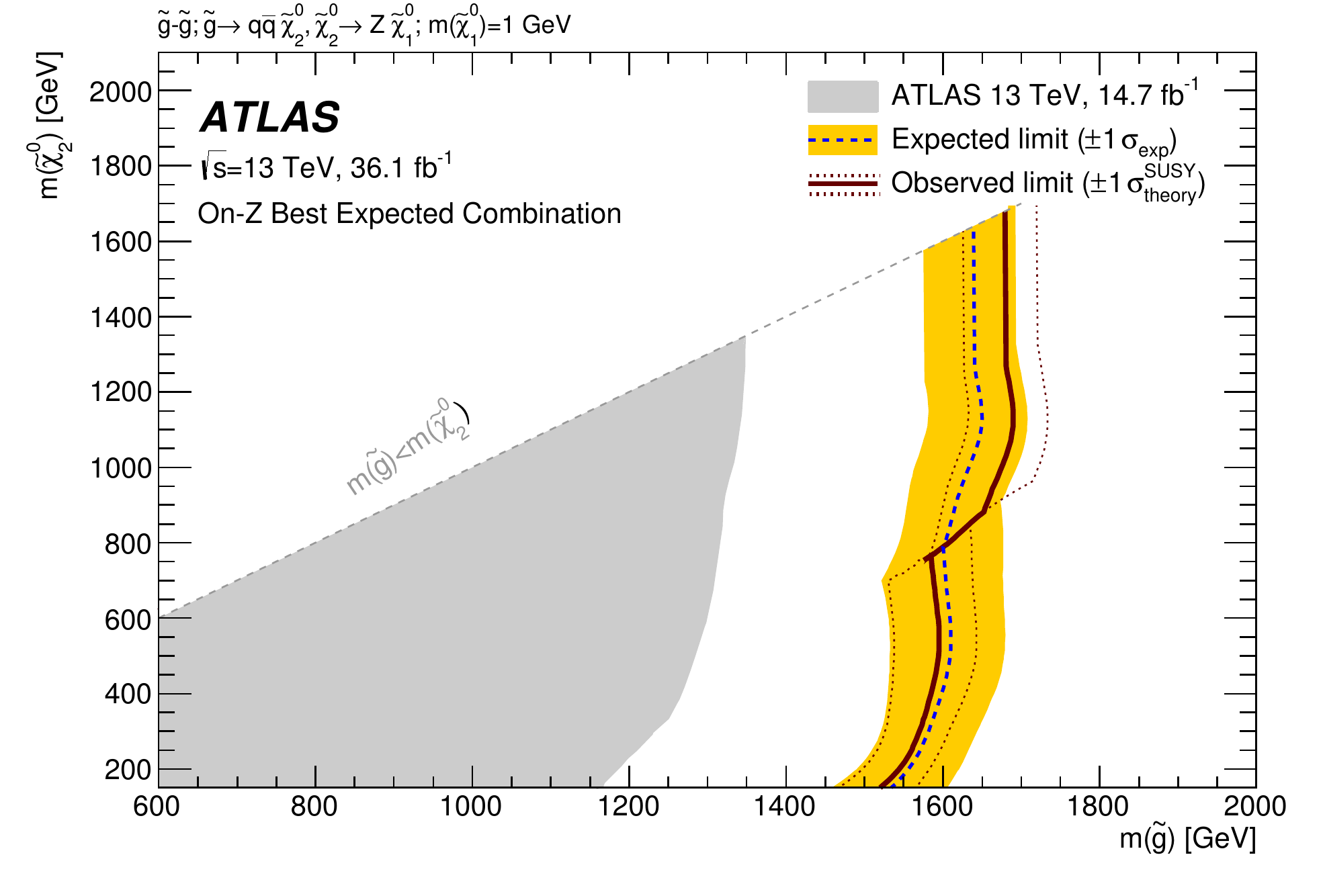}
\includegraphics[width=.8\textwidth]{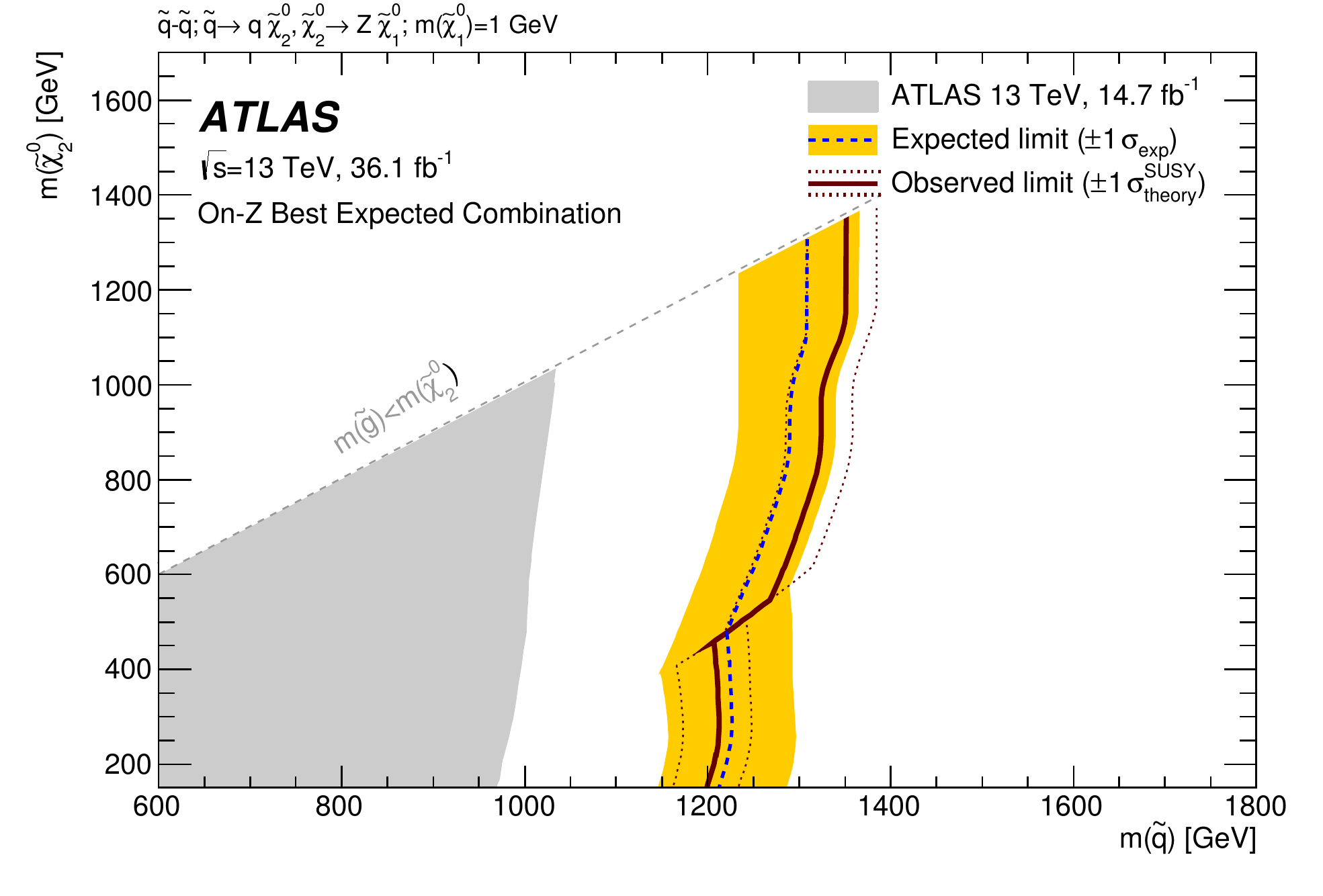}
\caption{
Expected and observed exclusion contours derived from the best-expected-sensitivity combination of results in the on-$Z$ \mll\ windows of SR-medium and SR-high 
for the (top) $\tilde{g}$--$\chitwozero$ on-shell grid and (bottom) $\tilde{q}$--$\chitwozero$ on-shell grid. 
The dashed line indicates the expected limits at $95\%$ CL and the surrounding band shows the $1\sigma$ variation of the expected limit as a consequence of the uncertainties in the background prediction and the experimental uncertainties in the signal ($\pm1\sigma_\text{exp}$). 
The dotted lines surrounding the observed limit contours indicate the variation resulting from changing the signal cross-section within its uncertainty ($\pm1\sigma^\text{SUSY}_\text{theory}$). 
The shaded area indicates the observed limit on this model from Ref.~\cite{SUSY-2016-05}. \label{fig:excl_SMGGN2_1}
}

\end{figure}

\begin{figure}[htbp]
\centering
\includegraphics[width=.8\textwidth]{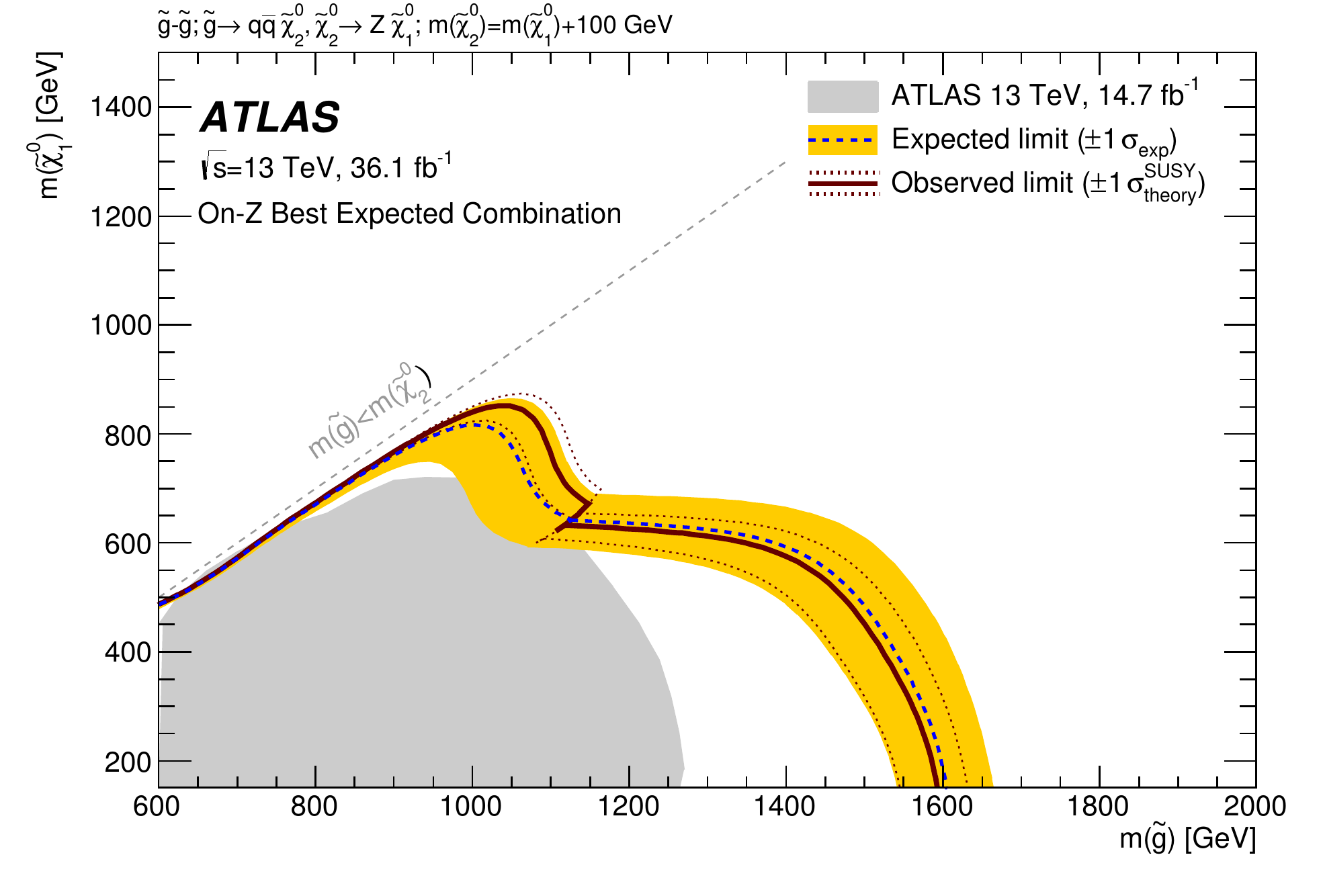}
\caption{
Expected and observed exclusion contours derived from the best-expected-sensitivity combination of results in the on-$Z$ \mll\ windows of SR-medium and SR-high 
for the $\tilde{g}$--$\chionezero$ on-shell grid. 
The dashed line indicates the expected limits at $95\%$ CL and the surrounding band shows the $1\sigma$ variation of the expected limit as a consequence of the uncertainties in the background prediction and the experimental uncertainties in the signal ($\pm1\sigma_\text{exp}$). 
The dotted lines surrounding the observed limit contour indicate the variation resulting from changing the signal cross-section within its uncertainty ($\pm1\sigma^\text{SUSY}_\text{theory}$). 
The shaded area indicates the observed limit on this model from Ref.~\cite{SUSY-2016-05}.
\label{fig:excl_SMGGN2}
}
\end{figure}

\FloatBarrier

\section{Conclusion}
\label{sec:conclusion}
This paper presents a search for new phenomena in final states containing a \SF\ \OS\ electron or muon pair,
jets and large missing transverse momentum using \lumi\ of $\sqrt{s}=13$~\TeV\ $pp$ collision data collected during 2015 and 2016 by the ATLAS detector at the LHC.
For the high-\pt\ and \lowpt\ searches combined, a set of 41 \mll\ windows are considered, 
with different requirements on \MET, \mttwo, \ptll\ and $\HT$, to be sensitive to signals with different kinematic endpoint values in the dilepton invariant mass distribution.
The data are found to be consistent with the Standard Model expectation.
The results are interpreted in simplified models of gluino-pair production and squark-pair production, and exclude gluinos (squarks) with masses as large as 1.85~\TeV\ (1.3~\TeV).
Models with mass splittings as low as 20~\GeV\ are excluded due to the sensitivity to compressed scenarios offered by the \lowpt\ SRs.

\section*{Acknowledgements}


We thank CERN for the very successful operation of the LHC, as well as the
support staff from our institutions without whom ATLAS could not be
operated efficiently.

We acknowledge the support of ANPCyT, Argentina; YerPhI, Armenia; ARC, Australia; BMWFW and FWF, Austria; ANAS, Azerbaijan; SSTC, Belarus; CNPq and FAPESP, Brazil; NSERC, NRC and CFI, Canada; CERN; CONICYT, Chile; CAS, MOST and NSFC, China; COLCIENCIAS, Colombia; MSMT CR, MPO CR and VSC CR, Czech Republic; DNRF and DNSRC, Denmark; IN2P3-CNRS, CEA-DRF/IRFU, France; SRNSFG, Georgia; BMBF, HGF, and MPG, Germany; GSRT, Greece; RGC, Hong Kong SAR, China; ISF, I-CORE and Benoziyo Center, Israel; INFN, Italy; MEXT and JSPS, Japan; CNRST, Morocco; NWO, Netherlands; RCN, Norway; MNiSW and NCN, Poland; FCT, Portugal; MNE/IFA, Romania; MES of Russia and NRC KI, Russian Federation; JINR; MESTD, Serbia; MSSR, Slovakia; ARRS and MIZ\v{S}, Slovenia; DST/NRF, South Africa; MINECO, Spain; SRC and Wallenberg Foundation, Sweden; SERI, SNSF and Cantons of Bern and Geneva, Switzerland; MOST, Taiwan; TAEK, Turkey; STFC, United Kingdom; DOE and NSF, United States of America. In addition, individual groups and members have received support from BCKDF, the Canada Council, CANARIE, CRC, Compute Canada, FQRNT, and the Ontario Innovation Trust, Canada; EPLANET, ERC, ERDF, FP7, Horizon 2020 and Marie Sk{\l}odowska-Curie Actions, European Union; Investissements d'Avenir Labex and Idex, ANR, R{\'e}gion Auvergne and Fondation Partager le Savoir, France; DFG and AvH Foundation, Germany; Herakleitos, Thales and Aristeia programmes co-financed by EU-ESF and the Greek NSRF; BSF, GIF and Minerva, Israel; BRF, Norway; CERCA Programme Generalitat de Catalunya, Generalitat Valenciana, Spain; the Royal Society and Leverhulme Trust, United Kingdom.

The crucial computing support from all WLCG partners is acknowledged gratefully, in particular from CERN, the ATLAS Tier-1 facilities at TRIUMF (Canada), NDGF (Denmark, Norway, Sweden), CC-IN2P3 (France), KIT/GridKA (Germany), INFN-CNAF (Italy), NL-T1 (Netherlands), PIC (Spain), ASGC (Taiwan), RAL (UK) and BNL (USA), the Tier-2 facilities worldwide and large non-WLCG resource providers. Major contributors of computing resources are listed in Ref.~\cite{ATL-GEN-PUB-2016-002}.

\printbibliography
\clearpage
 
\begin{flushleft}
{\Large The ATLAS Collaboration}

\bigskip

M.~Aaboud$^\textrm{\scriptsize 34d}$,
G.~Aad$^\textrm{\scriptsize 99}$,
B.~Abbott$^\textrm{\scriptsize 124}$,
O.~Abdinov$^\textrm{\scriptsize 13,*}$,
B.~Abeloos$^\textrm{\scriptsize 128}$,
S.H.~Abidi$^\textrm{\scriptsize 165}$,
O.S.~AbouZeid$^\textrm{\scriptsize 143}$,
N.L.~Abraham$^\textrm{\scriptsize 153}$,
H.~Abramowicz$^\textrm{\scriptsize 159}$,
H.~Abreu$^\textrm{\scriptsize 158}$
% ...
\bigskip
\\

% ...
$^{13}$Institute of Physics, Azerbaijan Academy of Sciences, Baku; Azerbaijan.\\
% ...
$^{34}$$^{(a)}$Facult\'e des Sciences Ain Chock, R\'eseau Universitaire de Physique des Hautes Energies - Universit\'e Hassan II, Casablanca;$^{(b)}$Centre National de l'Energie des Sciences Techniques Nucleaires (CNESTEN), Rabat;$^{(c)}$Facult\'e des Sciences Semlalia, Universit\'e Cadi Ayyad, LPHEA-Marrakech;$^{(d)}$Facult\'e des Sciences, Universit\'e Mohamed Premier and LPTPM, Oujda;$^{(e)}$Facult\'e des sciences, Universit\'e Mohammed V, Rabat; Morocco.\\
% ...
$^{99}$CPPM, Aix-Marseille Universit\'e, CNRS/IN2P3, Marseille; France.\\
% ...
$^{124}$Homer L. Dodge Department of Physics and Astronomy, University of Oklahoma, Norman OK; United States of America.\\
% ...
$^{128}$LAL, Universit\'e Paris-Sud, CNRS/IN2P3, Universit\'e Paris-Saclay, Orsay; France.\\
% ...
$^{143}$Santa Cruz Institute for Particle Physics, University of California Santa Cruz, Santa Cruz CA; United States of America.\\
% ...
$^{153}$Department of Physics and Astronomy, University of Sussex, Brighton; United Kingdom.\\
% ...
$^{158}$Department of Physics, Technion, Israel Institute of Technology, Haifa; Israel.\\
$^{159}$Raymond and Beverly Sackler School of Physics and Astronomy, Tel Aviv University, Tel Aviv; Israel.\\
% ...
$^{165}$Department of Physics, University of Toronto, Toronto ON; Canada.\\
% ...
$^{a}$ Also at Borough of Manhattan Community College, City University of New York, New York City; United States of America.\\
$^{b}$ Also at Centre for High Performance Computing, CSIR Campus, Rosebank, Cape Town; South Africa.\\
$^{c}$ Also at CERN, Geneva; Switzerland.\\
$^{d}$ Also at CPPM, Aix-Marseille Universit\'e, CNRS/IN2P3, Marseille; France.\\
$^{e}$ Also at Departament de Fisica de la Universitat Autonoma de Barcelona, Barcelona; Spain.\\
% ...
$^{*}$ Deceased

\end{flushleft}


\end{document}